\documentclass[11pt]{cernrep}
\usepackage{graphicx,epsfig}
\newcommand{\be}{\begin{equation}}
\newcommand{\ee}{\end{equation}}
\newcommand{\bea}{\begin{eqnarray}}
\newcommand{\eea}{\end{eqnarray}}

\def\lsim{\mathrel{\mathpalette\@versim<}}
\def\gsim{\mathrel{\mathpalette\@versim>}}
 \def\@versim#1#2{\lower0.2ex\vbox{\baselineskip\z@skip\lineskip\z@skip
       \lineskiplimit\z@\ialign{$\m@th#1\hfil##$\crcr#2\crcr\sim\crcr}}}
\catcode`@=12 

\newcommand{\ben}{\begin{enumerate}}
\newcommand{\een}{\end{enumerate}}

\def\frac#1#2{{{#1}\over {#2}}}
\def\gsim{\mathrel{\rlap{\lower4pt\hbox{\hskip1pt$\sim$}}
    \raise1pt\hbox{$>$}}}         
\def\lsim{\mathrel{\rlap{\lower4pt\hbox{\hskip1pt$\sim$}}
    \raise1pt\hbox{$<$}}}         

\newcommand{\draft}[1]{}

\def \n0{N_j^{(0)}}

\def\lapprox{\lower .7ex\hbox{$\;\stackrel{\textstyle <}{\sim}\;$}}
\def\gapprox{\lower .7ex\hbox{$\;\stackrel{\textstyle >}{\sim}\;$}}

\usepackage{graphicx}
\usepackage{afterpage}
\usepackage{epsfig,cite}
\usepackage{amssymb}
\usepackage{amsmath}
\usepackage{dsfont}
\usepackage{multirow}
\usepackage{url,hyperref}

\usepackage{latexsym}
\usepackage{wasysym}
\usepackage{bm}
\usepackage{rotating}
\usepackage{booktabs}

\newcommand{\eVdist}{\kern-0.06667em}
\newcommand{\gev}{{\,\text{Ge}\eVdist\text{V\/}}}
\newcommand{\tev}{{\,\text{Te}\eVdist\text{V\/}}}

\begin{document}

  \title{\boldmath
 The PDF4LHC report on PDFs and LHC data: \\[0.05cm] Results from Run I and
  preparation for Run II}

  \author{
    Juan Rojo$^1$,
    Alberto Accardi$^{2,3}$,
Richard D.~Ball$^{4,5}$,
Amanda Cooper-Sarkar$^6$,
Albert de Roeck$^{5,7}$,
Stephen Farry$^{8}$,
James Ferrando$^9$,
Stefano Forte$^{10}$,
Jun Gao$^{11}$,
Lucian Harland-Lang$^{12}$,
Joey Huston$^{13}$,
Alexander Glazov$^{14}$,
Maxime Gouzevitch$^{15}$,
Claire Gwenlan$^6$,  
Katerina Lipka$^{14}$,
Mykhailo Lisovyi$^{16}$,
Michelangelo Mangano~$^5$,
Pavel Nadolsky$^{17}$, 
Luca Perrozzi$^{18}$,
Ringaile Pla\v cakyt\. e$^{14}$,
Voica Radescu$^{16}$,
Gavin P. Salam$^{5}$\footnote{*On leave from CNRS, UMR 7589, LPTHE, F-75005, Paris, France}
~and
Robert Thorne$^{12}$
}

\institute{
 ~$^1$ Rudolf Peierls Centre for Theoretical Physics,
  University of Oxford,\\
  1 Keble Road, Oxford OX1 3NP, UK.\\
  ~$^2$ Hampton University, Hampton, Virginia 23668, U.S.A. \\
 ~$^3$ Jefferson Lab, Newport News, Virginia 23606, U.S.A.\\
 ~$^4$ The Higgs Centre for Theoretical Physics, University of Edinburgh,\\
JCMB, KB, Mayfield Rd, Edinburgh EH9 3JZ, Scotland.\\
~$^5$  PH Department, CERN, CH-1211 Geneva 23, Switzerland.\\
 ~$^6$ Particle Physics, Department of Physics,
University of Oxford,\\ 1 Keble Road, Oxford OX1 3NP, UK.\\
~$^7$ Antwerp University, B–2610 Wilrijk, Belgium.\\
~$^{8}$ Department of Physics,
University of Liverpool, L69 7ZE Liverpool, UK.\\
~$^9$ SUPA, School of Physics and Astronomy, University of Glasgow, Glasgow, G12 8QQ, Scotland.\\
~$^{10}$ TIF Lab, Dipartimento di Fisica, Universit\`a di Milano and \\
INFN, Sezione di Milano,
Via Celoria 16, I-20133 Milano, Italy.\\
~$^{11}$ High Energy Physics Division, Argonne National Laboratory,
Argonne, Illinois 60439, U.S.A.\\
~$^{12}$ Department of Physics and Astronomy, University College London, \\Gower Street, London WC1E 6BT, UK. \\
~$^{13}$ Department of Physics and Astronomy, Michigan State University,
East Lansing, MI 48824 U.S.A.\\
~$^{14}$ Deutsches Elektronen-Synchrotron (DESY), Notkestrasse 85, D-22607 Hamburg,
Germany.\\
~$^{15}$ Universit\'e de Lyon, Universit\'e Claude Bernard Lyon 1, CNRS-IN2P3,\\
Institut de
Physique Nucl\'eaire de Lyon, Villeurbanne, France.\\
~$^{16}$ Physikalisches Institut, Universit\"at Heidelberg, Heidelberg, Germany. \\
~$^{17}$ Department of Physics, Southern Methodist University,
Dallas, TX 75275-0181, U.S.A.\\
~$^{18}$ Eidgenoessische Technische Hochschule (ETH), H\"onggerberg
8093 Z\"urich, Switzerland.\\[0.3cm]
 {\bf OUTP-15-11P, LCTS/2015-14, GLAS-PPE/2015-01, DESY 15-088, JLAB-THY-15-2064, CERN-PH-TH-2015-150}
}

\maketitle

\clearpage

\begin{abstract}
  
The accurate determination of the Parton Distribution Functions (PDFs)
of the proton is an essential ingredient of the Large Hadron Collider (LHC)
program.
PDF uncertainties impact a wide range of processes,
from Higgs boson characterisation and precision Standard
Model measurements to New Physics searches.
A major recent development in modern PDF analyses has been to exploit
the wealth of new information contained in precision measurements from the LHC Run I, as well as 
progress in tools and methods to include these data in PDF fits. 
In this report we summarise the information that PDF-sensitive
measurements at the LHC have provided so far,
and review the prospects for further constraining PDFs
with data from the recently started Run II.
This document aims to provide useful input to the LHC collaborations
to prioritise their PDF-sensitive measurements at Run II, as
well as a comprehensive reference for the PDF-fitting collaborations.

\end{abstract}

\tableofcontents

\clearpage
\section{Introduction and motivation}

The initial state of hadronic collisions is the domain of the
parton distribution functions (PDFs) of the
 proton, see Refs.~\cite{Forte:2010dt,Forte:2013wc,Perez:2012um,Ball:2012wy,Watt:2011kp,DeRoeck:2011na,Jimenez-Delgado:2013sma} for
recent reviews.
Accurate PDFs are an essential ingredient for LHC phenomenology: 
PDF uncertainties 
limit the ultimate accuracy
of the Higgs boson couplings extracted from
LHC measurements~\cite{Dittmaier:2011ti,Watt:2011kp},
degrade the reach of searches for massive new BSM particles at the 
TeV scale~\cite{AbelleiraFernandez:2012cc,Borschensky:2014cia}
and are the dominant 
systematic uncertainties in the determination of fundamental
parameters such as the $W$ boson mass or $\sin^2\theta_{\rm eff}$,
key ingredients for global stress-tests
of the Standard Model~\cite{Bozzi:2011ww,ATL-PHYS-PUB-2014-015,Bozzi:2015hha,Aad:2015uau,Baak:2014ora}.
Because they are non-perturbative objects,
although their scale dependence is
determined by the perturbative DGLAP evolution equations,
they need to be extracted from global fits to hard-scattering data.
Various PDF fitting collaborations provide regular updates
of their QCD analysis: some of the latest PDF releases include
{\sc ABM12}~\cite{Alekhin:2013nda}, {\sc CT14}~\cite{Dulat:2015mca},
CJ12~\cite{Owens:2012bv},
GR14~\cite{Jimenez-Delgado:2014twa},
{\sc HERAPDF2.0}~\cite{Abramowicz:2015mha}, 
{\sc MMHT14}~\cite{Harland-Lang:2014zoa} and {\sc NNPDF3.0}~\cite{Ball:2014uwa}.

A major recent development in PDF fits has been the inclusion of
a wide variety of LHC data.
Some of the LHC processes that are now used 
were already part of global PDF fits, mostly measured at the Tevatron, but the
LHC data open a wider
new kinematical range, 
as in the case of
jet production~\cite{Watt:2013oha,Aad:2013lpa,Chatrchyan:2012bja} 
and inclusive electroweak boson production~\cite{Chatrchyan:2013mza,Aad:2012sb}.
On the other hand, other types of processes have only become available for PDF fits after 
their measurement at the LHC, like isolated direct photon production~\cite{d'Enterria:2012yj}, 
$W$ production in association with charm quarks~\cite{Chatrchyan:2013uja,Aad:2014xca,Alekhin:2014sya},
top quark pair production~\cite{Czakon:2013tha,Guzzi:2014wia}, open charm and bottom production in proton-proton collisions~\cite{Zenaiev:2015rfa,Aaij:2013mga}, low- and high-mass Drell-Yan production~\cite{Chatrchyan:2013tia,Aad:2013iua} and $W$ and $Z$ production in association with jets~\cite{Malik:2013kba,Farry:2015xha}, among  others.
Remarkably, some of these processes open completely new avenues for PDF fits:
 data on $W$+$c$ production provides
 a clean handle on the strange PDF~\cite{Stirling:2012vh,Chatrchyan:2013uja,Alekhin:2014sya,Aad:2014xca} complementary to that of the low energy neutrino data~\cite{Mason:2007zz}, top quark-pair production 
 allows for improved constraints on the large-$x$ gluon complementary to jets~\cite{Czakon:2013tha,Guzzi:2014wia,Beneke:2012wb} and forward heavy-flavour production probes the gluon distribution at
 small-$x$~\cite{Zenaiev:2015rfa,Gauld:2015yia}.

 The fact that the LHC provides measurements at  different center-of-mass energies
 also allows one to construct novel observables with
 useful PDF sensitivity,
 the ratios and double ratios of cross-sections at different values of
 $\sqrt{s}$
 where several theoretical and experimental uncertainties cancel~\cite{Mangano:2012mh}.
 This concept has already been validated by measurements by
  ATLAS of the ratio
  of jet cross-sections between 7 and 2.76 TeV~\cite{Aad:2013lpa}, and by CMS of 
  the ratio of Drell-Yan cross-sections
  between 8 and 7 TeV~\cite{CMS:2014jea}.
  Other similar ratios, this time between 13 and 8 TeV, will become
  possible with the availability of Run II data.
  Another important example of the relevance of LHC data for parton
  distributions is given
  by the close interplay between PDFs and the tunes of soft and
  semi-hard QCD models in the context of LO and NLO event generators,
  where LHC measurements have been shown to provide invaluable 
  constraints~\cite{Skands:2014pea,ATL-PHYS-PUB-2014-021}.
  Finally, PDFs are an important
  source of theoretical systematic uncertainty to the extraction
  of Standard Model parameters at the LHC, for instance the recent
  direct measurements of the strong coupling constant in the
  TeV region~\cite{Khachatryan:2014waa,Chatrchyan:2013txa}.

  Another important development for
  PDF studies in the recent years has been the availability of
PDF fits being carried out also by
the ATLAS and CMS collaborations themselves.
Thanks to the know-how acquired from the HERA data analyses,
and the availability of a public tool for PDF fits,
{\sc\small HERA\-fitter}~\cite{Alekhin:2014irh}, both collaborations
have developed an extensive program
of PDF determinations from their
own measurements~\cite{Chatrchyan:2013mza,Aad:2013lpa,Aad:2012sb,Khachatryan:2014waa}.
The aim of these studies is not to provide an alternative to global fits,
but rather to 
study the constraining power of 
their 
new measurements
on PDFs, ensure that all information in correlated
systematics is suitably provided, and to perform checks of the data
prior to publication using the QCD analysis as a diagnostic toolbox.
  In addition, the {\sc\small HERAfitter} developers are also performing
  a number of independent PDF studies~\cite{Camarda:2015zba,::2014uva},
  providing useful input to the PDF community.

Given the importance of the LHC data in modern global PDF fits,
and the recent restart of the LHC at 13 TeV, it is now timely
to summarize what have we learned for PDFs from Run I data, and
to set the stage for the corresponding measurements at Run II.
One of the aims of this document is thus to review the constraints on PDFs that measurements at the Large Hadron Collider during the
Run I data-taking have provided.
With this motivation, we summarize the
relevant measurements
from ATLAS, CMS and LHCb,
and discuss the available phenomenological studies that quantify the
information on PDFs provided by these datasets.
Then we move on to discuss the prospects for PDF-sensitive analysis
at Run II.
We explore how the increase in center-of-mass energy and luminosity
can provide new opportunities for PDF studies beyond those available
at Run I.
We also quantify some of the constraints on PDFs that Run II
can provide by means of a profiling analysis using
$W$, $Z$ and $t\bar{t}$ simulated pseudo-data as input,
as well as study the impact of Run II inclusive jet data
in the CT framework.

The outline of this document is the following.
In Sect.~\ref{sec-pdffits} we summarize the status of some of the latest PDF releases,
with the emphasis on the role of LHC data.
In Sect.~\ref{sec:overview} we review recent studies that have quantified the PDF sensitivity 
of LHC measurements.
In Sect.~\ref{sec:runI} a more detailed overview of the relevant measurements and their
corresponding  constraints on PDFs provided by the ATLAS, the CMS and the LHCb collaborations 
during Run I is given.
In Sect.~\ref{sec:prospects}, we discuss the prospects for PDF-sensitive measurements in
Run II, including  a profiling analysis using $W$, $Z$ and $t\bar{t}$ simulated  pseudo-data, and a study of inclusive jet production in the CT global
analysis framework.
In Sect.~\ref{sec:recommendations} we present practical
recommendations for the presentation and delivery
of LHC measurements to be used in global PDF analysis.
We conclude in Sect.~\ref{sec:outlook} with an outlook on
the program of constraining PDFs with LHC data for the
coming years.

This report summarizes the discussions that have taken place at various forums, in particular 
at the regular PDF4LHC meetings, during the last months.
It is also indebted to the productive discussions that
took place at the {\it{``Parton Distributions for the LHC''}} workshop that took place between 
the 15th and the 22nd of February 2015 at the Benasque Center for Science 
{\it Pedro Pascual}.\footnote{\url{http://benasque.org/2015lhc/}}

We would like to mention that,
in parallel with the studies summarized in this report, an update 
of the benchmark comparisons between different PDF sets, as well as between different 
methods to combine them~\cite{Gao:2013bia,Carrazza:2015aoa,Carrazza:2015hva}, is also being performed, 
with the aim of updating the current PDF4LHC recommendations~\cite{Alekhin:2011sk,Botje:2011sn} for
PDF usage at the LHC Run II.
The results of these benchmark comparisons will be presented in a
separate report.

\section{PDF analysis at the dawn of the LHC Run II}
\label{sec-pdffits}

We begin this document with
a succinct review of the status of PDF fits at the dawn of the Run II of the LHC.
This is specially timely since most PDF groups have provided major updates of their fits in time to be used along the
Run II data in both theory predictions and in Monte Carlo simulations used in the data analysis.
In this section we summarize the recent developments of these various groups,
and emphasize the role that LHC data plays on each of these analyses.
The reader is encouraged to consult the original publications
for additional information about the updated PDF fits.
All the PDF sets discussed below are available from the {\tt LHAPDF6}~\cite{Buckley:2014ana} interface.
Further, in this section we also review the development of new tools for PDF analysis.

While an extensive comparison between these
updated PDF sets will be presented in the companion
PDF4LHC recommendations paper, here
for completeness we also show
for illustrative purposes some
comparisons between recent PDF
sets and the corresponding parton luminosities.

\subsection{CT14}

CT14 provides parton distribution functions at LO, NLO and NNLO~\cite{Dulat:2015mca}. These global PDF fits include LHC data for the first time, 
from ATLAS, CMS and LHCb, to go along with the data sets used for CT10~\cite{Gao:2013xoa}.
One important, recently released data set from the Tevatron has also been added, 
that of the D0 $W$ electron asymmetry measurement using the full Run 2 data sample~\cite{D0:2014kma}.
This data set provides important constraints on $u$ and $d$ quarks at high $x$. 
From the LHC experiments, the chosen data sets are vector boson ($W$,
$Z$) production cross sections and asymmetries,   
for which NNLO predictions are available; and inclusive jet cross sections, for which the complete NNLO calculation 
is not yet available, but the estimated impact of the NNLO
contributions is small compared to the current experimental uncertainties.
The 7 TeV LHC $W$ and $Z$ data allow us to perform better separation of $u$ and $d$ (anti-)quark PDFs at $x \sim 0.02$, 
and also provide an independent constraint on the strangeness PDF, $s(x,Q)$. The LHC jet cross sections have a potential to probe
the gluon PDF in a much wider $x$ than the Tevatron ones.

There are a total of 2947 data points included in the NNLO fit, with data from 33 experiments.
FastNLO and Applgrid interfaces~\cite{Carli:2010rw,Wobisch:2011ij}
have been used for quick calculations of NLO matrix elements in the global fits, supplemented by NNLO $K$-factors (for the NNLO fit).
ResBos~\cite{Balazs:1995nz,Balazs:1997xd,Landry:2002ix,Guzzi:2013aja} has been used for the calculation of the NNLO $K$-factors for $W/Z$ and $W$ asymmetry data. 

The PDF parametrization is more flexible than that of CT10. The PDFs
are expressed as a linear combination of Bernstein polynomials, having an advantage
that each basis polynomial 
peaks within a single $x$ region. This serves for reduction of correlations among the parameters. PDF error sets with a total of 28 eigenvectors 
are provided at both NLO and NNLO. Correlated systematic errors from the experiments are included, when available, 
and have impact on some properties, such as the gluon PDF at $x>0.1$.
A central value of $\alpha_S(m_Z^2)$ of 0.118 has been assumed in the
global fits at NLO and NNLO, and the PDF sets at alternative values of
$\alpha_S(m_Z^2)$ in an expanded range are also provided. Similar to the CTEQ6 analysis~\cite{Pumplin:2002vw}, two versions of the LO PDFs are supplied,  
one with 1-loop  evolution of $\alpha_S$, with an input value of $\alpha_S(m_Z^2)=0.130$; and the other with 2-loop evolution and $\alpha_S(m_Z^2)=0.118$.

In general, the CT14 PDFs are similar to those from CT10, albeit with a somewhat smaller strange-quark distribution and a softer gluon at high $x$. 
Furthermore, CT14 and CT10 differ in the $u$ and $d$ quark distributions at moderate to large $x$, due to the inclusion of new data, both from the LHC and from the Tevatron, and new parametrization forms. Particular attention is paid to the behavior of the $d/u$ and $\bar u/\bar d$ ratios in the limit as $x$ approaches 1. 
The CT14 parameterizations are more likely to predict finite constant values for $d/u$ and $\bar u/\bar d$ as $x\rightarrow 1$, besides the limits of 
zero or infinity that were preferred with the previous parametrization choices.  [This change affects only extrapolations to 
very large $x$ values that are not covered by the data.] At large $x$, $u_v(x)$ and $d_v(x)$ both vary as $(1-x)^{a_2}$,  
with the same value of $a_2$ for both (but allowing for different normalizations). This is consistent with the expectations of spectator counting rules.

\subsection{CTEQ-JLAB (CJ12)}

The CTEQ-Jefferson Lab (CJ) global PDF fits are based on the world
data on charged lepton DIS on proton and deuterium targets (including recent Jefferson Lab data),  lepton pair production with a
proton beam on proton and deuterium targets, $W$ asymmetries in
$\overline p p$ collisions, and jet production data from the Tevatron.
For DIS data, the fits include subleading ${\cal O}(1/Q^2)$ corrections, such as target mass and higher twist effects, and nuclear corrections for
the deuterium target data.
The CJ fits
incorporate data down to an invariant final state mass of $W^2 = 3$~GeV$^2$, weaker than the typical $\sim 12$ GeV$^2$ cut considered in the CT14, MMHT14, and NNPDF3.0 analysis, and therefore including much more DIS data from SLAC and Jefferson Lab extending to higher $x$ values.
The resulting fits~\cite{Accardi:2009br, Accardi:2011fa, Owens:2012bv}
have culminated in the release of the CJ12 PDF sets, valid in the
range $10^{-5} \lesssim x \lesssim 0.9$ and available on the CJ collaboration
web page\footnote{\url{http://www.jlab.org/cj}} as well as through the {\tt LHAPDF6} interface.
The fits were performed at next-to-leading order (NLO) in the zero
mass variable flavor number scheme, with $\alpha_S$ fixed to the
world average value. Full heavy quark treatment and fits of the strong coupling constant will be included in the upcoming CJ15 PDF release, and fits of the relevant LHC data will be considered in a subsequent analysis.

The CJ PDFs have been shown to be stable with the weaker cuts on
$W$ and $Q^2$, and the increased DIS data sample (of about 1000
additional points) has led to significantly reduced uncertainties,
up to $\sim$50\% on the $d$ quark PDF at large $x \gtrsim 0.6$,
where precise data have otherwise been until recently scarce~\cite{Accardi:2009br}. 
Since a precise $d$ quark flavor separation at high $x$ depends largely on DIS on deuterium targets, corrections for
nuclear Fermi motion and binding effects are included by convoluting
the nucleon structure functions with a smearing function computed
from the deuteron wave function.  As the $u$ quark is well
constrained by data on proton targets, the $d$ quark becomes
directly sensitive to the nuclear corrections.
The effect is a large suppression at high $x$, and a mild but
non-negligible increase at intermediate $x$~\cite{Accardi:2009br},
still inside the ``safe'' region defined by the larger $W$ cut
discussed above.  These findings have subsequently been confirmed 
by Ball {\it et al.}~\cite{Ball:2013gsa} and Martin {\it et al.}~\cite{Harland-Lang:2014zoa}.

The uncertainties on the $d$ quark PDF from theoretical modeling of
nuclear corrections (which we refer to as ``nuclear uncertainties'')
have been quantified in Refs.~\cite{Accardi:2011fa, Owens:2012bv}.
These range from mild, corresponding to the hardest of the deuteron
wave functions (WJC-1) coupled to a 0.3\% nucleon off-shell
correction, to strong, corresponding to the softest wave function
(CD-Bonn) and a large, 2.1\% nucleon off-shell correction; the
central value corresponds to the AV18 deuteron wave function with
a 1.2\% off-shell correction~\cite{Owens:2012bv}.  The resulting PDFs
are labeled ``CJ12min'', ``CJ12max'', and ``CJ12mid'', respectively.
This analysis demonstrates the usefulness of the deuterium data, even in
the presence of the nuclear uncertainties that its use introduces.

A further source of theoretical uncertainty was investigated in
Refs.~\cite{Accardi:2011fa, Owens:2012bv}, where a more flexible
parametrization was used for the valence $d_v$ quark at large-$x$,
with an admixture of the valence $u_v$ PDF,
\be
  d_v(x) \rightarrow d'_v(x) = 
    a_0^{d} \Big[ d_v(x) / a_0^{d} + b\, x^c u_v(x) \Big]\, ,
\label{eq:newd}
\ee
where $a_0^{d}$ is the $d$ quark normalization, and $b$ and $c$ are
two additional parameters. The result is that the $d/u$ ratio at
$x \to 1$ can now span the range $[0,\infty]$ rather than being 
limited to either 0 or $\infty$ as in all previous PDF fits.
A finite, nonzero value of this ratio is in fact expected from
several non-perturbative models of nucleon structure
\cite{Melnitchouk:1995fc, Holt:2010vj, Jimenez-Delgado:2013sma}.
It is also required from a purely practical point of view because
it avoids potentially large parametrization biases on the fitted $d$ quark PDF, as explained in more detail in
Ref.~\cite{Accardi:2013pra}.
An analogous extended $d$-quark parametrization has been more recently considered also in the
CT14 fits~\cite{Dulat:2015mca}.

The ratios of the $d$ to $u$ PDFs for the three CJ12 sets are constrained up to
$x \approx 0.8$ by the enlarged data set,
and when extrapolated to $x=1$ give the limiting value
\be
  d/u \,\xrightarrow[\,x\rightarrow 1\,]{} \, 0.22
	\pm 0.20 \, \text{({\small PDF})}
	\pm 0.10 \,\text{(nucl)}\, ,
\ee
where the first error is from the PDF fits and the second is from
the nuclear correction models.
These values encompass the full
range $\approx 0-0.5$ of available theoretical predictions~\cite{Melnitchouk:1995fc, Holt:2010vj}.  
The impact of the very recent high-precision data on $W \to e+\nu_e$ and reconstructed $W$ asymmetry from the D0 collaboration is being investigated in the context of the CJ15 fits, where the off-shell corrections are fitted to data (instead of being a-priori selected in a theoretically reasonable range) resulting in a further substantial reduction of the nuclear and statistical uncertainty.

The impact and relevance of precise large-$x$ quark PDFs on forward rapidity observables at Tevatron and LHC, as well as on production of large mass particles, has been studied by Brady {\it et al.}~\cite{Brady:2011hb}.
For example, the nuclear uncertainty from the CJ12 analysis becomes relevant for $W$ production at rapidity greater than 2 at the Tevatron, and greater
than 3 at the LHC.
For particles of heavier mass, such as the
putative $W'$ and $Z'$ bosons predicted in scenarios beyond the
standard model, the production cross section becomes sensitive to higher-$x$ PDFs and the nuclear
uncertainty may become larger than 20\% above the lower mass limit of $\approx 2.5$~TeV set by recent LHC data.
This illustrates how nuclear and other 
large-$x$ theoretical uncertainties may significantly affect the
interpretation of signals of new particles and the determination
of their properties, which requires a precise calculation of
background QCD processes, and motivates further dedicated efforts (such as the upcoming CJ15 analysis) to reduce these
uncertainties as much as possible.
Conversely, measurements of large rapidity observables at LHCb can have an impact on the determination of large-$x$ PDFs, as discussed
in Sects.~\ref{sec:overview} and~\ref{sec:lhcbconstraints} in this report.

\subsection{HERAPDF2.0}

The HERAPDF fits are based only on data collected at the HERA $ep$ collider. 
During HERA-I and HERA-II running approximately 1 fb$^{-1}$ of data were 
collected divided roughly equally between $e^+ p$ and $e^- p$ scattering.
All of the published measurements on inclusive neutral-current (NC) and charged-current (CC)
scattering have now been combined into a single coherent data set taking 
into account correlated systematic
uncertainties~\cite{Abramowicz:2015mha}.
This combination includes 
data taken at different proton beam energies,
$E_p = 920, 820, 575, 460$ GeV.
The data cover the ranges $ 6\times 10^{-7} < x < 0.65$, $ 0.045 < Q^2 < 50,000$
GeV$^2$. The combination led to significantly reduced uncertainties, 
below $1.5\%$ over the kinematic range $3 < Q^2 < 500$~GeV$^2$. 
This combination supersedes the previous combination of only HERA-I data~\cite{Aaron:2009aa}.

The availability of precision NC and CC data over such a large kinematic range 
allows the extraction of PDFs using only $ep$ data without the need for heavy 
target corrections. The difference between the NC $e^+p$ and $e^-p$ cross 
sections at high $Q^2$ constrains the valence PDFs. The high-$x$ CC data 
separate valence quark flavours. The CC $e^-p$ data allow the extraction of 
the $d_v$ PDF without assuming strong isospin symmetry. The lower-$Q^2$
 NC data constrain the sea PDF directly and through their scaling violations 
they constrain the gluon PDF. A further constraint on the gluon comes from 
the data at different beam energies, which probe the longitudinal structure 
function $F_L$.
The HERAPDF2.0 is based on the new data combination~\cite{Abramowicz:2015mha} and 
supersedes HERAPDF1.0~\cite{Aaron:2009aa} and 1.5 which were based on previous partial combinations.
The HERAPDF2.0 is available at LO, NLO and NNLO on {\tt LHAPDF6}.
The experimental uncertainties are presented as 14 pairs of Hessian eigenvectors evaluated by the standard criterion 
of $\Delta\chi^2=1$. For the NLO and NNLO PDFs, 13 further variations are supplied to cover uncertainties due to model 
assumptions and assumptions on the form of the parametrization. For the NLO and NNLO PDFs the standard value of 
$\alpha_S(m_Z^2) = 0.118$ but the PDFs are also supplied for values of $0.110 < \alpha_S(m_Z^2) < 0.130$ in steps of $0.001$.
Needless to say, this final HERA inclusive combination will be the backbone of all
future PDF analyses, similarly as the HERA-I combination is the backbone of all available
modern PDF sets.

Several further variations of the HERAPDF2.0 are also supplied: HERAPDF2.0HiQ2 for which only data 
with $Q^2 >10$ GeV$^2$ are used to avoid possible bias from low-$x$, low-$Q^2$ effects; HERAPDF2.0AG
for which the gluon takes form which is imposed to be positive definite for all $x$ for which 
$Q^2> 3.5$ GeV$^2$; HERAPDF2.0FF3A and FF3B, which use two different versions of the Fixed Flavour 
Number Scheme for heavy quarks; and finally HERAPDF2.0Jets which uses additional HERA data 
on jet production as well as the HERA combined charm data. The charm data mostly serve to 
constrain the uncertainty on the charm-quark mass parameter and this information is already used 
in the main HERAPDF2.0 PDFs, whereas the jet data put further constraints on the gluon PDF, such that
a simultaneous fit for $\alpha_S(m_Z^2)$ and the PDFs can be performed, resulting in a competitive 
determination of $\alpha_S(m_Z^2)$.

Let us briefly discuss also the prospects of future PDF-sensitive
measurements from HERA.
To begin with, the legacy HERA data on charm and beauty
structure functions will be combined to provide further constraints in the heavy flavour sector
and on the small-$x$ gluon PDF.
In addition, the final data on prompt photon production will provide information about
QED contributions to PDFs, in particular thanks to
photon-initiated processes~\cite{Martin:2004dh,Martin:2014nqa}.
Also, the legacy HERA data on vector meson production and diffractive di-jet production will
elucidate physics at low-$x$ and thus address the question as to whether it is
appropriate to include such data in PDF fits using the current conventional DGLAP formalism, or
if instead a BFKL type of approach is necessary.

\subsection{MMHT2014}

The MMHT2014 PDF sets were released in December 2014~\cite{Harland-Lang:2014zoa}.
They are 
the first major update based on this framework since the MSTW2008
PDFs~\cite{Martin:2009iq}. However, the updates incorporate the improvements
to the parametrization and deuteron corrections already
presented in Ref.~\cite{Martin:2012da}. 
This study showed that the new parameterizations, which use Chebyshev
polynomials in $(1-2\sqrt{x})$ rather than simple powers of $\sqrt{x}$
and up to 7 free parameters for a particular PDF, can reproduce functions 
obtained from
a much greater number of parameters up to a small fraction of percent over 
a wide range in $x$. The more flexible deuteron corrections improve the fit 
quality and result in a shape similar to the models in {\it e.g.}~\cite{Owens:2012bv}.
The new PDF sets also use the optimal variable flavour 
number scheme of~\cite{Thorne:2012az}, updated heavy nucleus corrections 
\cite{deFlorian:2011fp}, a modified central value and uncertainty 
for the branching ratio $B_{\mu}=B(D\to \mu)$ used in the determination of 
the strange quark from dimuon data, and use the multiplicative rather than 
additive definition for correlated systematic uncertainties~\cite{Ball:2009qv}.

The data used in the fits have been very significantly updated from
the MSTW2008 analysis, with relevant data sets published before the 
beginning of 2014 included, as summarized in Sect.~\ref{sec:runI}.
In particular, the combined HERA total cross section data 
\cite{Aaron:2009aa} and combined charm data~\cite{Abramowicz:1900rp} are now
used, along with some updates on Tevatron $W$ production. Moreover, a variety 
of LHC data, including $W,Z$ and $\gamma^{\star}$ data from ATLAS, CMS and
LHCb, inclusive jet data from ATLAS and CMS and total top quark-pair production
from ATLAS and CMS (and the Tevatron combined result~\cite{Aaltonen:2013wca}) 
have now been included. 
Although they have 
not been used to determine the PDFs, data on $W$$+$$c$ production and differential 
top quark-pair production have been checked against the QCD predictions using these PDFs and give good 
agreement. NLO calculations are produced for LHC data using~\cite{Carli:2010rw,Wobisch:2011ij} and
$K$-factors employed at NNLO. The LHC inclusive jet data are currently not 
used at NNLO, leading to a very slight increase in the uncertainty for the 
high-$x$ gluon at NNLO compared to NLO. 

The resulting PDFs are made available together with 25 eigenvector pairs of uncertainties given 
at $68\%$ confidence level at LO, NLO and NNLO and correspond to $\alpha_S(m_Z^2))$ values of 0.130 at LO, 0.118 and 0.120
at NLO and 0.118 at NNLO. The increase in the number of eigenvectors from 
20 in the MSTW2008 sets is related to the increased flexibility of the PDFs, 
partially made possible by extra constraints coming from new LHC processes. 
The value of $\alpha_S(m_Z^2))$ is left free in fits in the first instance, resulting in 
best fits near 0.135 at LO, 0.120 at NLO and 0.118 at NNLO. Therefore, the 
choice of the values for the eigenvector sets, though 0.118 is available as
well as 0.120 at NLO since this is close to the world average for 
$\alpha_S(m_Z^2))$, and a set with this value may be required by users. Each 
eigenvector set is accompanied by a central set with $\alpha_S(m_Z^2))$ values
with $\pm 0.001$ in order to enable uncertainties due to variations of $\alpha_S(m_Z^2))$
to be calculated.

A dedicated study about the uncertainties in $\alpha_S(m_Z^2))$ in the MMHT14 analysis has been
presented in~\cite{Harland-Lang:2015nxa}, and the corresponding sets with a
wide variety of $\alpha_S(m_Z^2))$ values have also been released.
PDF sets with a variety of values of charm and bottom mass values will also
follow soon, as well as PDF sets in  the three and four flavour schemes.
The MMHT2014 PDFs generally give similar results for the LHC observables as the MSTW2008 PDFs, and have comparable uncertainties. 
The main change in the MMHT2014 PDFs is in the small-$x$ valence quarks, related to the improved parameterization and deuteron 
corrections, and an increase in the uncertainty (and to a lesser extent the central value) of the strange quark. The latter is 
due to the quite generous uncertainty allowed on $B_{\mu}=B(D\to \mu)$ and an extra free parameter for 
the strange quark contributing to the eigenvectors. This combination of extra freedom in the strange PDF is then given an 
extra constraint by the LHC $W$- and $Z$-boson production data.

\subsection{NNPDF3.0}

The NNPDF3.0 sets were released in October 2014~\cite{Ball:2014uwa}.
As compared previous NNPDF global analysis~\cite{Ball:2008by,Ball:2009mk,Ball:2010de,Ball:2011mu,Ball:2011uy,Ball:2012cx}, NNPDF3.0
is the result of a extensive redevelopment of the NNPDF code, including constraints from new experimental data, theoretical calculations of new processes, and a major code re-organization. Furthermore, NNPDF3.0 is the first set of global PDFs with a fitting methodology validated through a closure test.

Regarding experimental data, NNPDF3.0 includes the fixed target, HERA, Tevatron and LHC data already included in NNPDF2.3, and in addition all the published HERA-II data from H1 and ZEUS, and a wide range of more recent ATLAS, CMS and LHCb data on jet production, weak boson production and asymmetries, Drell-Yan, $W$+charm and top quark pair production. A total of 4276 data points are fitted at NLO and 4078 at NNLO. The complete list of LHC measurements that have been included in NNPDF3.0 is summarized in Sect.~\ref{sec:runI}.

Concerning theory calculations, all collider processes have been computed using fast
NLO interfaces~\cite{amcfast,Wobisch:2011ij,Carli:2010rw}, supplemented by NNLO and electroweak $K$-factors when required.
Inclusive jets are treated at NNLO using the approximate threshold
calculation~\cite{deFlorian:2013qia,Carrazza:2014hra}, validated
on the exact calculation in the $gg$ channel~\cite{Ridder:2013mf}.
Heavy quark mass effects are computed in the FONLL General-Mass variable-flavor number scheme~\cite{Forte:2010ta}, with
the main difference being that at NLO it is the FONLL-B scheme that is used, rather than FONLL-A as previously, since this 
provides a better description of the low-$Q^2$ charm production data.

All the fitting code has been rewritten from {\sc\small Fortran} to {\sc\small C++} and
{\sc\small Python}, making it robust and modular, so that the modification of the theoretical calculations, the addition of new datasets, or the generation of entirely new sets of pseudo-data for use in closure testing, can be done easily and quickly with no need to modify the rest of the code. Improved positivity constraints and dynamical preprocessing exponents have also been implemented.

As both data and theory improve, it becomes even more necessary to ensure that the fitting methodology is consistent and unbiased. To this end the NNPDF methodology has now been subjected to a closure test~\cite{Ball:2014uwa}. This is performed by generating pseudodata based on an assumed prior PDF (for example MSTW08) and a particular theory (for example NLO perturbative QCD with given $\alpha_S$, heavy quark scheme, etc). To make the test as realistic as possible, the pseudodata are generated using the experimental uncertainties of the current global dataset. In the context of the closure test, the pseudodata are
`perfect': in particular they are fully consistent both with each other and with the assumed theory.
A full fit to the pseudodata is thus a rigorous test of the fitting methodology: fitted PDFs should have manifestly unbiased central values, and statistically meaningful uncertainties and correlations (so that for example the fitted PDF agrees with the assumed prior at one sigma 68\% of the time). 

The success of the NNPDF3.0 closure test proves that the NNPDF3.0 PDF sets fitted to real data are unbiased and have statistically meaningful uncertainties and correlations. This in turn confirms that most modern datasets are consistent (both internally and in the context of the global dataset), in the sense that their systematic errors have been sensibly estimated, and furthermore that NNLO QCD is sufficient to describe the global dataset within a common universal framework. In particular the NNPDF3.0 fits to real data show no sign of tension between deep inelastic and hadronic data.

The NNPDF3.0 PDFs are available on {\tt LHAPDF6} as sets of 100 replicas. Baseline fits are available for $\alpha_S(m_Z)=0.118$ at LO, NLO and NNLO, and also at $\alpha_S(m_Z)=0.115,0.117,0.119,0.121$ at NLO and NNLO. A LO fit with $\alpha_S(m_Z)=0.130$ is also provided. The baseline fits are provided with 5 active flavours: alternative fits with $N_f = 3,4,6$ active flavours are also available. NNPDF also provide fits to reduced datasets, for studies by the LHC experimental collaborations: HERA-only, HERA+ATLAS, HERA+CMS, no-LHC, and no-jets. All these are available at NLO and NNLO, for $\alpha_S(m_Z)=0.117$, $0.118$, $0.119$. The fits with $\alpha_S(m_Z)=0.118$ are also provided with 1000 replicas, for use in reweighting studies.
In addition, improved delivery tools, such as  reducing the number of
replicas~\cite{Carrazza:2015hva}, or by provision of
Hessian eigenvectors~\cite{Carrazza:2015aoa}, have also become available recently.

\subsection{PDF analysis tools}

When performing a QCD analysis to determine PDFs there are various assumptions and choices to be made concerning, for example, 
the functional form of the input parametrization, the treatment of heavy quarks and their mass values, alternative theoretical calculations or representations of the fit quality estimator, $\chi^2$, and for different ways of treating correlated systematic uncertainties. It is useful to discriminate or quantify the effect of a chosen anstaz within
a common framework and the {\tt HERAFitter}, an open source QCD fit analysis project~\cite{HERAFitter,Alekhin:2014irh}, is optimally designed for such tests.

{\tt HERAfitter} incorporates results from a wide range of experimental measurements in lepton-proton deep inelastic scattering, proton-proton and proton-antiproton collisions. 
These are complemented with a variety of theoretical options for calculating PDF-dependent cross section predictions corresponding to the measurements.
The framework covers a large number of the existing methods (e.g. {\tt fastNLO} and {\tt APPLgrid}, described later in this section) 
and schemes used for PDF determination. The data and theoretical predictions are confronted 
by means of numerous methodological options for performing PDF fits and plotting tools to help visualize the results.
For example, recently the {\tt HERAFitter} framework has been used to study the consistency of the legacy measurements of the $W$-boson
charge asymmetry and of the $Z$-boson production cross sections from Tevatron with the NLO QCD theoretical predictions, which are found 
in good agreement~\cite{Camarda:2015hea} and illustrate the importance of the Tevatron data to constrain the
$d$-quark and the valence PDFs.
In summary, with sufficient options to reproduce the majority of the different theoretical choices made in global PDF fits, {\tt HERAFitter} is a valuable tool for benchmarking and understanding differences in the phenomenology of PDF fits by different groups and it can be used to study the impact of new precision measurements at hadron colliders.

Precise measurements require accurate theoretical predictions in order to maximize their impact in PDF fits.
Perturbative calculations become more complex and time-consuming at higher orders due to the increasing number 
of relevant Feynman diagrams. The direct inclusion of computationally demanding higher-order calculations
into iterative fits is thus not possible currently. However, a full repetition of the perturbative calculation 
for small changes in input parameters is not necessary at each step of the iteration. 
Two methods have been developed which take advantage of this to solve the problem: the $K$-factor technique and the {\it fast grid} technique. 

In the $K$-factor method, the ratio of the prediction of a higher-order pQCD calculation, usually time-consuming, to a lower-order calculation 
using the same PDF, are estimated once for a given PDF, stored into a table of $K$-factors, and applied multiplicatively to 
the theory prediction derived from the fast lower-order calculation throughout the iterative process in minimising the $\chi^2$. Hence, 
 this technique avoids iteration of the higher-order calculation at each step. This procedure, however, neglects the fact that the $K$-factors 
are PDF dependent, and as a consequence, they have to be re-evaluated for the newly determined PDF at the end of the fit until input and output 
$K$-factors have converged (typically 2-3 iterations are needed). This method has been used for the NNLO QCD fits to the Drell-Yan measurements.

In the {\it fast grid} method, a generic PDF can be approximated by a set of interpolating functions with a sufficient number of support points.
The accuracy of this approximation is checked and optimized such that the approximation bias is negligible compared to the experimental and theoretical accuracy.
Hence, this method can be used to perform the time consuming higher-order calculations only once for the set of interpolating functions.
Further iterations of the calculation for a particular PDF set are fast, involving only sums over the set of interpolators
multiplied by factors depending on the PDF. This approach can be used to calculate the cross sections of processes involving 
one or two hadrons in the initial state and to assess their renormalization and factorization scale variation.

There are three projects most commonly used to exploit the described techniques: 
{\tt FastNLO}~\cite{Adloff:2000tq,Kluge:2006xs}, {\tt APPLgrid}~\cite{Carli:2005ji,Carli:2010rw}
and {\tt aMCfast}~\cite{amcfast}.
The packages differ in their interpolation and optimization strategies, but 
they all
construct tables with
grids for each bin of an observable in two steps: in the first step, the accessible phase space in the parton momentum fractions $x$ and
the renormalization and factorization scales $\mu_r$ and $\mu_f$ is explored in order to optimize the table size. In the second step
the grid is filled for the requested observables. Higher-order cross sections can then be obtained very efficiently from the pre-computed grids while varying
externally provided PDF sets, $\mu_r$ and $\mu_f$, or $\alpha_S(\mu_R)$. This approach can
be extended to arbitrary processes.
This requires an interface between the
higher-order theory programs and the fast interpolation frameworks.

The open-source project {\tt fastNLO}~\cite{fastNLO:HepForge}, has been interfaced to the {\tt NLOjet++} program~\cite{Nagy:1998bb}
for the calculation of jet production in DIS~\cite{Nagy:2001xb} as well as 2- and 3-jet
production in hadron-hadron collisions at NLO~\cite{Nagy:2003tz,Nagy:2001fj}. Threshold corrections at 2-loop
order, which approximate NNLO for the inclusive jet cross section for $pp$ and $p\bar p$, have also been included into the framework~\cite{Wobisch:2011ij} 
following~\cite{Kidonakis:2000gi}.
The latest version of the {\tt fastNLO} convolution program~\cite{Britzger:2012bs} allows for the
creation of tables in which renormalization and factorization scales
can be varied as a function of two pre-defined observables.
More recently, the differential calculation of top-pair production in hadron collisions at approximate NNLO~\cite{Guzzi:2014wia} has been interfaced to 
{\tt fastNLO}~\cite{dis2014Fast}.

In the {\tt APPLgrid} package~\cite{APPLgrid:HepForge}, in addition to jet cross sections for $pp(p\bar p)$ and DIS processes, calculations 
of Drell-Yan production and other processes are also implemented using an interface to the
{\tt MCFM} parton level generator~\cite{Campbell:1999ah,Campbell:2000je,Campbell:2010ff}.
Variation of the renormalization and factorization scales is possible a posteriori
when calculating theory predictions with the {\tt APPLgrid} tables,
using the {\tt HOPPET} program~\cite{Salam:2008qg},
and
independent variation of $\alpha_S$ is also allowed.
The {\tt aMCfast} project is based on using the same {\tt APPLgrid} interpolation
methods within the fully automated {\tt MadGraph5\_aMC@NLO}~\cite{Alwall:2014hca} framework
to achieve the automation of fast NLO QCD calculations for
PDF fits, for arbitrary processes.
Work in progress in the {\tt aMCfast} code
is directed towards achieving the same automation
for NLO calculations matched to parton showers and to the inclusion
in PDF fits of generic NLO electroweak corrections.

\begin{figure}[t]
\begin{center}
\includegraphics[width=0.49\textwidth]{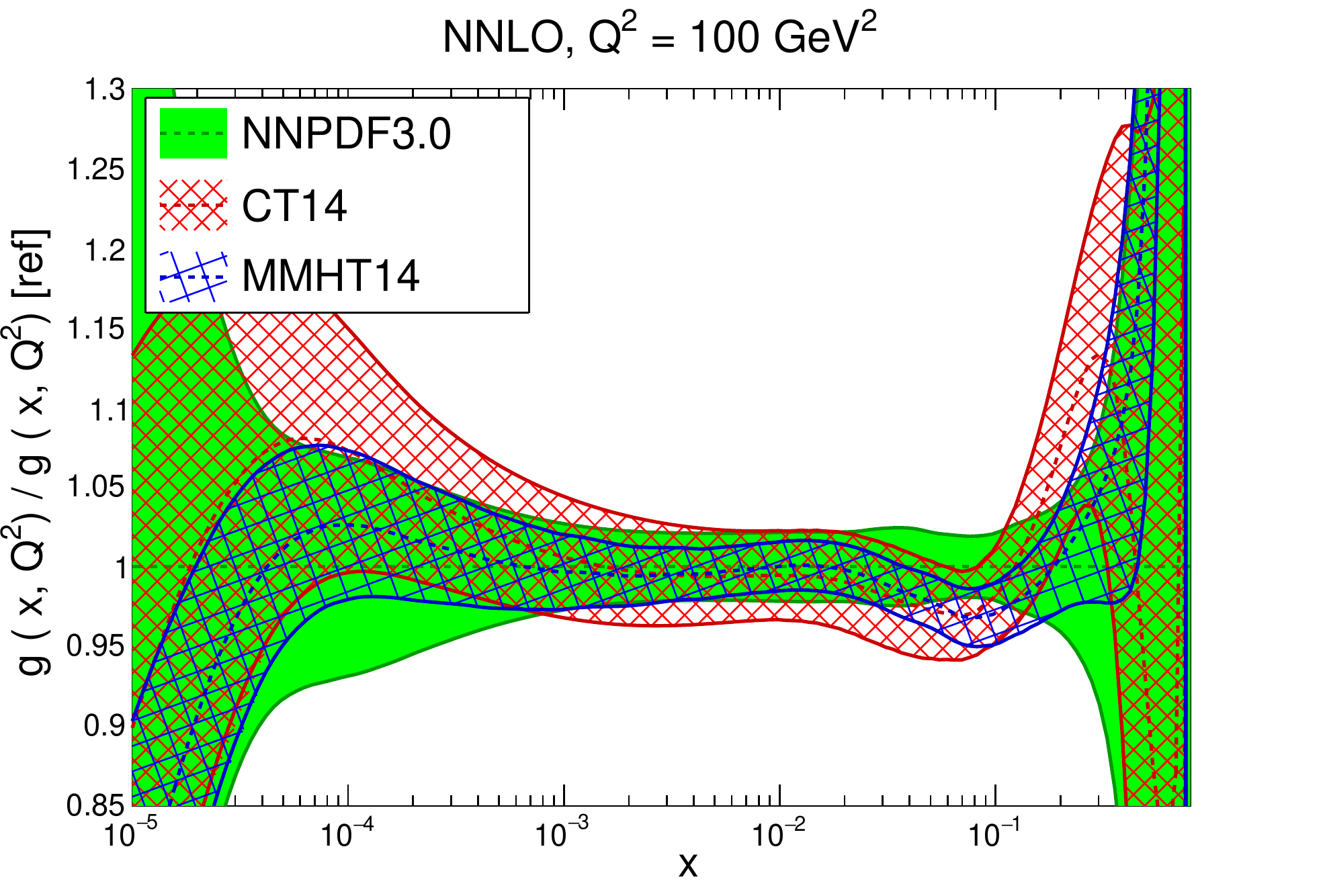}
\includegraphics[width=0.49\textwidth]{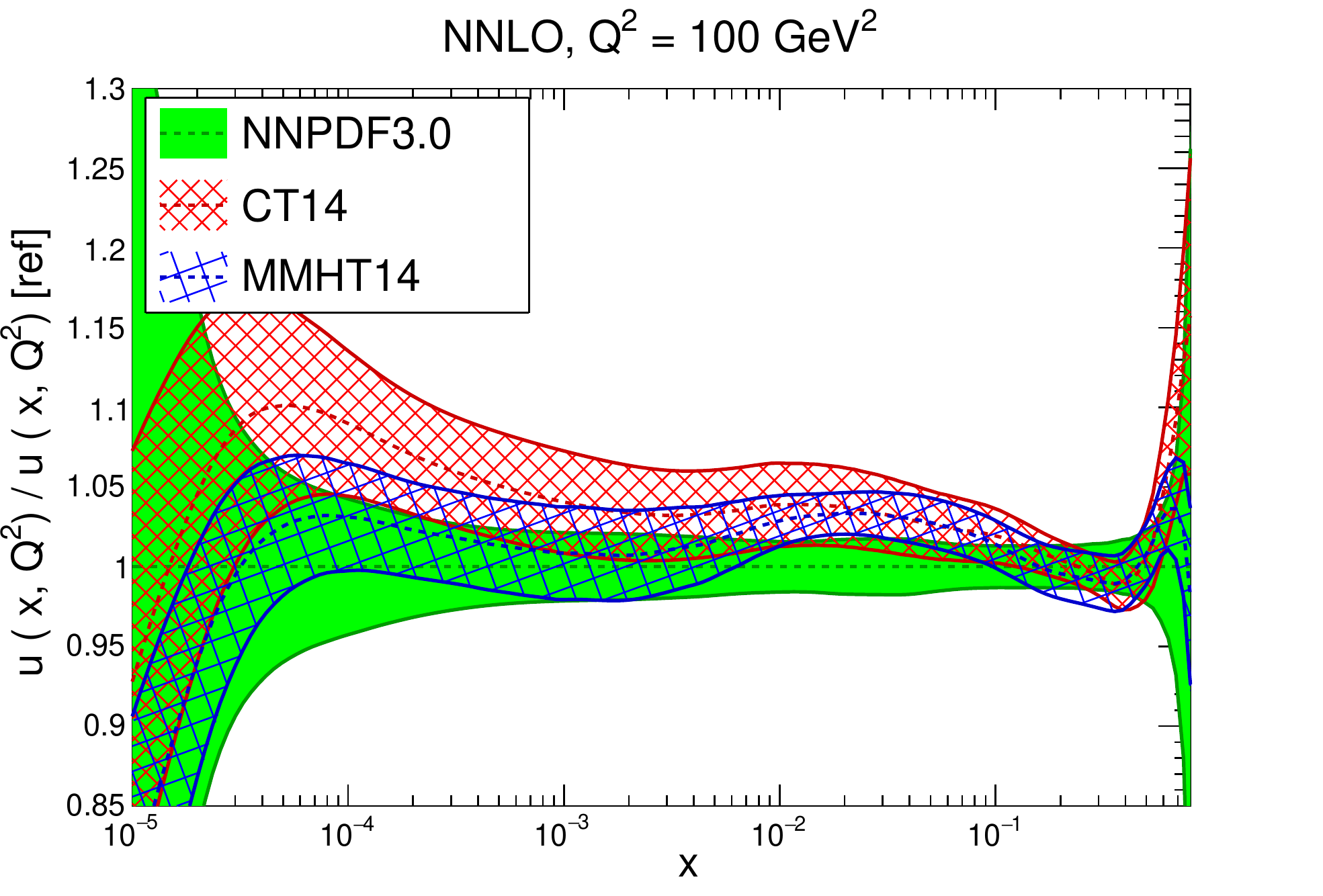}
\includegraphics[width=0.49\textwidth]{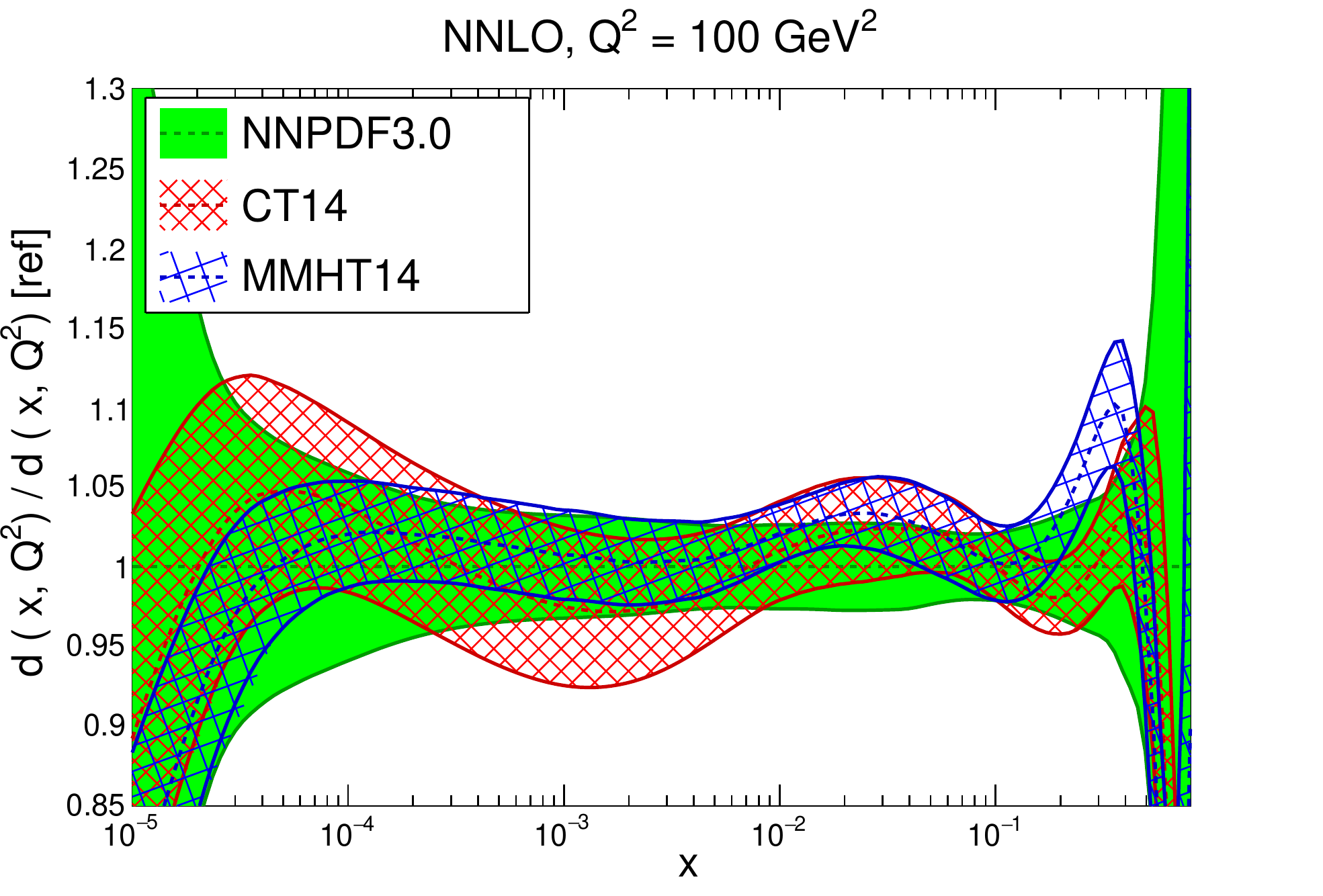}
\includegraphics[width=0.49\textwidth]{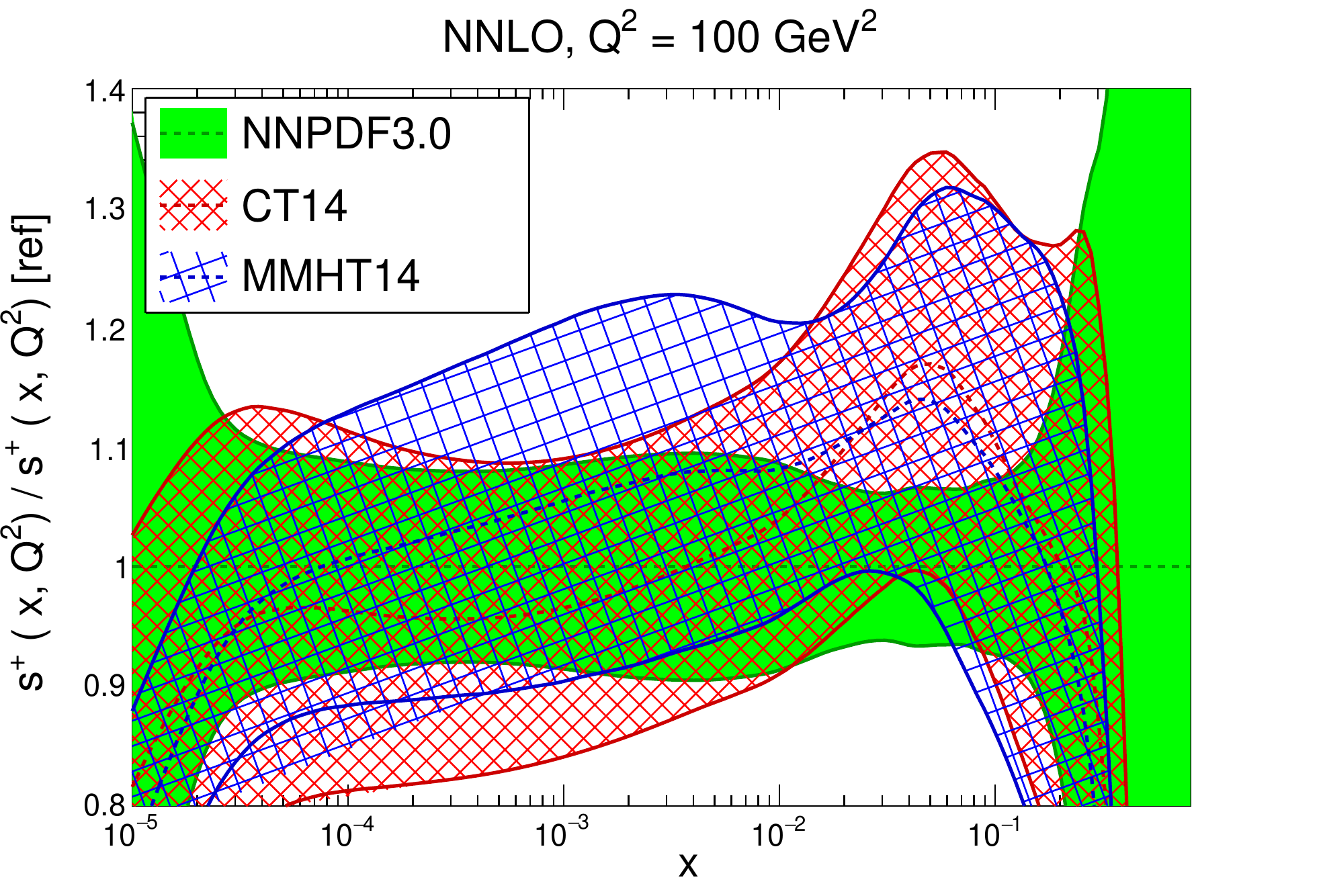}
\end{center}
\vspace{-0.3cm}
\caption{\small \label{fig:pdfs}
  Comparison of PDFs at $Q^2=10^2$ GeV$^2$ between the NNPDF3.0,
  CT14 and MMHT14 sets, all of them at NNLO, with $\alpha_S(m_Z^2))=0.118$.
  From top to bottom, and from left to right,
  we show the gluon, the up quark,
  the down quark, and the total strangeness PDFs.
  Results are shown normalized to the central value of NNPDF3.0.
}
\end{figure}

\begin{figure}[t]
  \begin{center}
    \includegraphics[width=0.49\textwidth]{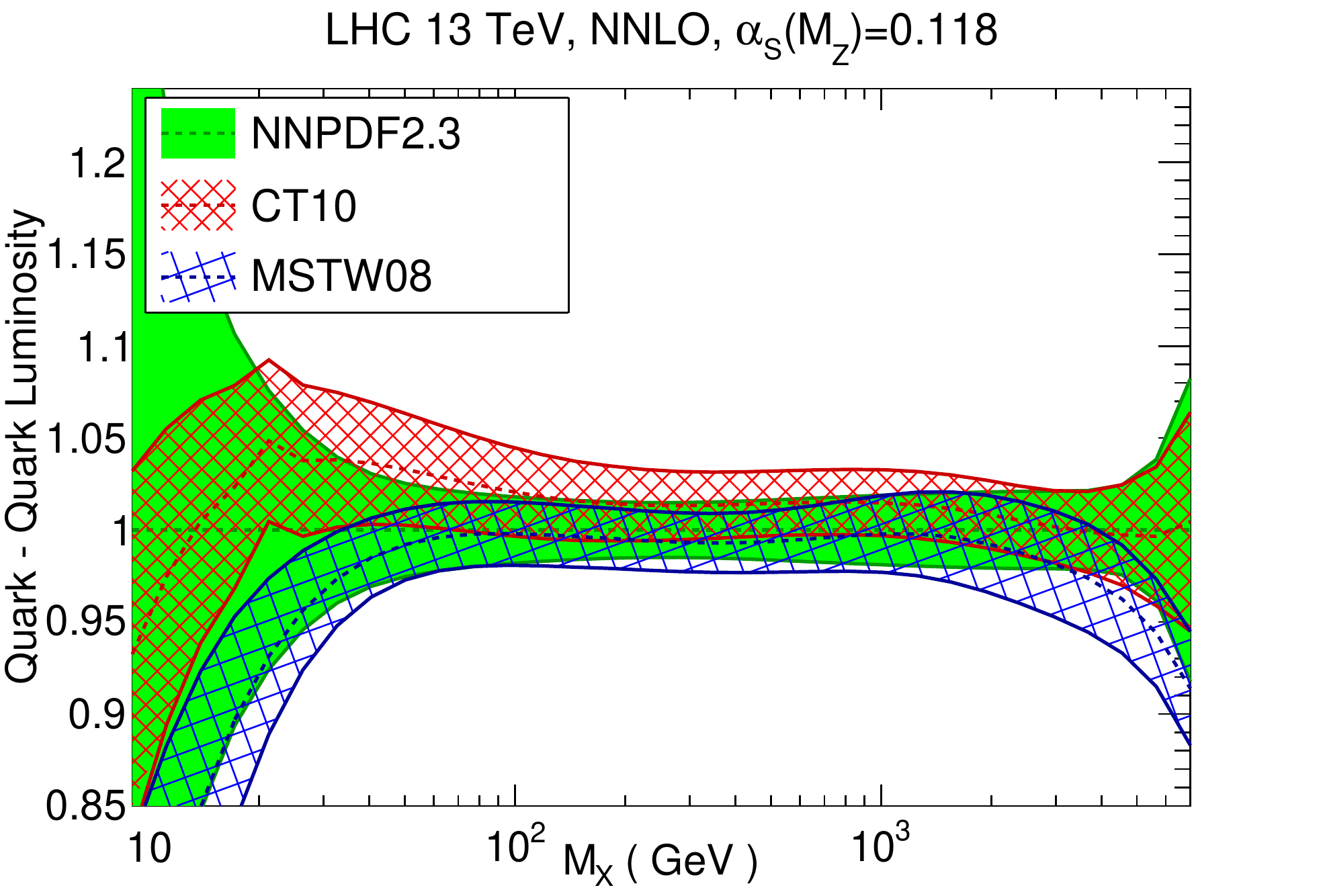}
\includegraphics[width=0.49\textwidth]{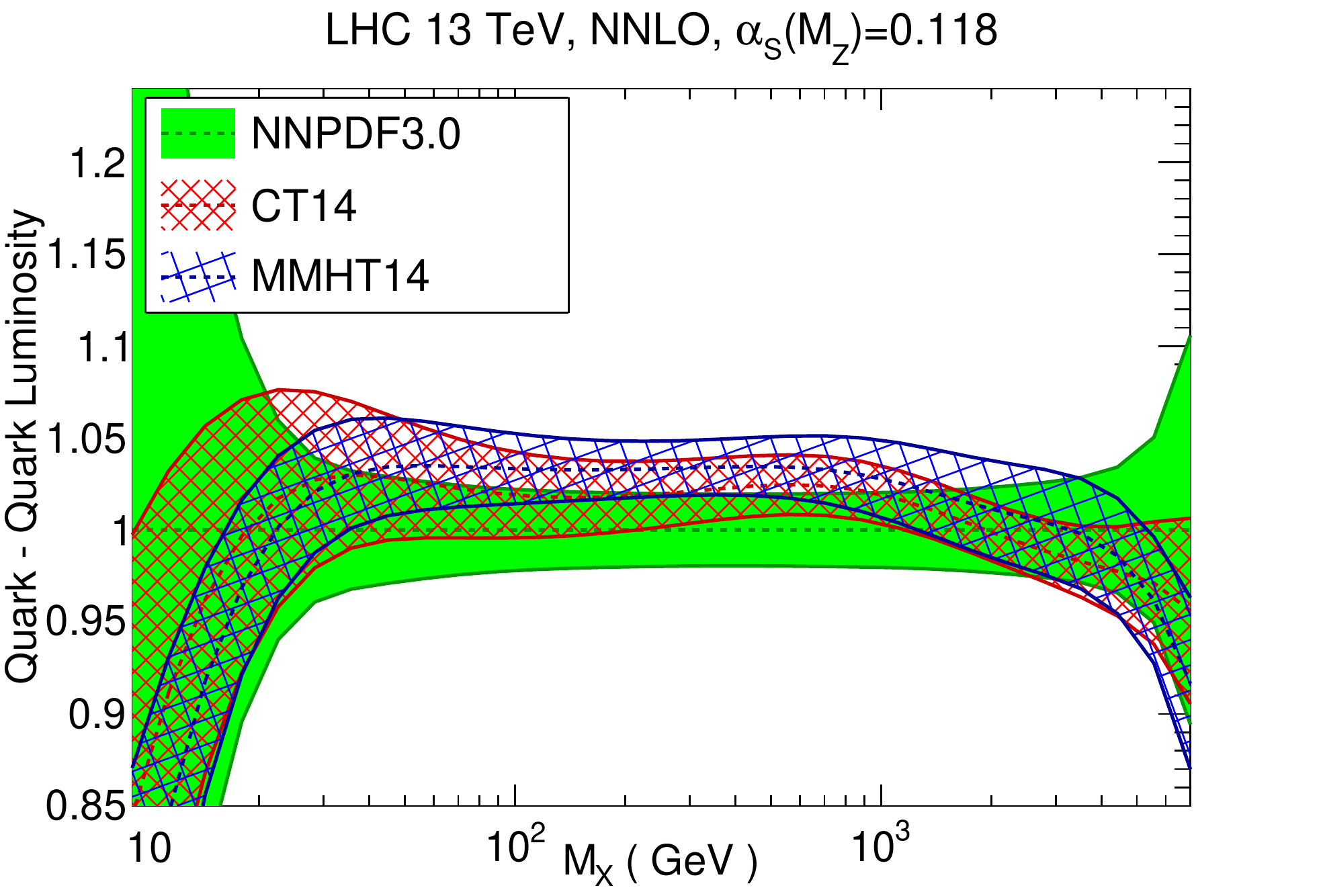}
    \includegraphics[width=0.49\textwidth]{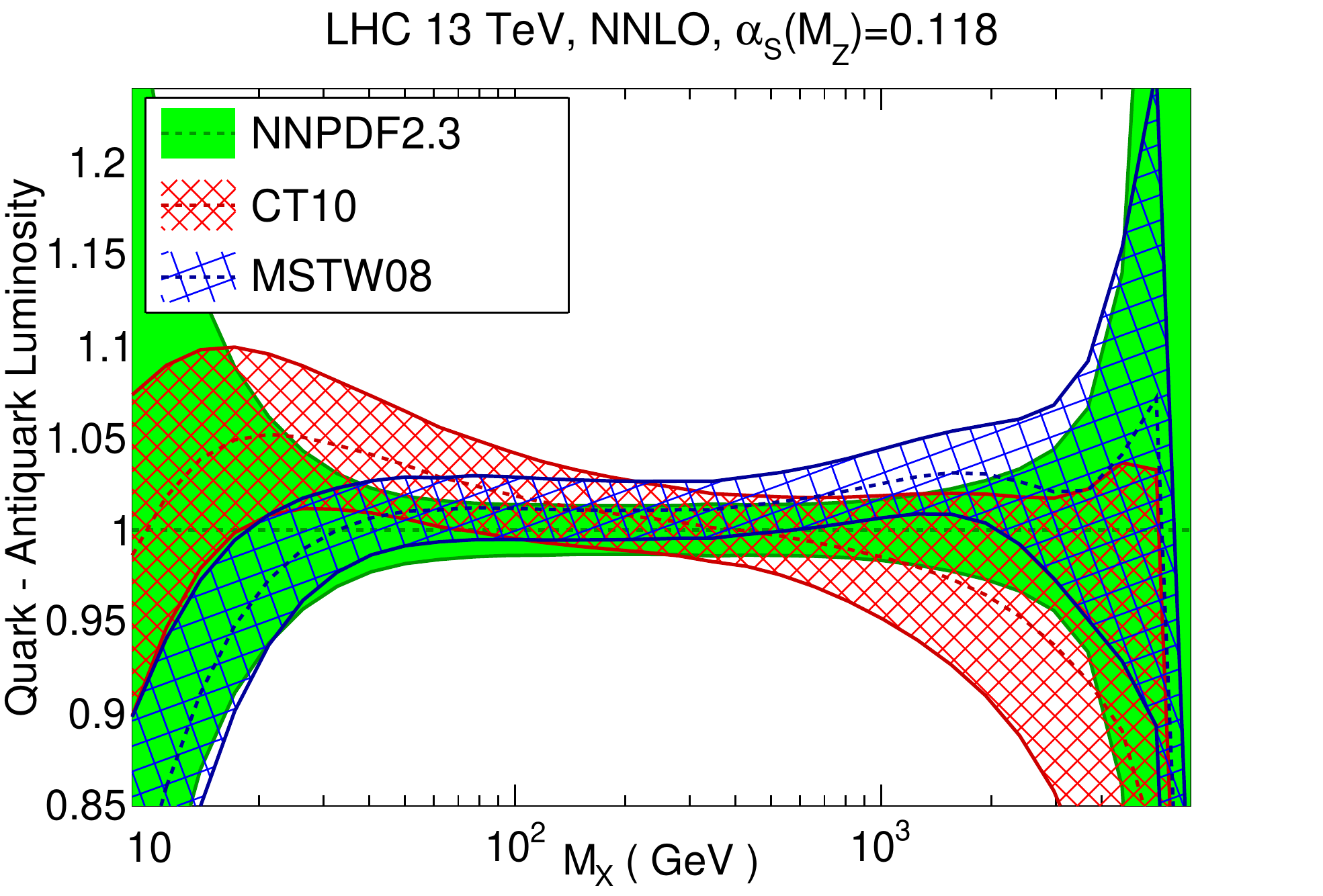}
\includegraphics[width=0.49\textwidth]{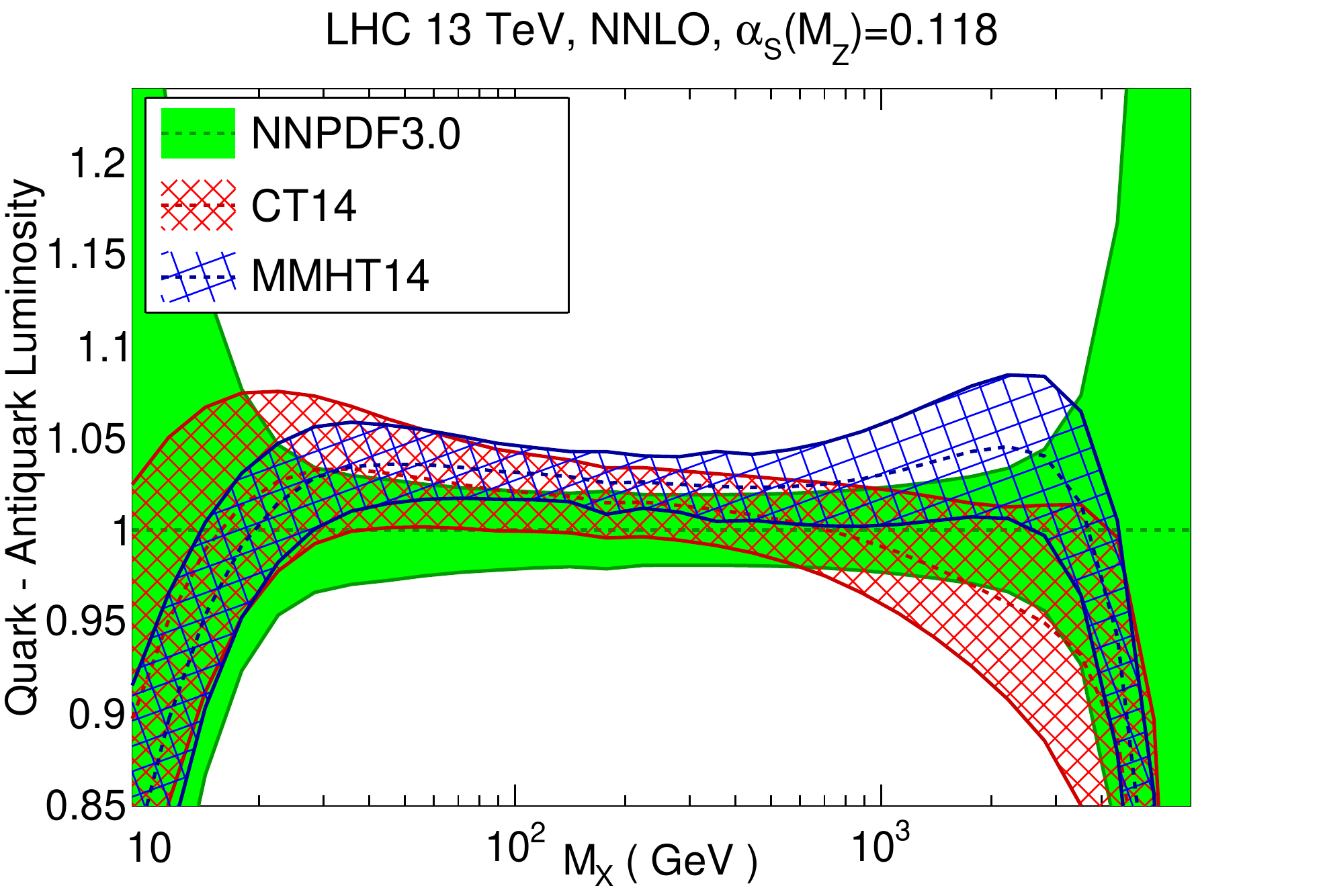}
\end{center}
\vspace{-0.3cm}
\caption{\small \label{fig:lumis}
  Comparison of the quark-quark (upper plots) and
  quark-antiquark (lower plots) PDF luminosities between
  NNPDF2.3, MSTW08 and CT10 (left column) and the more recent
  NNPDF3.0, MMHT14 and CT14 (right column) PDF sets.
  The comparison has been performed at NNLO and
  $\alpha_S(m_Z^2))=0.118$ for the LHC with a center of mass energy of
  13 TeV, as a function of the invariant mass of the final state system $M_X$.
  Results are shown normalized to the central value of the NNPDF sets.
}
\end{figure}

\begin{figure}[t]
  \begin{center}
    \includegraphics[width=0.49\textwidth]{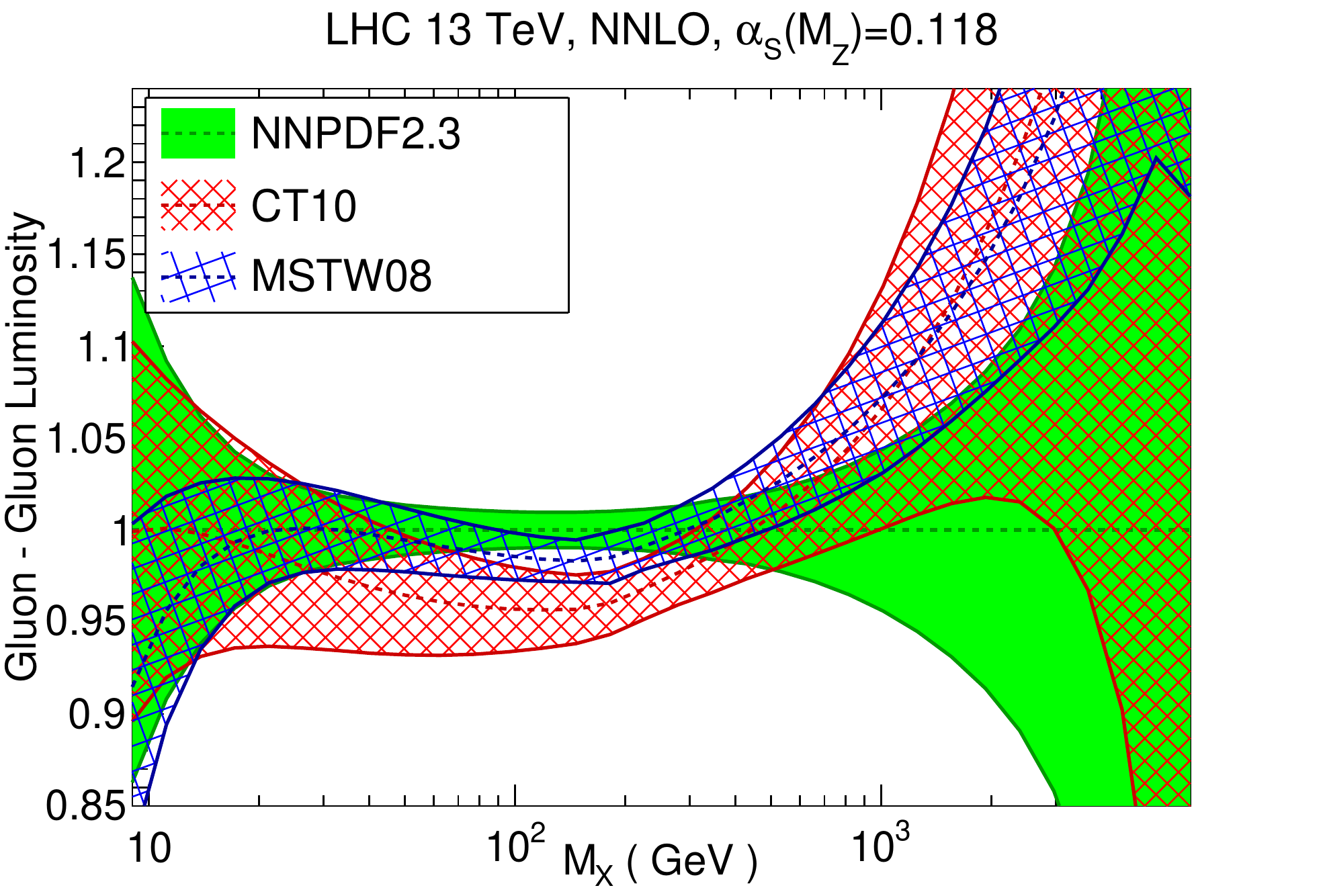}
\includegraphics[width=0.49\textwidth]{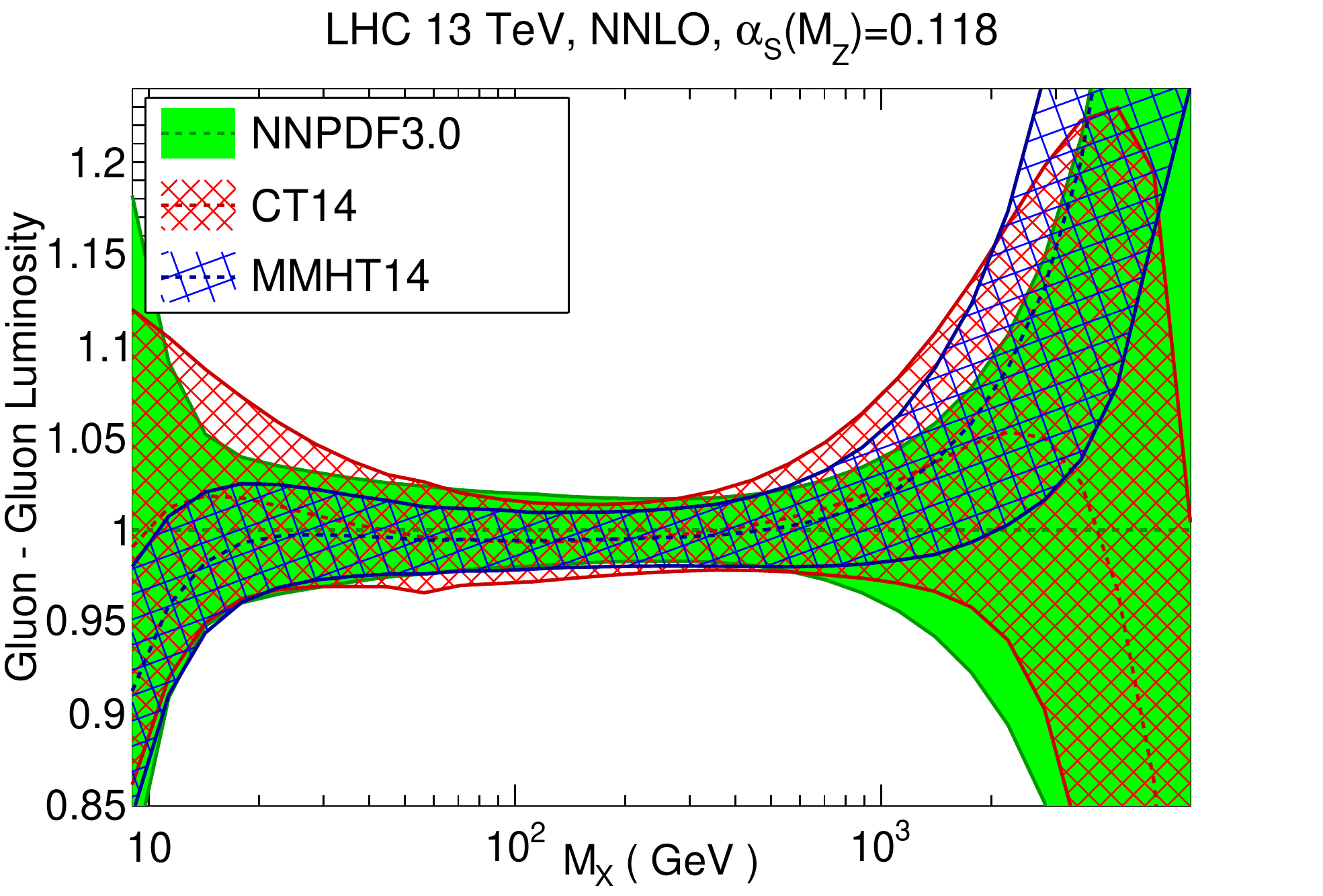}
\includegraphics[width=0.49\textwidth]{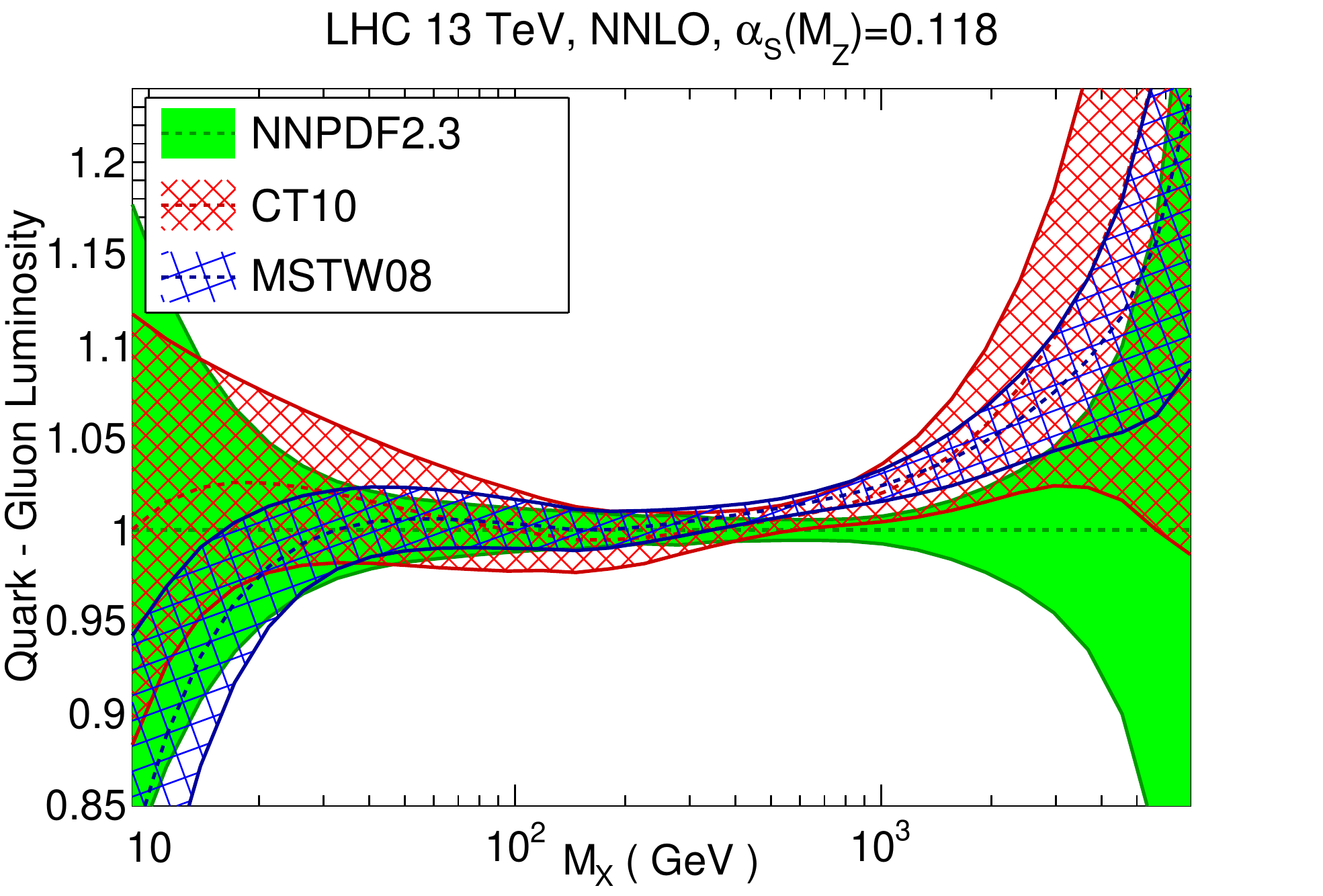}
\includegraphics[width=0.49\textwidth]{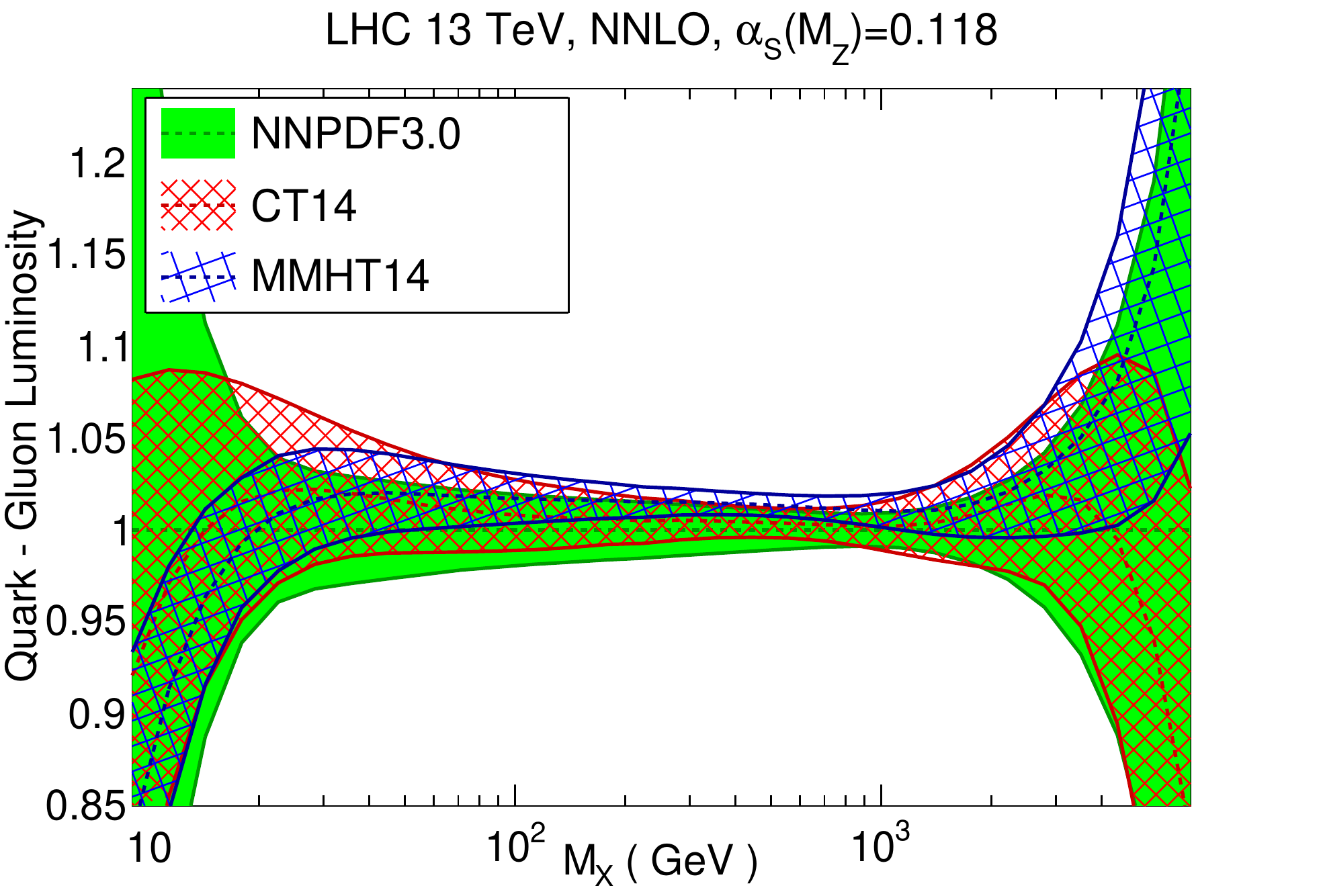}
\end{center}
\vspace{-0.3cm}
\caption{\small \label{fig:lumis2}
  Same as Fig.~\ref{fig:lumis} now for the gluon-gluon (upper plots) and
  quark-gluon (lower plots) PDF luminosities
}
\end{figure}

\subsection{Comparison of PDFs and parton luminosities}

To conclude this section, we compare some of the recent releases from the various PDF groups in terms of PDFs and parton luminosities.
For this purpose, the {\tt APFEL-Web} online
PDF plotting
interface~\cite{Bertone:2013vaa,Carrazza:2014gfa} has been used.
First, we compare the NNPDF3.0, CT14 and MMHT14 NNLO sets in
Fig.~\ref{fig:pdfs},
at a scale of $Q^2=100$ GeV$^2$.
 From top to bottom, and from left to right,
  we show the gluon, the up quark,
  the down quark, and the total strangeness PDFs.
  Results are shown normalized to the central value of NNPDF3.0.

  The comparisons in Fig.~\ref{fig:pdfs} indicate that there
  is reasonable agreement at the level of the one standard
  deviation of the PDF
  uncertainties between the three groups.
  In some cases the agreement is only marginal, for
  instance for the $d$ PDF at large-$x$.
  In general, it is at small and large values of $x$,
  in regions with limited kinematical coverage, that
  the differences between the three fits are more marked.
  For some PDF combinations, the size of the PDF
  uncertaintiy can also show differences between the three groups,
  such in the total strangeness $s^+$ PDF at intermediate values
  of $x$.

Next, we turn to a comparison of PDF luminosities for the LHC at a center of mass energy of 13 TeV.
To illustrate the differences between the previous releases from
NNPDF, CT and MSTW/MMHT,
in Fig.~\ref{fig:lumis} we compare the quark-quark and quark-antiquark
luminosities in NNPDF2.3, CT10 and MSTW08 with the same results
from NNPDF3.0, CT14 and MMHT14.
The corresponding comparison for the gluon-gluon and
quark-gluon luminosities is shown in Fig.~\ref{fig:lumis2}.
 The comparison has been performed at NNLO and
  $\alpha_S(m_Z^2)=0.118$,
  as a function of the invariant mass of the final state system $M_X$.
  Results are shown normalized to the central value of the NNPDF sets.

  Comparing the newer and the older
  PDF sets, we notice that
  in general there has been improved agreement between
  the three sets in a number of phenomenologically important regions,
  like the $gg$ luminosity at intermediate values 
of the final-state invariant mass $M_X$.
For the four luminosities that are compared here, the three PDF sets agree at the one-sigma level or better in all the relevant range of $M_X$ values.
The differences are larger at large invariant masses, a key region for massive New Physics
searches at the LHC, where also the intrinsic PDF uncertainties for each group
are  substantial due to the lack of experimental constraints.
We also find that in some cases, like the quark-quark luminosity,
the agreement is only marginal, driven by the differences
at the level of $u$ and $d$ PDFs observed in
Fig.~\ref{fig:pdfs}.

\section{Overview of PDF-sensitive measurements at the LHC}
\label{sec:overview}

In this section we review  LHC processes relevant for PDF constraints, summarized in Table~\ref{tab:processes}.
Emphasis is put on information not previously accounted for in PDF fits. For each process, the sensitivity to specific PDF flavours is briefly described
and the probed ranges of $x$ and $Q^2$ are listed.
The corresponding measurements 
performed by the ATLAS, CMS and LHCb collaborations,
when already available, are presented in the next section.

\begin{table}[t]
  \centering
  \small
\begin{tabular}{c|c|c|c|c}
\toprule
\textsc{reaction} & \textsc{observable} & \textsc{pdfs}
& $x$ & $Q$ \\
\midrule
\multirow{2}*{$pp\to W^\pm + X$} &
\multirow{2}*{$d\sigma(W^{\pm})/dy_l$} &
\multirow{2}*{$q,\bar{q}$} &
\multirow{2}*{$10^{-3}\lesssim x \lesssim 0.7$} &  
\multirow{2}*{$\sim M_W$} \\
& & & & \\
\multirow{2}*{$pp\to \gamma^*/Z + X$} &
\multirow{2}*{$d^2\sigma( \gamma^*/Z)/dy_{ll}dM_{ll}$} &
\multirow{2}*{$q,\bar{q}$} &
\multirow{2}*{$10^{-3}\lesssim x \lesssim 0.7$} &  
\multirow{2}*{$5~{\rm GeV}\lesssim Q \lesssim 2$ TeV } \\
& & & & \\
\multirow{2}*{$pp\to \gamma^*/Z + {\rm jet} + X$} &
\multirow{2}*{$d\sigma( \gamma^*/Z)/dp_T^{ll}$} &
\multirow{2}*{$q,g$} &
\multirow{2}*{$10^{-2}\lesssim x \lesssim 0.7$} &  
\multirow{2}*{$200~{\rm GeV}\lesssim Q \lesssim 1$ TeV } \\
& & & & \\
\multirow{2}*{$pp\to {\rm jet}+X$} &
\multirow{2}*{$d\sigma({\rm jet})/dp_Tdy$} &
\multirow{2}*{$q,g$} &
\multirow{2}*{$ 10^{-2}\lesssim x \lesssim 0.8 $} &  
\multirow{2}*{$20~{\rm GeV}\lesssim Q \lesssim 3$ TeV } \\
& & & & \\
\multirow{2}*{$pp\to {\rm jet}+{\rm jet}+X$} &
\multirow{2}*{$d\sigma({\rm jet})/dM_{jj}dy_{jj}$} &
\multirow{2}*{$q,g$} &
\multirow{2}*{$ 10^{-2}\lesssim x \lesssim 0.8 $} &  
\multirow{2}*{$500~{\rm GeV}\lesssim Q \lesssim 5$ TeV } \\
& & & & \\
\multirow{2}*{$pp\to t\bar{t}+X$} &
\multirow{2}*{$\sigma(t\bar{t}),d\sigma(t\bar{t})/dM_{t\bar{t}}$, ....} &
\multirow{2}*{$g$} &
\multirow{2}*{$ 0.1\lesssim x \lesssim 0.7 $} &  
\multirow{2}*{$350~{\rm GeV}\lesssim Q \lesssim 1$ TeV } \\
& & & & \\
\multirow{2}*{$pp\to c\bar{c}+X$} &
\multirow{2}*{$d\sigma(c\bar{c})/dp_{T,c}dy_c$} &
\multirow{2}*{$g$} &
\multirow{2}*{$10^{-5}\lesssim x \lesssim 10^{-3} $} &  
\multirow{2}*{$1~{\rm GeV}\lesssim Q \lesssim 10$ GeV } \\
& & & & \\
\multirow{2}*{$pp\to b\bar{b}+X$} &
\multirow{2}*{$d\sigma(b\bar{b})/dp_{T,c}dy_c$} &
\multirow{2}*{$g$} &
\multirow{2}*{$10^{-4}\lesssim x \lesssim 10^{-2} $} &  
\multirow{2}*{$5~{\rm GeV}\lesssim Q \lesssim 30$ GeV } \\
& & & & \\
\multirow{2}*{$pp\to W+c$} &
\multirow{2}*{$d\sigma(W+c)/d\eta_l$} &
\multirow{2}*{$s,\bar{s}$} &
\multirow{2}*{$ 0.01\lesssim x \lesssim 0.5 $} &  
\multirow{2}*{$\sim M_W$ } \\
& & & & \\
\bottomrule
\end{tabular}
\caption{\small Summary  of 
LHC processes sensitive to PDFs.
For each process, we quote the corresponding measured distribution, the PDFs that are probed, and the approximate ranges of $x$ and $Q^2$ that can be 
accessible using available Run I data.
These ranges have been obtained assuming the Born kinematics.
}
\label{tab:processes}
\end{table}

\subsection{Jet production}

\label{sec:jetproduction}

Jet production allows one to constrain quarks and gluons at medium and large-$x$, for $x\gsim 0.005$~\cite{Rojo:2014kta,Dulat:2015mca}, a region where the
constraints from deep-inelastic scattering data are only indirect.
Inclusive jet production has been used in PDF fits since the first measurements at the Tevatron.
Nowadays, a number of precise LHC measurements of 
inclusive jet, dijet and trijet production are available.
In addition, jet production provides a unique possibility for direct determinations of the strong coupling $\alpha_S(Q)$ in the TeV range way 
above any other existing measurements, providing information on BSM physics.

Jet production can be presented in a number of complementary ways, the most traditional are the measurements
of the inclusive jet and dijet cross-sections, but measurements of three-jet and multi-jet cross-sections have also became
available recently.
The impact of ATLAS and CMS jet data on PDFs has been quantified in a number of studies, both from global PDF
fitting groups~\cite{Ball:2014uwa,Watt:2013oha,Dulat:2015mca}
and from the LHC collaborations themselves~\cite{Khachatryan:2014waa,Aad:2013lpa}.
In addition to the gluon, also information on the large-$x$ quarks can be obtained, 
since the quark-quark scattering mechanism dominates
at the highest values of the jet transverse momentum, due to the steeper fallout of the gluon PDF at large-$x$.

While the NLO calculation for inclusive and dijet production has been available for more than 20 years, only recently the first partial results
on the NNLO calculation have become available~\cite{Ridder:2013mf,Currie:2013dwa}.
While the full calculation is not yet available, it has been proposed that a subset of jet data can still be consistently included in NNLO fits by using the approximate NNLO 
threshold calculation.
This strategy, presented in~\cite{Carrazza:2014hra}, has been used 
to include LHC jet data in the NNPDF3.0 fit.

\begin{figure}[h]
\begin{center}
\includegraphics[width=0.49\textwidth]{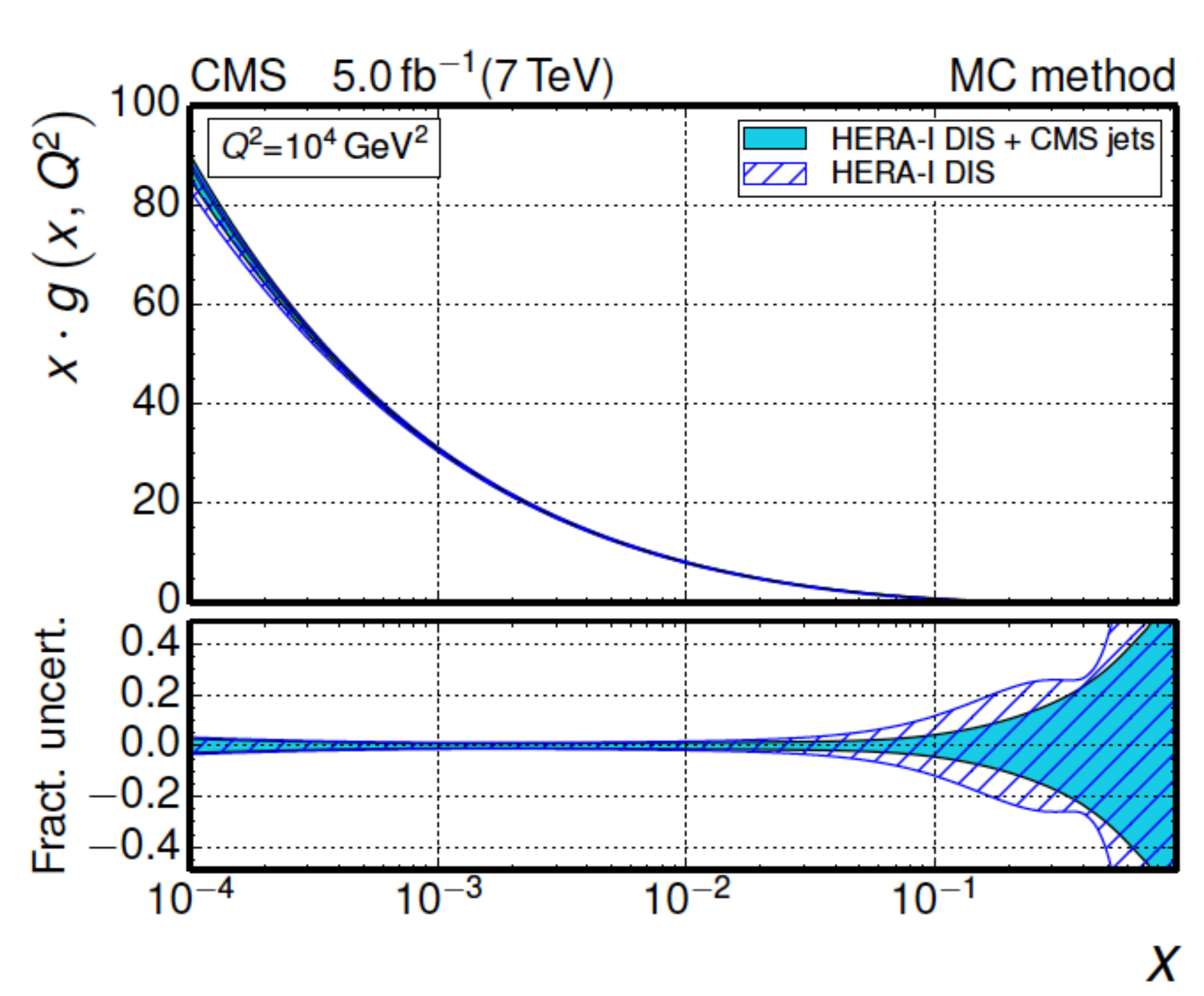}
\includegraphics[width=0.49\textwidth]{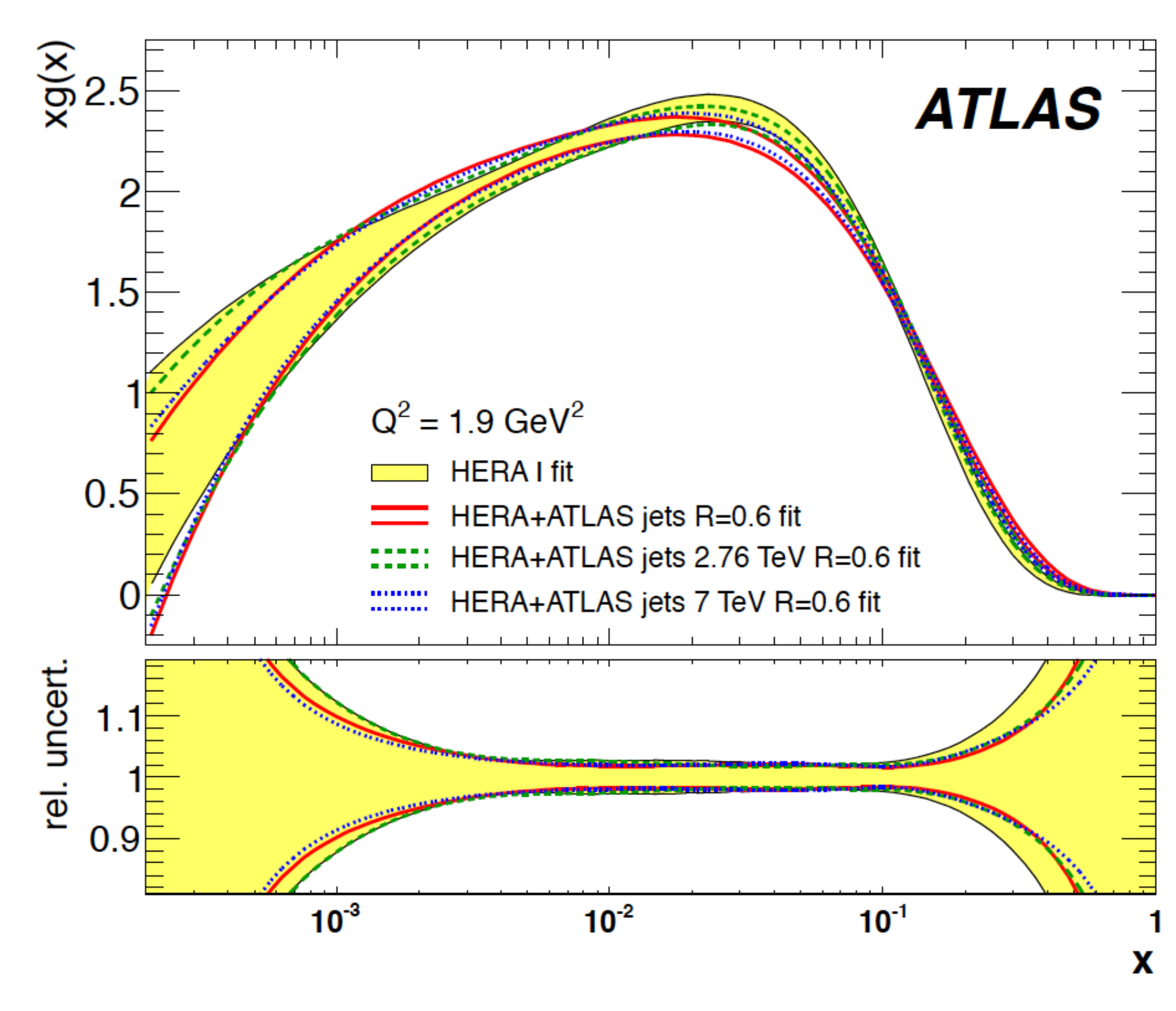}\\
\end{center}
\vspace{-0.4cm}
\caption{\small \label{fig:gluonjets} Left plot: impact of the
  CMS 2011 inclusive jet data on the gluon PDF,
  from Ref.~\cite{Khachatryan:2014waa}.
  Right plot: impact of the ATLAS measurement on the ratio
  of inclusive jet cross-sections between 7 TeV and 2.76 TeV
  on the gluon PDF, from Ref.~\cite{Aad:2013lpa}.
}
\end{figure}

In Fig.~\ref{fig:gluonjets} we illustrate the impact of the
  CMS 2011 inclusive jet data on the large-$x$ gluon PDF,
  from Ref.~\cite{Khachatryan:2014waa}.
  In the same figure we also show the constraints that
  the ATLAS measurement on the ratio
  of inclusive jet cross-sections between 7 TeV and 2.76 TeV imposes
  on the gluon PDF, from Ref.~\cite{Aad:2013lpa}.
  In both cases, the PDF fits have been performed
  using the {\sc\small HERAfitter} framework.

\subsection{Prompt photon production}
\label{sec:promptphoton}

Direct photon production measurements from fixed-target experiments were
recognised as a useful probe of the gluon PDFs a long time ago~\cite{Vogelsang:1995bg},
but their use was very limited,
because of inconsistencies between the various experiments, especially after the competitive Tevatron jet production data were published.
More recently, the use of LHC isolated photon data in the PDF fits was advocated in Ref.~\cite{d'Enterria:2012yj}, where a reanalysis of all available fixed target and collider
isolated direct-photon production data was performed, finding good consistency with NLO QCD calculations.
In Ref.~\cite{d'Enterria:2012yj} it was also shown that direct photon data can potentially constrain the gluon PDF in an intermediate range of $x$, 
around $\sim 0.01$,
which is the region relevant for Higgs-boson production via gluon fusion.
The main obstacle for the full inclusion of direct photon data into PDF fits is the large scale uncertainties that affect the NLO QCD calculation.
The possibility to use isolated photon production in association with additional jets has also been explored~\cite{Carminati:2012mm},
however a substantial reduction of the experimental uncertainties, with respect to that of available measurements, would
be needed before this data could be used effectively in the PDF fits.
While LHC photon data has still not been directly included in global PDF fits, a systematic comparison between different PDF sets and direct photon
data was presented by ATLAS~\cite{ATL-PHYS-PUB-2013-018}. 

\subsection{Inclusive $W$ and $Z$ production and asymmetries}
\label{sec:inclusiveWZ}

Inclusive production of $W$ and $Z$ bosons,
presented in the form of total cross sections, differential distributions in leptonic rapidities, and corresponding asymmetries, has
been important in the global PDF fits
since the first such measurements were made
at the Tevatron.
As compared to inclusive DIS, where only flavor symmetric components $q+\bar{q}$ can be constrained, inclusive $W$ and $Z$ production provides
a clean handle on quark flavour separation.
At the LHC, the kinematical range in terms of the underlying 
$x$ has substantially increased as compared to the Tevatron,
reaching both smaller and larger values of $x$.
To pin down the PDF quark flavor separation, a number of measurements have been presented by ATLAS, CMS and LHCb, as will be discussed
in more detail in Sect.~\ref{sec:runI}.
In addition, as shown by ATLAS, once the rapidity distributions of $W$ and $Z$ bosons are measured simultaneously accounting for the
correlated systematics between the various
distributions~\cite{Aad:2011dm}, an additional handle
on the strangeness content of the nucleon can be provided~\cite{Aad:2012sb}.

\subsection{High and low mass Drell-Yan production}
\label{sec:DY}

Data from fixed-target Drell-Yan experiments, such as E605~\cite{Moreno:1990sf}
and
E866~\cite{Towell:2001nh}, have been included in global PDF fits since many years.
However, these data are affected by some drawbacks since they miss information on the systematic correlations, and because $\sqrt{s}$ is small, thus leading to potentially large perturbative and non-perturbative corrections to fixed-order calculations.
This has provided the motivation to perform, for the first time, the measurements of the off-peak Drell-Yan processes at a hadron collider.
At low mass, Drell-Yan provides interesting constraints on the low-$x$ quarks and gluons (occurring in gluon radiation from the quarks), as
well as tests of perturbative QCD, like the possible
breakdown of DGLAP evolution~\cite{Marzani:2008uh}.
The high-mass region, instead, provides information from the high-$x$
quarks and anti-quarks, which are affected by substantial uncertainties (the latter in particular).
To maximize the impact of the data in PDF fits, it is essential to
use the most updated NNLO QCD and NLO EW theory calculations~\cite{dynnlo2,Gavin:2012sy}.

In addition, both the low and high-mass Drell-Yan production are sensitive
to the photon PDF $\gamma(x,Q^2)$: indeed, since $\gamma\gamma \to l^+l^-$ production
is $t$-channel, as opposed to the $s$-channel quark-induced diagrams
$q\bar{q} \to l^+l^-$, photon-initiated contributions become comparable to quark-initiated
at low and high invariant di-lepton masses.
Therefore, off-peak Drell-Yan production provides important constraints
on the photon PDF~\cite{Boughezal:2013cwa} for PDF fits which account for
QED corrections~\cite{Ball:2013hta,Martin:2004dh}.

Finally, the high-mass region provides a crucial validation of theory calculations in a region which is instrumental
for new physics searches.

\subsection{The transverse momentum of $W$ and  $Z$ bosons}
\label{sec:VPT}

The transverse momentum, $p_T$, of the $W$ and  $Z$ bosons is a key observable for hadron collider phenomenology.
At low $p_T$, it is used to validate Monte Carlo predictions, analytically resummed calculations, and is important for many precision measurements
like the $W$ mass.
At large values of the transverse momentum, we would expect that fixed-order theory provides a reasonable description of the data.
For large $p_T$, the transverse momentum distribution of $W$ and  $Z$ bosons uniquely depends on the combination
$\alpha_S \times q \times g$, where the fraction of gluon-initiated contributions increases with $p_T$.
Therefore, one might want to use the high-$p_T$ spectrum of $W$ and $Z$ bosons as a direct probe of the gluon PDF.
This option seems particularly robust for the case of the $Z$ $p_T$, where
a high precision measurement
can be performed in terms of leptonic
variables only~\cite{Aad:2014xaa}.

One possible issue for
the inclusion of these measurements
in PDF fits is that available data on the $Z$ boson transverse momentum
from ATLAS~\cite{Aad:2014xaa}
and CMS~\cite{Khachatryan:2015oaa} exhibit a $\mathcal{O}(10\%)$ discrepancy with pure
NLO calculations in the region around 100 GeV, where the accuracy of
the experimental measurement is around $\mathcal{O}(1\%)$.
In this respect, having the full NNLO results for
the $Z$ $p_T$ will shed light on the origin of this discrepancy,
though available results for the NNLO $W$+jets
calculation~\cite{Boughezal:2015dva}
suggest that higher-order corrections on top of NLO might not
be enough to explain the differences observed  between theory and data.

The measurements of the ratios of $W$ and $Z$ cross sections as a function of boson $p_T$ would provide additional information on PDFs. 
As motivated in Ref.~\cite{Malik:2013kba}, various ratios of $W$ and $Z$
cross sections at high $p_T$
provide a handle on the proton's flavour decomposition, while cancelling various theoretical uncertainties like higher order QCD and EW effects.
In Fig.~\ref{fig:cc-gluon} we show
the transverse momentum distribution of $Z$ bosons at the
LHC 8 TeV computed at NLO using various sets of PDFs, shown as
ratio to the MSTW08 prediction~\cite{Malik:2013kba}.

In addition, a number of measurements of $W$ and $Z$ boson production in association with jets has been performed at the LHC.
The main motivation for these measurements is to validate Monte Carlo event generators, but given the fact that the underlying
dynamics are the same as those that generate the vector-boson $p_T$, it is conceivable that these can also be used for PDF fits.
However, these measurements will be affected by larger theoretical uncertainties (due to the higher final state multiplicity)
and experimental uncertainties (due to the presence of jets) than the inclusive $W$ and $Z$ $p_T$ measurement, and thus
might not be competitive with the latter.

\subsection{$W$ production in association with charm quarks}
\label{sec:wcharm}

Production of $W$ bosons in association with charm quarks has been
proposed for a long time as a direct probe of the strangeness content
of the proton~\cite{Baur:1993zd}.
Indeed, before the LHC start-up, constraints on strangeness from global fits
were provided mostly by low-energy neutrino data, in particular by the measurements of charm production through di-muon final
states~\cite{Goncharov:2001qe,Ball:2009mk}.
At the LHC, independent constraints on the strangeness can now
be provided
by the measurement of the $W$$+$$c$ process cross section~\cite{Stirling:2012vh}, and also the differences
between $s^+$ and $s^-$ content can be potentially
be investigated with cross section ratios such as $(W^+$$+$$c)/(W^-$$+$$c)$.
As we will discuss in the next Section, this measurement has been recently 
published by both ATLAS and CMS, and is already part of several global PDF fits.

\begin{figure}[t]
\begin{center}
\includegraphics[width=0.65\textwidth]{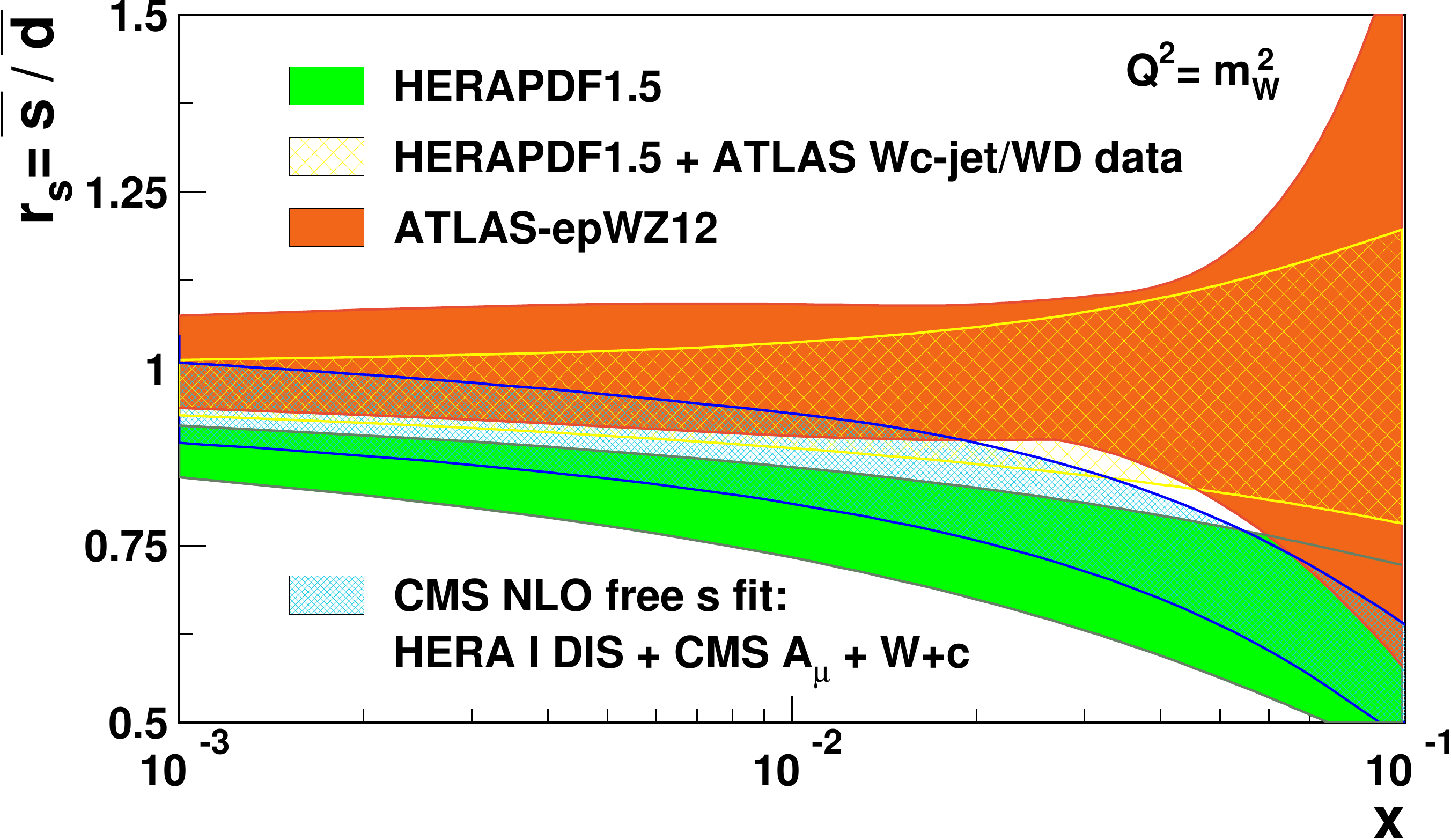}
\end{center}
\vspace{-0.4cm}
\caption{\small \label{fig:strange}
  The ratio of the $s$ and $d$ PDFs, as a function
  of $x$, compared in different analysis.
  We show the ATLAS results based on $Z$ and $W$-boson production measurements~\cite{Aad:2012sb} and
  on the associated production of $W$ with charmed hadrons~\cite{Aad:2014xca}
  as well as the CMS result~\cite{Chatrchyan:2013mza}, 
  based on the $W$$+$$c$ production measurement of Ref.~\cite{Chatrchyan:2013uja}.
  For comparison, the HERAPDF1.5 result  is also shown, where the constraints on the strange quark distribution are obtained from the neutrino-scattering experiments.
}
\end{figure}

A topic that has attracted sizable attention recently is whether the LHC
$W$$+$$c$ data suggest a symmetric strange sea, opposite to neutrino charm
data which clearly indicates a strangeness suppression.
In Fig.~\ref{fig:strange} we show the strangeness fraction in the quark sea, obtained by ATLAS and CMS by using inclusive $W$ and $Z$ measurements and the $W$$+$$c$ data.
In addition we show the HERAPDF1.5 result, where the constraints on the
strange quark distribution are
obtained from the neutrino-scattering experiments.
While CMS data prefer a suppressed strangeness, and the ATLAS measurements indicate
a symmetric light quark sea, both results are consistent within uncertainties.
Moreover, recent global analyses combining both fixed-target and collider data
sensitive to the strangeness~\cite{Alekhin:2014sya,Ball:2014uwa} demonstrated the
general consistency of the LHC data among each other and with the measurements of neutrino experiments in the $x$-range accessible by the LHC measurements.
Future, higher precision data from Run II will shed more light
on this issue.

\subsection{Top quark pair production}
\label{sec:topprod}

Top quarks are abundantly produced at the LHC, 
which can be considered a real ``top factory" 
due to the high center of mass energy and luminosity.
As opposed to the Tevatron, where top quark pairs are produced predominantly
via quark-anti-quark annihilation, at the LHC they are produced
mostly in the gluon-gluon channel.
Therefore, they provide potentially useful information on the gluons for
$x \ge 0.1$, a region which is only covered by
jet production in PDF global fits.
In addition, for differential distributions sensitive to large-$x$
PDFs, such as the $t\bar{t}$ invariant mass distribution or
the tail of the $p_T^t$ distribution, there is also sensitivity
to quarks and anti-quarks.

While NLO calculations are affected by large scale uncertainties, the completion of the full NNLO calculation for 
total production cross-sections~\cite{Czakon:2013goa}
and for differential distributions~\cite{Czakon:2014xsa,Abelof:2015lna}
will allow for consistent use of the top quark-pair data in the fits at NNLO.
Furthermore, their availability allows for more precise extractions of fundamental QCD parameters, like top-quark mass 
and $\alpha_S$~\cite{Chatrchyan:2013haa}.
Since the exact differential NNLO calculation is not yet available in a form suitable for QCD analyses, 
its approximate version~\cite{Guzzi:2014wia}, featuring the methods of threshold resummation, might be used.

Up to now, a number of studies has quantified the sensitivity of top quark pair production data to the gluon PDFs
using the total top-quark pair production cross-sections, 
showing that available data from ATLAS and CMS already provide powerful constrains on the large-$x$ gluon~\cite{Czakon:2013tha,Beneke:2012wb}.
Among other collaborations that include top data in their fits,
ABM has explored their impact 
showing that it can lead to a shift in the gluon PDF up to
one-sigma~\cite{Alekhin:2013nda} in units of the PDF uncertainties.
The impact of total cross-sections in PDF fits is only moderate, 
but the full constraining power of top quark data 
will be assessed using the differential distributions.
A first study on this respect, based on the approximate NNLO from threshold-resummed calculation, has been presented in Ref.~\cite{Guzzi:2014wia}.

\subsection{Charm and bottom pair production}
\label{sec:heavyquarkprod}
Production of heavy quark pairs in hadron collisions is a powerful test of perturbative QCD. 
While 
top pair production at the LHC is nowadays included in PDF fits,
this is not the case for charm and bottom quarks.
On the other hand, their differential $p_T$ and rapidity distributions ($d^2\sigma/dp_Tdy$)
are directly sensitive to the small-$x$ gluon PDF at low scale.
The downside is a large scale dependence of theoretical predictions
(currently available only at NLO), which, however, may be mitigated by
analyzing ratios of differential rates in various experimental bins,
instead of their absolute values~\cite{Zenaiev:2015rfa}.

\begin{figure}[t]
\begin{center}
\includegraphics[width=0.42\textwidth]{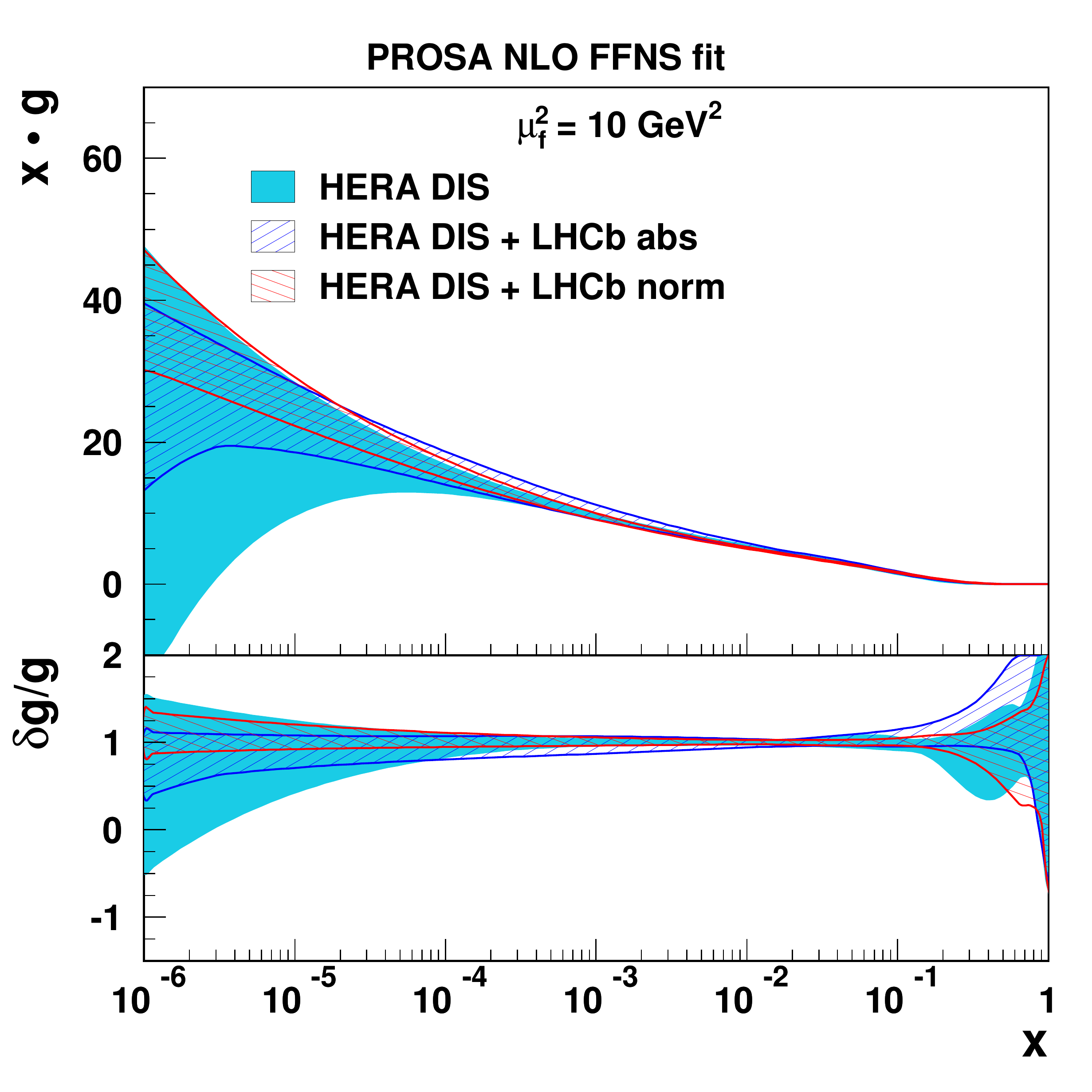}
\includegraphics[width=0.57\textwidth]{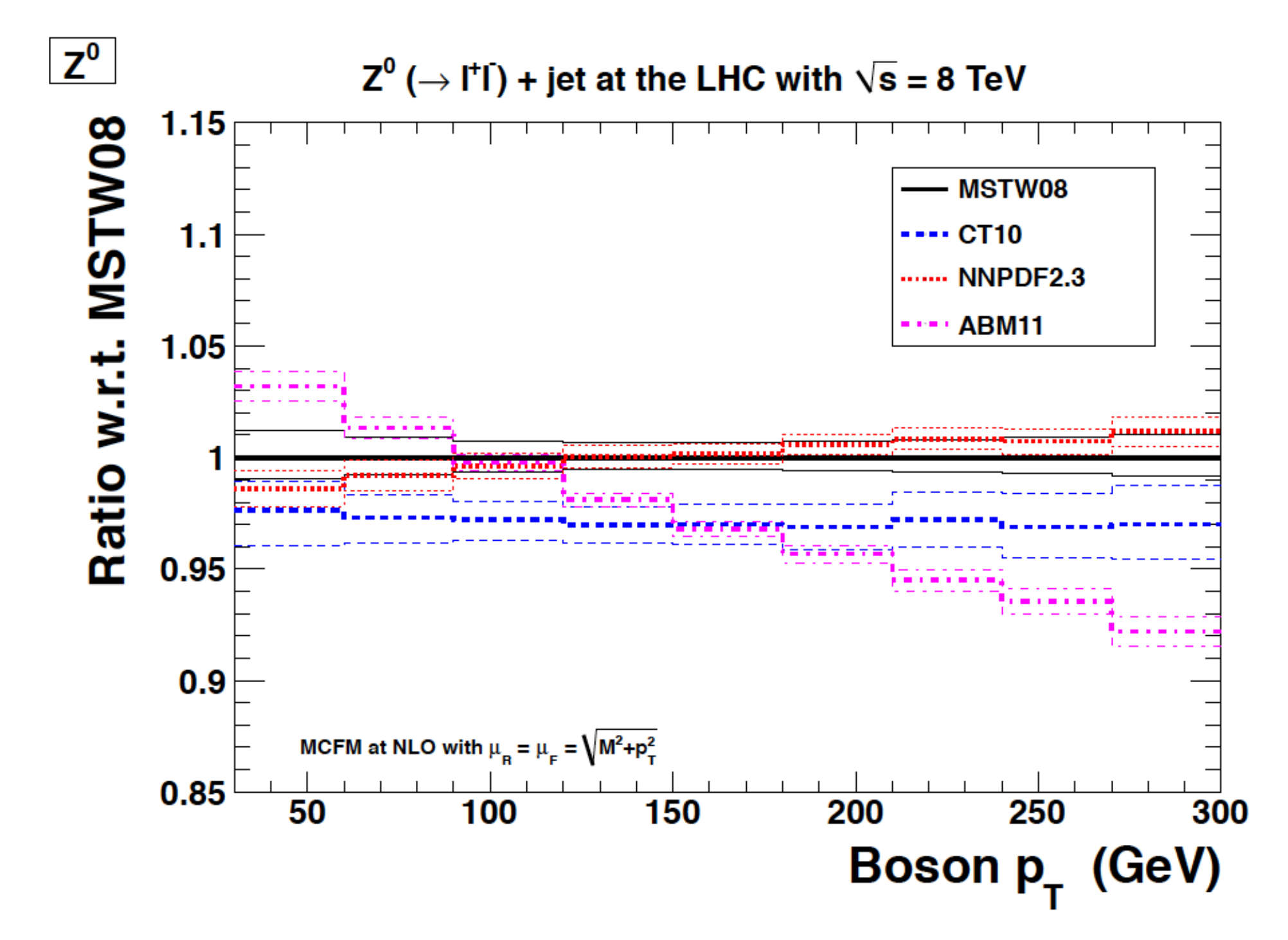}\\
\end{center}
\vspace{-0.7cm}
\caption{\small \label{fig:cc-gluon} Left plot: gluon distribution (top) and its relative uncertainty (bottom), shown as a function of $x$ at the factorization scale of 10 GeV$^2$,
  comparing the results of the NLO QCD analysis~\cite{Zenaiev:2015rfa} of HERA-I DIS data only (filled band) and that including either absolute or normalized cross sections of heavy-flavor production at LHCb (dashed bands).
Right plot: the transverse momentum distribution of $Z$ bosons at the
LHC 8 TeV computed at NLO using various sets of PDFs, shown as
ratio to the MSTW08 prediction, from Ref.~\cite{Malik:2013kba}
}
\end{figure}

The constraints are expected to be particularly powerful in
measurements in the LHCb
acceptance.
The LHCb detector is suitable to precisely tag and measure the properties of heavy quark
mesons. Its forward coverage allows one to access
the small-$x$ region that cannot be accessed by ATLAS and CMS.
A first study in this direction has been performed by the PROSA Collaboration~\cite{Zenaiev:2015rfa},
which evaluated
the impact of recent measurements of heavy-flavour production at LHCb 
using the {\sc\small HERA\-Fitter} framework
in a QCD analysis at NLO in a fixed-flavour number scheme.
Significant reduction of the gluon and the sea-quark distribution uncertainties
is found down to $x\sim 5 \times 10^{-6}$, as illustrated in Fig.~\ref{fig:cc-gluon}.
A related analysis of the constraints of the LHCb 7 TeV charm production
data using the NNPDF reweighting method has been presented
in~\cite{Gauld:2015yia}.
Further developments of the underlying theory and the generalization of the 
NNLO calculations for top quark-pair
production~\cite{Czakon:2014xsa,Guzzi:2014wia} 
for the case of charm and beauty production is highly desirable
to exploit the full constraining power of these data. 

\subsection{Central exclusive production of heavy-flavours}
\label{sec:centralexclusive}

As mentioned above,
the production of charm and bottom 
quarks at the LHC provides useful constraints on the low-$x$ gluon. 
Related processes, which can potentially give 
similar information on 
the gluon, are the central exclusive production of $J/\psi$ and related 
mesons, such as the $\Upsilon$ or $\psi(2S)$. These processes have the advantage 
of providing very clean experimentally final states. 
Calculations for these processes have been presented in Refs.~\cite{Jones:2013pga,Jones:2013eda}
within the diffractive photo-production formalism and have been compared 
to the LHCb data~\cite{Aaij:2013jxj,Aaij:2014iea}. 
While this theoretical approach is distinct from the standard collinear factorization picture, the predictions may be extended so as 
to be related to the collinear gluon PDF, allowing such data to be incorporated consistently into PDF fits.
The full NLO contribution has been recently obtained and the scale dependence can be reduced by using 
an appropriate choice of the factorization scale~\cite{Jonesfut}.

\subsection{Ratios of cross-sections for different center-of-mass energies}
\label{sec:CMratio}

The  availability of LHC data 
at different center of mass energies ($\sqrt{s}$), allows us to construct novel types
of observables, namely the ratios and double ratios of cross-sections measured
at different $\sqrt{s}$~\cite{Mangano:2012mh}.
The main advantage of this approach is that
several experimental and theoretical systematics 
cancel to some approximation.
Predictions reach higher accuracy 
because they are less sensitive to
higher orders,
and  measurements  become more precise due to the  reduction
of systematic uncertainties such as jet energy scale
and luminosities.
On the other hand,
the interest in these ratios 
relies on the fact that
the PDF dependence does not factorize out, because data at different $\sqrt{s}$ will
probe different range in $x$ and $Q^2$, as well as slightly different flavor content,
and therefore 
might be used
to constrain PDFs.
Their constraints can be complementary to those of the absolute cross-sections in many cases: in top quark production, 
for example, the dependence on the exact value of $m_t$ will completely cancel in the ratio.

From the experimental point of view, the crucial point is to quantify the degree of correlation of systematic uncertainties between 
measurements at different values of $\sqrt{s}$.
Up to now, this idea has been implemented in two cases: the ratio of inclusive jet cross-sections between 7 TeV and 2.76 TeV from
ATLAS~\cite{Aad:2013lpa}, and the ratio of Drell-Yan cross-sections between 7 and 8 TeV at CMS~\cite{CMS:2014jea}.
While these two measurements are important as proof-of-concept, their impact on PDF fits will be moderate 
due to the limited statistics of the 2.76 TeV sample (for the ATLAS
analysis) and the small lever arm between 7 and 8 TeV (in the CMS measurement).
However, during Run II, several ratios between 13 and 8 TeV measurements should be performed, 
and could 
become a valuable source of PDF constraints 
in the global analysis.

\section{Constraining PDFs with LHC data at Run I}
\label{sec:runI}

In this section we present an overview of 
LHC Run I measurements 
with potential sensitivity to PDFs,
along the lines
discussed in Sect.~\ref{sec:overview}.
We report here the specific
analyses that have been performed 
and, whenever applicable,
the studies
used to document their PDF constraining power.
This section is split by experiment: we begin
with the summary of the results from ATLAS, then
we move to CMS and finally to LHCb. A summary table,
including only measurements that are published or 
submitted to a journal, 
is provided for each experiment.
Measurements which are still in preliminary form, or have been superseded, are not
included in this report.

\subsection{Constraints from ATLAS}
\label{subsec:runI:atlas}

The measurement from the ATLAS collaboration are summarized in
Table~\ref{table:runIoverviewATLAS}.
\begin{table}[t]
\centering
\footnotesize
\begin{tabular}{l|l|c|c|c}
  \hline
\hline
\multicolumn{5}{c}{ATLAS} \\
\hline
\hline
Measurement & $\sqrt{s}$, year of data, $\mathcal{L}_{\rm int}$  & Motivation &
Reference  & PDF fits \\
\hline
$W,Z$ rapidity & 7 TeV, 2010, 36 pb$^{-1}$  & Sect.~\ref{sec:inclusiveWZ} & \cite{Aad:2011dm} & \cite{Alekhin:2013nda,Ball:2012cx,Aad:2012sb,Ball:2014uwa,Harland-Lang:2014zoa} \\
\hline
High mass Drell-Yan & 7 TeV, 2011, 4.9 fb$^{-1}$ & Sect.~\ref{sec:DY}  & \cite{Aad:2013iua}& \cite{Ball:2013hta,Ball:2014uwa,Harland-Lang:2014zoa}\\
Low mass Drell-Yan & 7 TeV, 2011+2010, 1.6 fb$^{-1}$+35 pb$^{-1}$ & Sect.~\ref{sec:DY}  & \cite{Aad:2014qja} & - \\
$Z$ $A_{FB}$ & 7 TeV, 2011, 4.8 fb$^{-1}$ & Sect.~\ref{sec:DY} & \cite{Aad:2015uau}  &  - \\
\hline
$W$+charm production & 7 TeV, 2011, 4.6 fb$^{-1}$ & Sect.~\ref{sec:wcharm} & \cite{Aad:2014xca}  & \cite{Aad:2014xca} \\
$W$+beauty production & 7 TeV, 2010, 35 pb$^{-1}$ & Sect.~\ref{sec:wcharm} & \cite{Aad:2011kp}  & - \\
$W$+beauty production & 7 TeV, 2011, 4.6 fb$^{-1}$ & Sect.~\ref{sec:wcharm} & \cite{Aad:2013vka}  & - \\
$Z$+beauty production & 7 TeV, 2010, 36 pb$^{-1}$ & Sect.~\ref{sec:wcharm} & \cite{Aad:2011jn}  & - \\
$Z$+beauty production & 7 TeV, 2011, 4.6 fb$^{-1}$ & Sect.~\ref{sec:wcharm} & \cite{Aad:2014dvb}  & - \\
\hline
$Z$ p$_{T}$ & 7 TeV, 2010, 40 pb$^{-1}$ & Sect.~\ref{sec:VPT} & \cite{Aad:2011gj}  & - \\
$Z$ p$_{T}$ & 7 TeV, 2011, 4.7 fb$^{-1}$ & Sect.~\ref{sec:VPT} & \cite{Aad:2014xaa}  & -  \\
$W$ p$_{T}$ & 7 TeV, 2010, 31 pb$^{-1}$ & Sect.~\ref{sec:VPT} & \cite{Aad:2011fp}  & \cite{Ball:2014uwa} \\
\hline
$Z$+jets & 7 TeV, 2010, 36 pb$^{-1}$ & Sect.~\ref{sec:VPT} & \cite{Aad:2011qv}  & - \\
$Z$+jets & 7 TeV, 2011, 4.6 fb$^{-1}$ & Sect.~\ref{sec:VPT} & \cite{Aad:2013ysa}  & - \\
$W$+jets & 7 TeV, 2010, 36 pb$^{-1}$ & Sect.~\ref{sec:VPT} & \cite{Aad:2012en}  & - \\
$W$+jets & 7 TeV, 2011, 4.6 fb$^{-1}$ & Sect.~\ref{sec:VPT} & \cite{Aad:2014qxa}  &  - \\
$R_{\rm jets}$ ($W$+jets/$Z$+jets) & 7 TeV, 2011, 4.6 fb$^{-1}$ & Sect.~\ref{sec:VPT} & \cite{Aad:2014rta}  & - \\
\hline
Inclusive jets & 7 TeV, 2010, 37 pb$^{-1}$ & Sect.~\ref{sec:jetproduction} & \cite{Aad:2011fc} &\cite{Ball:2012cx,Ball:2014uwa,Harland-Lang:2014zoa} \\
Inclusive jets & 7 TeV, 2011, 4.5 fb$^{-1}$ & Sect.~\ref{sec:jetproduction} & \cite{Aad:2014vwa} & -  \\
Inclusive jets (+ 7 TeV ratio) & 2.76 TeV, 2010, 0.2 pb$^{-1}$ & Sect.~\ref{sec:jetproduction},~\ref{sec:CMratio} & \cite{Aad:2013lpa} & \cite{Aad:2013lpa,Ball:2014uwa,Harland-Lang:2014zoa}  \\
Dijets & 7 TeV, 2010, 37 pb$^{-1}$ & Sect.~\ref{sec:jetproduction} & \cite{Aad:2011fc} & - \\
Dijets & 7 TeV, 2011, 4.6 fb$^{-1}$ & Sect.~\ref{sec:jetproduction} & \cite{Aad:2013tea} & - \\
Trijets & 7 TeV, 2011, 4.5 fb$^{-1}$ & Sect.~\ref{sec:jetproduction} & \cite{Aad:2014rma} & - \\
\hline
$\gamma$ inclusive production & 7 TeV, 2010, 35 pb$^{-1}$ & Sect.~\ref{sec:promptphoton} & \cite{Aad:2011tw} & - \\
$\gamma$ inclusive production & 7 TeV, 2011, 4.6 fb$^{-1}$ & Sect.~\ref{sec:promptphoton} & \cite{Aad:2013zba} & \cite{ATL-PHYS-PUB-2013-018} \\
$\gamma$+jets & 7 TeV, 2010, 37 pb$^{-1}$ & Sect.~\ref{sec:promptphoton} & \cite{ATLAS:2012ar} & - \\
\hline
$t\bar{t}$  incl (single lepton, dilepton) & 7 TeV, 2010, 2.9 pb$^{-1}$ & Sect.~\ref{sec:topprod} & \cite{Aad:2010ey} & \cite{Harland-Lang:2014zoa} \\
$t\bar{t}$  incl (dilepton) & 7 TeV, 2010, 35 pb$^{-1}$ & Sect.~\ref{sec:topprod} & \cite{Aad:2011yb} & \cite{Harland-Lang:2014zoa}  \\
$t\bar{t}$  incl (single lepton) & 7 TeV, 2010, 35 pb$^{-1}$ & Sect.~\ref{sec:topprod} & \cite{Aad:2012qf} & \cite{Harland-Lang:2014zoa}  \\
$t\bar{t}$  incl (dilepton) & 7 TeV, 2011, 0.70 fb$^{-1}$ & Sect.~\ref{sec:topprod} & \cite{Aad:2012aa} & \cite{Harland-Lang:2014zoa,Ball:2014uwa} \\
$t\bar{t}$  incl (e/$\mu$ + $\tau$) & 7 TeV, 2011, 2.05 fb$^{-1}$ & Sect.~\ref{sec:topprod} & \cite{Aad:2012mza} & \cite{Harland-Lang:2014zoa}  \\
$t\bar{t}$  incl (tau+jets) & 7 TeV, 2011, 1.67 fb$^{-1}$ & Sect.~\ref{sec:topprod} & \cite{Aad:2012vip} & \cite{Harland-Lang:2014zoa}  \\
$t\bar{t}$  incl (e$\mu$ b-tag jets) & 7+8 TeV, 2012, 24.9 fb$^{-1}$ & Sect.~\ref{sec:topprod} & \cite{Aad:2014kva} &  \cite{Ball:2014uwa} \\
$t\bar{t}$ differential & 7 TeV, 2011, 2.05 fb$^{-1}$ & Sect.~\ref{sec:topprod} & \cite{Aad:2012hg} & - \\
$t\bar{t}$ differential & 7 TeV, 2011, 4.6 fb$^{-1}$ & Sect.~\ref{sec:topprod} & \cite{Aad:2014zka} & - \\
\hline
$WW$, $Z\to\tau\tau$, $t\bar{t}$ xsec & 7 TeV, 2011, 4.6 fb$^{-1}$ & Sect.~\ref{sec:inclusiveWZ}  & \cite{Aad:2014jra}  &  - \\
\hline
\end{tabular}
\caption{\small Overview of published 
  PDF-sensitive measurements from the LHC Run I
  from the ATLAS experiment,
where we provide  the
center-of-mass energy, year of data, and the integrated
luminosity, its motivation in terms of PDF sensitivity, the publication
reference  
and the references where these measurements
have been used to quantify PDF constraints.
 \label{table:runIoverviewATLAS}}
\end{table}
 
The available ATLAS jet production measurements, relevant for PDF studies, 
are the inclusive and dijet differential cross sections
from the 2010 dataset~\cite{Aad:2011fc}, 
the ratio of 2.76 to 7 TeV inclusive jet cross sections~\cite{Aad:2013lpa}  
and, more recently, the inclusive~\cite{Aad:2014vwa}, dijet~\cite{Aad:2013tea} 
and trijet~\cite{Aad:2014rma} differential cross sections from the
2011 dataset. 
The jets are reconstructed using the anti-k$_T$ clustering algorithm\cite{Cacciari:2008gp}, 
and the cross section measurements are available separately for two 
radius parameters: $R=0.4$ and $R=0.6$. The different radii have different 
sensitivity to final state radiation and underlying event effects. Non-perturbative 
corrections are supplied in the relevant publications, as well as electroweak corrections 
for the 2011 inclusive and dijet measurements. The cross sections for either $R=0.4$ or $R=0.6$ jets  
can be included in a PDF fit, though not both at the same time since the measurements are correlated.

A PDF fit to demonstrate the sensitivity to the large-$x$ gluon has been performed,  
adding the ratio of 2.76 TeV over 7 TeV inclusive jet data to the HERA I data,
and comparing to a baseline fit determined using only the HERA data~\cite{Aad:2013lpa}. 
While the measurement has limited statistics, it provides a powerful proof-of-principle 
of the enhanced PDF sensitivity of such ratio measurements, 
due to the partial cancellation of the dominant theoretical 
and 
experimental 
systematic uncertainties, as discussed in the previous section and
in Ref.~\cite{Mangano:2012mh}. The data were shown to be sensitive to the large-$x$ gluon, 
both reducing the uncertainty and favoring a larger gluon at high $x$ compared to 
the fit including only HERA I data.

The jet measurements performed using 2011 data
have substantially higher precision 
compared to previous measurements, and span a wider
kinematic range, extending up to 2 TeV in inclusive jet $p_T$, 
and up to 5 TeV in dijet or trijet invariant mass.
They
are particularly interesting 
since,
for the first time, the full statistical and systematic 
correlations between various measurements are provided,
allowing their simultaneous inclusion
in a PDF fit, 
thus enhancing 
the
constraining information. Jet measurements using 8 TeV data are also ongoing.

Measurements of top quark pair production provide complementary 
information on the gluon PDF at high $x$, and at the same time are 
sensitive to the strong coupling and the top mass. 
The ATLAS collaboration has released many measurements of the top quark pair production  
cross section~\cite{Aad:2010ey,Aad:2011yb,Aad:2012qf,Aad:2012aa,Aad:2012mza,Aad:2012vip,Aad:2014kva} 
with increasing precision and using a variety of final states. Several of these measurements have been used 
in recent global fits~\cite{Ball:2014uwa,Harland-Lang:2014zoa}.
Differential cross sections, using the 2011~\cite{Aad:2012hg,Aad:2014zka} dataset, have also been measured. 
In the most recent of these measurements, using the full 2011 dataset,  
the normalized differential cross sections are measured as a function of the top quark transverse momentum, 
and of the top quark pair 
mass, transverse momentum and rapidity.
Normalizing the cross sections 
reduces the dependence on higher order QCD corrections, 
though it also slightly degrades the sensitivity to the gluon PDF overall normalization.
When compared to a range of PDFs, 
the measured data distributions tend to be softer than the predictions, 
indicating either their power to constrain the gluon, and/or the importance of
higher order QCD and/or electroweak corrections. 
Further measurements at 8 TeV are also ongoing. These, as well as future precision measurements at higher energy, 
could place more significant constraints on the gluon. 

Additional constraints on the gluon PDF at medium and large-$x$
could be provided by isolated prompt photon data. Compared to jets, 
prompt photons feature a cleaner experimental environment,
though current measurements are of limited precision. 
ATLAS has measured isolated prompt photon production
cross sections in the 2010~\cite{Aad:2011tw} and 2011~\cite{Aad:2013zba} datasets. 
The latter provide differential cross sections 
as a function of photon transverse energy and pseudo-rapidity, 
in central and forward pseudo-rapidity regions.
The ATLAS collaboration
has studied the sensitivity of these data to PDFs~\cite{ATL-PHYS-PUB-2013-018}, 
providing a quantitative $\chi^2$ comparison of the agreement
with NLO QCD predictions for a range of PDFs. 
The results show some tension between data and theory for several current PDFs,
indicating potential to constrain the shape and uncertainty of 
the gluon, though theoretical scale uncertainties are also large.
Measurements of prompt photons in association with jets~\cite{ATLAS:2012ar} are also available.

Inclusive electroweak boson production data provide constraints on quarks and quark flavor
separation.
ATLAS has published $W$ and $Z$ rapidity distributions using the 2010 dataset,
including full experimental correlations~\cite{Aad:2011dm}. 
This has become a standard dataset, now widely used in the global PDF fits.
In a separate ATLAS analysis, these data were used in a full PDF fit using the
{\tt HERAfitter} framework,
in order to quantify the sensitivity to the strange quark content of the proton\cite{Aad:2011dm}.  
The result favors a non-suppressed strangeness. The data also provide constraints on the valence quarks. 
The ATLAS inclusive $W$ and $Z$ analysis of 2011 data is also close to publication.
The larger dataset will allow for more differential
measurements, such as the double differential Z/$\gamma^*$ cross sections binned in 
dilepton mass and rapidity, and $W^\pm$ cross sections in bins 
of lepton pseudo-rapidity, as well as double differentially 
in lepton pseudo-rapidity and lepton transverse momentum.
Importantly, these data have the potential to provide constraints on valence quark PDFs, and will shed 
further light on the strangeness content of the proton. 

ATLAS has also published low~\cite{Aad:2014qja} and high\cite{Aad:2013iua}
mass Drell-Yan measurements, 
providing additional and complementary information to the measurements
around the $Z$ mass peak.
The ATLAS low mass Drell-Yan cross sections are measured as a 
function of dilepton invariant mass with coverage $12 < M_{ll} < 26$ GeV for the 2010 dataset,
and $26 < M_{ll} < 66$ GeV on a subset of the 2011 dataset.
High-mass Drell-Yan measurements can provide important constraints on large $x$ 
quarks and anti-quarks. The ATLAS analysis,
based on the 2011 dataset, 
reaches dilepton invariant masses
up to 1.5~TeV
and has been included in recent PDF fits~\cite{Ball:2014uwa,Harland-Lang:2014zoa}.
In addition, in the presence of QED corrections, this measurement
can also be used to constrain the photon content of the proton, 
and has been included in the NNPDF2.3QED fit~\cite{Ball:2013hta}.
A higher precision measurement at 8 TeV is in preparation.
ATLAS has also studied the forward-backward asymmetry in Drell-Yan events \cite{Aad:2015uau}, using 2011 data, 
which may provide sensitivity to PDFs. In the same paper the effective weak mixing angle was extracted. 

ATLAS has explored the production of vector bosons in association with heavy flavours. 
The production of $W$ in association with charm quarks has been measured differentially 
in charged lepton pseudo-rapidity, using the 2011 dataset~\cite{Aad:2014xca}.  
These data give a direct handle on the strange content of the proton, 
and the ratio of $W^+$$+$$\bar{c}$ to $W^-$$+ c$ is also  
sensitive to the strange asymmetry, $s - \bar{s}$.
In the same analysis, the data are analysed and found to be consistent with a range of PDFs, 
and indicate a preference for PDFs with an $SU(3)$ symmetric light quark sea,
consistent with the result found with the 
$W$ and $Z$ rapidity distributions. Work is underway to include such data 
fully and consistently in a PDF fit for the first time. 
Measurements of vector boson in association with beauty 
are also available~\cite{Aad:2011kp,Aad:2013vka,Aad:2011jn,Aad:2014dvb}. These 
are especially useful to test pQCD calculational schemes~\cite{Maltoni:2012pa}.

Vector boson production in association with jets provides further sensitivity to the 
gluon and sea quark PDFs. ATLAS has released a
number of measurements of $W+$ jets and $Z+$ jets 
using 2010 and 2011 data~\cite{Aad:2011qv,Aad:2012en,Aad:2013ysa,Aad:2014qxa}. 
In addition, ATLAS has measured the 
$W+$ jets to $Z+$ jets ratio~\cite{Aad:2014rta}, which is complementary to the
individual  measurements, and is especially interesting due to the 
large cancellation of experimental systematic uncertainties and non-perturbative QCD effects.
Vector boson transverse momentum distributions are also sensitive to the gluon and 
quark PDFs over a wide range of $x$. ATLAS has measured
the W $p_{T}$~\cite{Aad:2011fp} distribution with 2010 data, 
and the Z $p_T$ distribution with both 2010~\cite{Aad:2011gj} and 2011~\cite{Aad:2014xaa} data.

\clearpage

\subsection{Constraints from CMS}
The results from the CMS collaboration sensitive to PDFs are summarized in Table~\ref{table:runIoverviewCMS}. 

High-precision measurements of the cross-sections of multi-jet production in proton-proton collisions have been performed by the CMS collaboration and the 
systematic correlations have been investigated. Also, the potential of several jet measurements to constrain the PDFs and determine the strong coupling has been 
demonstrated. 

\begin{table}[t]
\centering
\footnotesize
\begin{tabular}{l|l|c|c|c}
  \hline
\hline
\multicolumn{5}{c}{CMS} \\
\hline
\hline
Measurement & $\sqrt{s}$, $\mathcal{L}_{\rm int}$  & Motivation &
Reference  & Used in PDF\\
 &  &  &  & or $\alpha_S$ fits\\
\hline

\hline
High and low mass Drell-Yan & 7 TeV,  5 fb$^{-1}$ & Sect.~\ref{sec:DY} & \cite{Chatrchyan:2013tia} & \cite{Harland-Lang:2014zoa, Rojo:2014kta}\\
High and low mass Drell-Yan & 8 TeV,  20 fb$^{-1}$ & Sect.~\ref{sec:DY} & \cite{CMS:2014jea} & --\\
Drell-Yan AFB & 7 TeV, 5 fb$^{-1}$ & Sect.~\ref{sec:DY} & \cite{Chatrchyan:2012dc} & -- \\
\hline
$W$ asymmetry &  7 TeV, 36 pb$^{-1}$ & Sect.~\ref{sec:inclusiveWZ} & \cite{Chatrchyan:2011jz} &  --\\
$W$ $e$ asymmetry &  7 TeV, 880 pb$^{-1}$ & Sect.~\ref{sec:inclusiveWZ} &  \cite{Chatrchyan:2012xt} & -- \\
$W$ $\mu$ asymmetry &  7 TeV, 4.7 fb$^{-1}$ & Sect.~\ref{sec:inclusiveWZ} & \cite{Chatrchyan:2013mza} & \cite{Chatrchyan:2013mza, Rojo:2014kta}  \\
\hline
$W, Z$ production and rapidity  & 7 TeV, 3 pb$^{-1}$ & Sect.~\ref{sec:inclusiveWZ} & \cite{Khachatryan:2010xn} & -- \\
$W, Z$ inclusive production & 7 TeV, 36 pb$^{-1}$ &  Sect.~\ref{sec:inclusiveWZ}    & \cite{CMS:2011aa} & -- \\
$W, Z$ inclusive production & 8 TeV, 19 pb$^{-1}$ &  Sect.~\ref{sec:inclusiveWZ}   & \cite{Chatrchyan:2014mua} & -- \\
\hline
$Z$ $p_T$ and rapidity & 7 TeV, 36 pb$^{-1}$ & Sect.~ \ref{sec:VPT},\ref{sec:inclusiveWZ} & \cite{Chatrchyan:2011wt} & -- \\
$Z$ $p_T$ and rapidity & 8 TeV, 19.7 fb$^{-1}$ & Sect.~ \ref{sec:VPT},\ref{sec:inclusiveWZ} & \cite{Khachatryan:2015oaa} & -- \\
\hline
Inclusive jets & 7 TeV, 5 fb$^{-1}$ & Sect.~\ref{sec:jetproduction} & 
\cite{Chatrchyan:2012bja,Chatrchyan:2014gia} & \cite{Ball:2012cx,Harland-Lang:2014zoa,Khachatryan:2014waa}  \\
Dijets & 7 TeV, 5 fb$^{-1}$ & Sect.~\ref{sec:jetproduction} & 
\cite{Chatrchyan:2012bja} & --  \\
Three-jets & 7 TeV, 5 fb$^{-1}$ & Sect.~\ref{sec:jetproduction} & 
\cite{CMS:2014mna} & \cite{CMS:2014mna}  \\
Three-jets/Di-jets ratio & 7 TeV, 5 fb$^{-1}$ & Sect.~\ref{sec:jetproduction} & 
\cite{Chatrchyan:2013txa} & \cite{Chatrchyan:2013txa}  \\
\hline
$W$+charm  & 7 TeV, 5 fb$^{-1}$  & Sect.~\ref{sec:wcharm} & \cite{Chatrchyan:2013uja}  &
\cite{Ball:2012cx,Chatrchyan:2013mza,Alekhin:2014sya}\\
$Z$+beauty & 7 TeV, 5 fb$^{-1}$  & Sect.~\ref{sec:wcharm} & \cite{Chatrchyan:2014dha} & -- \\
\hline
$\gamma$ inclusive production & 7 TeV, 36 pb$^{-1}$ & Sect.~\ref{sec:promptphoton}  & \cite{Chatrchyan:2011ue} & \cite{d'Enterria:2012yj}\\
$\gamma$+jets & 7 TeV, 2.1 fb$^{-1}$ & Sect.~\ref{sec:promptphoton}  & \cite{Chatrchyan:2013mwa} & -- \\
\hline
$t\bar{t}$  inclusive & 7 TeV, 2.3 fb$^{-1}$ & Sect.~\ref{sec:topprod} & \cite{Chatrchyan:2012bra} & \cite{Czakon:2013tha,Guzzi:2014wia,Chatrchyan:2013haa}  \\
$t\bar{t}$ differential & 7 TeV, 5.0 fb$^{-1}$ & Sect.~\ref{sec:topprod} & \cite{Chatrchyan:2012saa} & \cite{Guzzi:2014wia}\\
$t\bar{t}$ inclusive & 8 TeV, 1.14 fb$^{-1}$ & Sect.~\ref{sec:topprod} & \cite{CMS:2011qva} & \cite{Czakon:2013tha}\\
$t\bar{t}$ inclusive & 8 TeV, 2.8 fb$^{-1}$ & Sect.~\ref{sec:topprod} & \cite{CMS:2012iba} & \cite{Czakon:2013tha}\\
$t\bar{t}$ inclusive & 8 TeV, 2.4 fb$^{-1}$ & Sect.~\ref{sec:topprod} & \cite{CMS:2012lba} & ~\cite{Guzzi:2014wia} \\
$t\bar{t}$ differential & 8 TeV, 19.7 fb$^{-1}$ & Sect.~\ref{sec:topprod} & \cite{Khachatryan:2015oqa} & -- \\
\hline
\end{tabular}
\caption{\small Same as Table~\ref{table:runIoverviewATLAS},
  for the CMS experiment.
  In the last column, we also indicate which of these measurements have been used
  as input for either a determination of PDFs or of
  the strong coupling $\alpha_s$.
  \label{table:runIoverviewCMS}
}
\end{table}

Jets are reconstructed with the same anti-$k_T$ clustering algorithm used by ATLAS.
A different value of radius parameter, $R=0.7$, is chosen for jet analyses performed with only jets in the final state.
This is motivated by the fact that a smaller cone is more sensitive to the final state radiation effects, 
which are not well described by the NLO predictions in pQCD.
However, in the case of the associated production of jets with vector bosons, the value of the jet radius $R=0.5$ is preferred.

The measurement of inclusive jet production cross-sections in $pp$ collisions at $\sqrt{s} = 7$ TeV 
based on the data collected in 2011, has been published in Ref.~\cite{Chatrchyan:2012bja}
as a function of jet kinematics.
Furthermore,
the correlations of the systematic uncertainties have been reanalyzed and the recommendations for usage of the 
measurement in the PDF fits published~\cite{Khachatryan:2014waa}.
Another analysis~\cite{Chatrchyan:2014gia}, designed to test the performance and result of different jet radii,
has measured the inclusive jets cross section ratio 
using the same data with two different radii parameters: 0.5 and 0.7.
In this latter paper, an inclusive jet cross section with $R = 0.5$ is also presented, as well as the cross section with $R = 0.7$ extrapolated towards lower $p_T$.

A comprehensive QCD analysis~\cite{Khachatryan:2014waa} of the inclusive jet cross-section measurement at 7 TeV has been performed by the CMS collaboration 
to demonstrate the impact of these data on the PDFs and to determine the strong coupling constant.
The impact 
of the inclusive jet measurement on the PDFs of the proton is investigated in detail  using the
{\tt HERAFitter} 
tool~\cite{Alekhin:2014irh}, using both the Hessian~\cite{Martin:2002aw}
and the Monte Carlo methods~\cite{Ball:2010de}.
When the CMS inclusive jet data are used together with the HERA-I DIS measurements, the uncertainty 
in the gluon distribution is reduced in particular at large $x$, and a significant reduction of the
parametric uncertainty is observed.
At the same time, a modest reduction of the uncertainties on $u$ and $d$ valence quark distributions is observed,
consistent with the dominance of $qq$ scattering of jet production at high $p_T$.
The inclusion of the 
CMS inclusive jet data also allows for a combined fit of $\alpha_S (m_Z)$ and of the PDFs, which is not possible 
with the HERA
data alone.
As summarized in Table~\ref{table:runIoverviewCMS},
these inclusive jets results are already used by several PDF collaborations.
Further inclusive jets measurements at CMS are still ongoing, and are expected to extend and complete the Run I legacy picture.

Two measurements of the three-jet cross section have been performed and optimized for the extraction of $\alpha_S$ running.
The first one~\cite{Chatrchyan:2013txa} has used the ratio between the three jets cross section and the dijet cross section, that is proportional to $\alpha_S$ at leading order.
This observable has a reduced dependence on the proton PDF and is used to partially decouple the measurement of $\alpha_S$ from the gluon density.
The second analysis~\cite{CMS:2014mna} makes usage of the three jet mass spectrum, which is proportional to $\alpha_S^3$ at leading order. 
In principle this observable is more sensitive than the previous one, 
but is also more dependent on the choice of the proton PDF and suffers from larger systematic uncertainties.
The three measurements together provide for the first time a stringent test of the strong coupling running in the region between 100 GeV and 2 TeV.
In particular, these are the first direct measurements of the strong coupling constant
at the TeV scale, that can be used to provide constraints on BSM scenarios~\cite{Becciolini:2014lya}.

The dijet cross-sections~\cite{Chatrchyan:2012bja} have been measured using the CMS data collected in 2011 at $\sqrt{s} = 7$ TeV.
This measurements exhibit a significant statistical correlation with the inclusive jets case.
Since no statistical correlation
matrix has been provided between the various CMS jet measurements,
it is not possible to use them at the same time in a PDF analysis.

A significant effort within the CMS collaboration has been devoted
to the precise measurements of the inclusive vector boson production. 
Three sets of measurements can be identified: the neutral and charged
Drell-Yan (DY) production with a particular attention dedicated to the $Z$ peak;
the charged lepton radial asymmetry in the $W$ production (hereafter referred as $W$ asymmetry); the high $p_T$ bosons production.
While the first two measurements are expected to be mainly sensitive to quark density,  the third one should provide additional constrain on the gluon density.

The inclusive measurements in electron and muon channels of the on-peak neutral and charged DY cross section have been performed at 7 and 8 TeV using the first low luminosity 
data in order to reduce the contamination from the pile-up~\cite{Khachatryan:2010xn, CMS:2011aa, Chatrchyan:2014mua}.
Subsequently a precise measurement of the double-differential cross sections as a function of lepton-pair mass and rapidity has been produced, normalized to the peak cross section~\cite{Chatrchyan:2013tia,CMS:2014jea}. A full correlation matrix between bins of the normalized measurement as well as the peak cross section has been provided. Moreover the 8 TeV analysis has been designed to simplify the measurement of the cross section ratio between 8 and 7 TeV.
This extremely precise result, with typical uncertainties at a percent level in the bulk of the cross section data, is ready to be used by the PDF extraction groups and its sensitivity to the parton densities still needs to be assessed. 

The tensor properties of the DY events have been studied using the forward-backward asymmetry in DY events~\cite{Chatrchyan:2012dc} 
and subsequently used to extract the
effective Weinberg angle\cite{Chatrchyan:2011ya}. The former measurement may also provide a certain sensitivity to the PDFs, which has not been studied yet.

The lepton-charge asymmetry measurements in $W$-boson production has been performed separately with muons and electrons 
that are sensitive to different experimental systematic effects. The most precise measurement 
available today~\cite{Chatrchyan:2013mza} remains the muon charge asymmetry measurement at CMS, performed with the full 7 
TeV data set, while the electron charge asymmetry is limited by the available $p_T$ single lepton trigger to 880~pb$^{-1}$ 
data sample~\cite{Chatrchyan:2012xt}. The sensitivity of the muon charge asymmetry to the valence quark density has been
studied in a QCD analysis at NLO in~\cite{Chatrchyan:2013mza}. A significant reduction of the uncertainty on the 
$d$-valence and $u$-valence distributions is observed with respect to a PDF fit in which only HERA-I inclusive DIS data are used.
The lepton-charge asymmetry is now a standard component of the PDF extraction by many global fit groups. Even more precise 
measurements of muon charge asymmetry at $\sqrt{s}=8$ TeV by the CMS collaboration is ongoing. 

The on-peak Z-boson production cross section has been measured double differentially in $p_T$ and $y$ 
in the muon channel with the experimental precision at a percent level~\cite{Khachatryan:2015oaa}. This measurement should provide an additional constraint on the gluon density.
The available predictions for the boson production with $p_{ T}(Z) \approx M_{Z}$ are available only at NLO, while an NNLO prediction would to be necessary to explore the full advantage of the experimental precision.

CMS has also explored the production of vector-boson associated with heavy quarks.
The cross sections of the associated production of the $W$ boson together with charm quark 
has been measured differentially as a function of charged lepton rapidity at 7 TeV~\cite{Chatrchyan:2013uja}. 
This measurement provides a direct probe of the 
strange-quark content of the proton sea, as demonstrated by the CMS collaboration in a QCD analysis~\cite{Chatrchyan:2013mza} 
at NLO, in which HERA-I DIS data, measurements of muon charge asymmetry and the cross sections of $W+$charm production are used. 
The strange-quark content, as determined by the analysis, is demonstrated to be consistent with results of the neutrino-scattering experiments.   
The $Z+b$ production at 7 TeV~\cite{Chatrchyan:2014dha} is measured single-differentially due to a lack of statistics, but a differential measurement is expected to be provided at 8 TeV.
Besides to their sensitivity to the PDFs, the measurements of gauge boson production in association with heavy quarks provide useful information about applicability of different heavy-quark schemes in the probed energy regime.

Measurements of the 
top-pair production at the LHC probe the gluon distribution at high $x$ and at the same time provide constraints on 
the top-quark mass and the strong coupling constant. For the first time, the value of $\alpha_s$ has been 
determined~\cite{Chatrchyan:2013haa} at NNLO using the inclusive $t\bar{t}$ production cross section measured by the CMS collaboration~\cite{Chatrchyan:2012bra}. The impact of the inclusive cross section of $t\bar{t}$ production on the gluon distribution is studied~\cite{Czakon:2013tha}, where the CMS measurements at $\sqrt{s}=7$ TeV~\cite{Chatrchyan:2012bra} and $\sqrt{s}=8$ TeV~\cite{CMS:2011qva,CMS:2012iba} are included. 
In the QCD analysis~\cite{Guzzi:2014wia}, the inclusive and differential cross sections of the top-quark pair production 
are included and a moderate reduction of the uncertainty on the gluon distribution at high $x$ is demonstrated. In this 
analysis, the CMS  measurements of total~\cite{Chatrchyan:2012bra,CMS:2012lba} and differential~\cite{Chatrchyan:2012saa} top-pair 
production cross sections are used. More significant improvement of the precision of the gluon distribution is expected with more 
precise data of the LHC at higher energies. 
It is important to notice that, for future PDF fits using the top-pair production 
measurements, the parton-level cross sections provided in the full phase space should be supplemented by the information about 
correlations of the statistic and systematic uncertainties, also between the data sets of different energies and between 
inclusive and (normalized) differential cross section measurements.

\subsection{Constraints from LHCb}
\label{sec:lhcbconstraints}

The LHCb experiment, thanks to its unique forward coverage, extends
the kinematical range covered by ATLAS and CMS and in particular
allows to explore in better detail the small-$x$ region~\cite{Thorne:2008am}.
Therefore, even for the same underlying  physical process, LHCb measurements are
fully complementary to those of ATLAS and CMS.
The corresponding overview of LHCb results
are summarized in Table~\ref{table:runIoverviewLHCb}. 

Measurements of $W$ and $Z$ production using muon final states have been performed with 37 pb$^{-1}$ of data collected in 2010~\cite{Aaij:2012vn}. 
These measurements, along with those of $Z$ production in the di-electron channel at 7 TeV~\cite{Aaij:2012mda}, 
have been incorporated by the CT, MMHT and NNPDF collaborations into their
latest PDF fits~\cite{Ball:2014uwa,Harland-Lang:2014zoa}. 
Updated measurements of the $W$ and $Z$ production cross-sections and their ratio have since been performed 
with the full 2011 dataset~\cite{Aaij:2014wba, Aaij:2015gna}.
Among these, Ref.~\cite{Aaij:2015gna} 
contains the most up-to-date and precise measurement of both the $W$ and $Z$ cross-sections. 
The precision is significantly improved due to the larger data sample, a better understanding of the detector effects, 
and an improved luminosity determination~\cite{Aaij:2014ida}. As regards the dataset collected in 2012 at a centre-of-mass energy of 8~TeV, 
$Z$ production has been measured in the di-electron channel~\cite{Aaij:2015vua}, 
with $W$ and $Z$ measurements in the more precise muon channels expected to follow in 2015.

Low-mass Drell-Yan measurements at LHCb are sensitive to $x$ values as low as 8~x~10$^{-6}$ at $Q^2=25$ GeV$^{2}$. 
A preliminary measurement has been performed by the collaboration at 7 TeV~\cite{LHCb-CONF-2012-013} and work is ongoing 
to finalize the result with the Run-I dataset. Measurements of the associated production of $Z$ bosons with $b$-quarks and $D$ mesons 
have been performed in~\cite{Aaij:2014gta,Aaij:2014hea} while more recent measurements of $W$ production in association with 
beauty and charm jets are also presented in~\cite{Aaij:2015cha}. 
In the latter measurement, the jets are identified using the algorithm outlined in~\cite{Aaij:2015yqa} 
achieving a 65\% (25\%) efficiency for identifying beauty (charm) jets with a corresponding light-jet mis-tag rate of 0.3\%. 
The first observation of top quark production in the forward region, relevant for constraining the large-$x$ gluon PDF, 
has been also presented in ~\cite{Aaij:2015mwa}.

Measurements of inclusive beauty and charm quark production have been performed \cite{Aaij:2013mga, Aaij:2013noa} 
using data collected in 2010 and 2011 at 7 TeV. The measurements exploit LHCb's particle identification and vertexing 
capabilities to fully reconstruct $B$ and $D$ mesons using hadronic decay modes. As discussed in Sect.~\ref{sec:heavyquarkprod}, 
heavy flavor production can be used to constrain the gluon distribution at low-$x$ and the impact of these results on the PDFs 
is under study by a number of groups~\cite{Zenaiev:2015rfa}.,

As discussed in Sect.~\ref{sec:centralexclusive}, precise measurements of $J/\psi$ and $\Upsilon$ photo-production 
can also lead to strong constraints on the low-$x$ gluon distribution~\cite{Jones:2013pga}. 
As these processes are characterized by events containing just two muon tracks and a large rapidity gap, 
LHCb is well suited to their detection due to its relatively low pile-up running conditions and partial backward coverage. 
Measurements have been made of central exclusive $J/\psi$ production at a centre-of-mass energy of 7 TeV~\cite{Aaij:2014iea} 
with $\Upsilon$ production in collisions at 7 and 8~TeV~\cite{Aaij:2015kea}.

%

\begin{table}[t]
\centering
\small
\begin{tabular}{l|l|c|c|c}
  \hline
\hline
\multicolumn{5}{c}{LHCb} \\
\hline
\hline
Measurement & $\sqrt{s}$, $\mathcal{L}_{\rm int}$  & Motivation &
Reference  & Used in PDF fits \\
\hline
\hline
$W, Z$ muon rap dist &  7 TeV, 1.0 fb$^{-1}$ & Sect.~\ref{sec:inclusiveWZ} &
~\cite{Aaij:2015gna} &   \cite{Ball:2014uwa,Harland-Lang:2014zoa} \\
\hline
$Z\to ee$ rap dist &  7 TeV, 0.94 fb$^{-1}$ & Sect.~\ref{sec:inclusiveWZ} &
\cite{Aaij:2012mda} &  \cite{Ball:2014uwa,Harland-Lang:2014zoa}\\
$Z\to ee$ rap dist &  8 TeV, 2.0 fb$^{-1}$ & Sect.~\ref{sec:inclusiveWZ} &
\cite{Aaij:2015vua} & -- \\
 \hline
$W+b/c$ & 7,8 TeV, 3.0 fb$^{-1}$ & Sect.~\ref{sec:wcharm} & \cite{Aaij:2015cha} & --  \\
\hline
$c\bar{c}$ production & 7 TeV, 15 nb$^{-1}$  & Sect.~\ref{sec:heavyquarkprod} &
\cite{Aaij:2013mga} &  \cite{Gauld:2015yia,Zenaiev:2015rfa}  \\
$b\bar{b}$ production & 7 TeV, 0.36 fb$^{-1}$  & Sect.~\ref{sec:heavyquarkprod} &
\cite{Aaij:2013noa} &  \cite{Zenaiev:2015rfa}  \\
\hline
Exclusive $J/\psi$ production & 7 TeV, 1.0 fb$^{-1}$  & Sect.~\ref{sec:centralexclusive} &
\cite{Aaij:2014iea} &  --  \\
Exclusive $\Upsilon$ production & 7, 8 TeV, 3.0 fb$^{-1}$  & Sect.~\ref{sec:centralexclusive} & 
 \cite{Aaij:2015kea} &  -- \\
\hline
\end{tabular}
\caption{\small Same as Table~\ref{table:runIoverviewATLAS},
  for the LHCb experiment.
  \label{table:runIoverviewLHCb}
}
\end{table}
%

\section{Prospects for LHC Run II measurements}
\label{sec:prospects}

In this section we present a general overview of the plans for the ATLAS, CMS and LHCb collaborations
concerning PDF-sensitive measurements for the LHC Run II, including a possible time-line.
In addition, we present the results of a profiling analysis which provides
an estimate of the impact on PDFs on a number of Run II measurements
for $W$ and $Z$ bosons and $t\bar{t}$ production, as well of an estimate of the
impact of Run II inclusive jet measurements performed in the frameworks of the CT global
analysis.
It is worth reminding that, in the near future, complementary measurements
relevant for PDF fits will be provided also by other experiments,
including HERA and JLab and among others, but their characteristics will not be discussed here.

\subsection{Prospects for the LHC experiments}
The LHC Run II will produce proton proton collisions at 13 TeV center-of-mass energy, 
with integrated luminosity up to 300 fb$^{-1}$.
Compared to Run I, the higher center-of-mass energy implies larger cross 
sections and extended kinematic reach for many processes of interest
like jets, Drell-Yan, prompt photons, $t\bar{t}$ and vector
bosons in association with heavy quarks. For example, an increase by a factor 2 
for the electroweak vector bosons, and a factor 4 for the $t\bar{t}$, 
are expected for the inclusive production cross sections at 13 TeV compared to 8 TeV.
Therefore, Run II data will provide complementary PDF sensitivity
with respect to the measurements performed during LHC Run I at 7 and 8 TeV.
Furthermore, the increased integrated luminosity will lead to a significant
reduction of the statistical and systematic uncertainties.

In addition to the total and fiducial cross sections,
a special role will be reserved for the measurements of cross section ratios,
also involving different center-of-mass energies, 
which provide more stringent PDF constraints 
thanks to substantial cancellation of systematic uncertainties,
provided a careful treatment of the correlations.

The time-line behind the program for measurements sensitive to PDF during the LHC Run II 
will be most likely based on the optimal usage of data collected with different running conditions, 
which are expected to change substantially over time.
The very first set of data delivered by the LHC is expected to contain limited pile-up (PU), 
typically below 5 interactions per bunch crossing. Depending on the integrated luminosity,
which could sum up to 30 pb$^{-1}$ or more,
these data could be used for a quick but precise determination of the benchmark cross sections for the inclusive $Z$ and $W$-boson production. 
In particular, a limited amount of PU significantly simplifies the extraction of the $W$ cross section, affected by the performance
of the missing transverse energy, and the use of low trigger threshold for the electron channel, otherwise affected by large fake rates.
If the statistical uncertainty of the sample allows, 
measurements of differential distributions as functions of the boson $p_T$ and $y$ could be performed on the same dataset.

A subsequent period of data taking with bunch spacing of 50 ns and integrated luminosity up to 1 fb$^{-1}$ is foreseen.
These data will be provided with pileup conditions very similar to those occurring at the end of Run I (PU=20) and will represent 
a perfect candidate to measure the cross section ratios at different center-of-mass energies.
The rest of the data, corresponding to the largest part of the integrated luminosity,
will be collected with a bunch spacing of 25 ns, with pileup rapidly increasing from 20 to 40, and probably more. 
These data will be used for the long term program of Run II,
where measurements with increases statistical accuracy and wider phase space coverage will be delivered.

\subsubsection{ATLAS and CMS}

Run II measurements of Drell-Yan production
as a function of the dilepton invariant mass distribution
will potentially improve their experimental precision,
providing information down to $x \sim 10^{-4}$
in the low mass 
region where PDF uncertainties are large.
High mass measurements will also benefit significantly
from the new conditions, substantially improving their statistical 
precision
and allowing extended coverage up to 3 TeV,
thus providing direct constraints on the
poorly known quark and antiquark PDFs at large $x$
and provide constraints on the photon PDF.

Further measurements of vector boson $p_{T}$ distributions, 
and of vector bosons in association with jets
(including their ratios) are planned, where
both the kinematic reach in $p_T$ and the experimental
uncertainties can be improved as compared to
the corresponding 8 TeV measurements.

Measurements of vector boson in association with heavy flavor production 
are also of significant interest for Run II.
In fact, measurements like $W$$+$$c$$/$$W$$+$$D^*$ performed by ATLAS, are statistically limited
and the new data can substantially
reduce the statistical uncertainty. A factor of 2 in this respect might be achieved already with 2015 data,
potentially allowing a widening of phase space (with the extended coverage
at low track $p_T$, provided by the newly inserted Insertable $B$-layer, IBL in ATLAS).
Given that the ATLAS inclusive $W$ and $Z$ in Run I have suggested 
an enhanced strangeness content of the proton,
supported by the current ATLAS Run I $W$$+$$c$ data,
it will be important to measure this process at Run II with the highest 
precision possible, to shed further light.
The same emphasis will be put by CMS on detailed study of heavy flavour production.
Both the collaborations will able to make higher precision measurements 
of vector bosons in association with bottom quarks, providing a 
means to explore different heavy flavor schemes, among other things.
While $Z$$+$$b$ is known to be a channel more sensitive to the flavour scheme used in PDF evolution than the PDF content itself, $W$$+$$c$ was demonstrated to provide an impact on strangeness content of the proton. Finally the $Z$ or $\gamma$$+$$c$ and $W$$+$$b$ channels are expected to provide for the first time constraints for the intrinsic charm content of the proton~\cite{Bednyakov:2013zta}.

Jet measurements at Run II will allow an extended kinematic reach 
up to inclusive jet transverse momenta of around 3.5 TeV. 
Again, ratio measurements at different center-of-mass energy, 
which will require careful consideration of correlated systematics 
between Run I and Run II data, can give a better control of dominant 
systematic uncertainties, as already demonstrated by the previous 
ATLAS measurement of the ratio of the 2.76 to 7 TeV inclusive jet cross 
sections (see Sect.~\ref{subsec:runI:atlas}). As for Run I,
measurements of dijet, trijet and multi-jet cross sections 
will also be possible, extending to higher scales
and potentially providing further constraints on PDFs and $\alpha_S$.
The CMS plans also include maintaining the effort and expertise to extend the tests of $\alpha_S$ running in the multi-TeV range. In particular the dijet production is expected to be measured triple-differentially in $m_{\rm jj}$, $y_{\rm j1}$ and $y_{\rm j2}$. This setup was proposed by the authors of Ref.~\cite{Currie:2013dwa} to take the best advantage of the NNLO calculations once these results become public  and can be used for the  $\alpha_S$ and PDF extraction.

Prompt photon production will also benefit from Run II, 
providing improved precision on the measurements, which
is required for these data to have significant PDF-constraining power.
At 13 TeV, the top quark pair production cross section is increased 
by a factor of 4.7 (3.3) compared to 7 (8) TeV. ATLAS will be able
to perform higher precision measurements of total and differential 
(normalized and absolute) cross sections,
as well as ratio measurements at different centre of mass energies,
which can help constrain and disentangle the high $x$ gluon PDF, and $\alpha_S$.

While the potential of $t\bar{t}$ differential production to constrain the gluon PDF was demonstrated with the Run I data, a statistically larger sample is required to make a sizeable impact on PDFs.
A number of differential distributions will be measured, in particular allowing to extend the coverage
of the gluon PDFs towards larger values of $x$.

Finally, both the ATLAS and CMS experiments 
foresee the measurement of cross section ratios
different center-of-mass energies, as well as double ratios of
different processes (e.g. $t\bar{t}$, $Z$, $W^+$, $W^-$).

\subsubsection{LHCb}
\label{sec-prospectslhcb}

The increased centre-of-mass energy extends the kinematic range of the experiment to lower $x$ values for $W$, $Z$ and low-mass Drell-Yan production. 
As shown in~\cite{Farry:2015xha},
LHCb measurements of the differential charged lepton asymmetry in $W+$ jet events  in  Run II have the potential to provide important PDF constraints. 
The forward acceptance of the LHCb detector, in addition to a significant $p_{\rm T}$ requirement on the jet, 
extends the sensitivity of the measurement to $x$ values of greater than 0.5, where reductions of up to 35\% 
on the $d$-quark PDF uncertainty are achievable. 
Larger cross-sections are also expected in  Run II for the production of $W$ and $Z$ bosons in association with heavy quarks, 
and more precise measurements can be expected. In particular, measurements of $W$ production 
in association with charm jets or $D$ mesons will provide information on the strange content of the proton complementary to that from ATLAS and CMS.

The greater centre-of-mass energy in  Run II will also result in a dramatic increase in the $t\bar{t}$ production cross-section in the LHCb fiducial region. Consequently measurements of $t\bar{t}$ production can be made with a much improved statistical precision.
Such a measurement, originally proposed in the context of the forward-backward asymmetry~\cite{Kagan:2011yx}, 
will provide important information on the large-$x$ gluon PDF~\cite{Gauld:2013aja}.

In addition to extending coverage to an even lower $x$-region, measurements of $b\bar{b}$ and $c\bar{c}$ production in  Run II will allow a determination of the production ratio of heavy quarks at different centre of mass energies. The relatively large theoretical uncertainties present in the predictions for these processes make the ratios particularly attractive as a partial cancellation is expected. As such, the ratios may provide more stringent constraints on the PDFs than the individual measurements.

The installation of a dedicated forward shower counter system (HERSCHEL) on 
the LHCb detector ahead of  Run II 
has the potential to improve the precision of measurements of central exclusive production 
by extending its
coverage into the very forward region. 
Current LHCb measurements of exclusive $J/\psi$ and $\psi(2S)$ production~\cite{Aaij:2014iea} 
contain large backgrounds arising from inelastic production where the dissociation of one or both protons is not detected. 
HERSCHEL allows such events to be rejected by identifying forward showers through the interaction of high rapidity particles with the beam pipe.
Consequently, a higher purity and precision can be expected also for these  Run II measurements.

\subsection{Constraining PDFs with Run II data: a profiling analysis}

\label{sec:profiling}

The upcoming Run II data will provide rich information on PDFs.
Compared to the Run I data, higher center-of-mass energy  
extends the probed kinematic range
while larger data samples should lead to reduced uncertainties. In the following a possible impact of the LHC data is estimated using 
the Hessian PDF profiling method which is implemented in the
{\tt HERAFitter} program. For this purpose, 
benchmark measurements,
 such as inclusive $W$, $Z$ and $t\bar{t}$ production are considered. 
An estimate of the data uncertainties is based on the existing Run I measurements which were published by the ATLAS and CMS
collaborations. The inclusive measurements are typically 
dominated by the systematic uncertainties 
already for the Run I based results, however several components of the systematic uncertainty may be reduced with increased
data statistics. Thus a simplified procedure is used to estimate 
the uncertainties of the  Run II measurements.   
Three possible scenarios are considered: baseline, when the data uncertainties are taken to be similar to those of  the Run I measurements; conservative,
when the data uncertainties are scaled up by factor of two; and aggressive, when the data uncertainties are reduced by factor of two.  

The study is an indication of the LHC Run II data sensitivity however it is not meant to be an exhaustive investigation.
For example, other measurements such as  off-peak neutral-current Drell-Yan production,
$W^{\pm}c^{\mp}$ charge asymmetry, and vector boson production in the forward region, 
which can be measured at the LHCb, are not considered.

\subsubsection{PDF profiling and theoretical predictions}
The impact of a pseudo-data set on a Hessian PDF set can be quantitatively
estimated with a profiling procedure~\cite{Paukkunen:2014zia,Camarda:2015zba}.\footnote{For Monte Carlo sets instead one should use the Bayesian reweighting method~\cite{Ball:2011gg,Ball:2010gb}.} The
profiling can be performed by minimizing a $\chi^2$ function 
comparing data and theory predictions
which includes both
the experimental uncertainties and the theoretical uncertainties
arising from PDF variations:
\begin{eqnarray}
\nonumber \lefteqn{\chi^2(\boldsymbol{\beta_{\rm exp}},\boldsymbol{\beta_{\rm th}}) = }  \\ 
&& \nonumber \sum_{i=1}^{N_{\rm data}} \frac{\textstyle \left( \sigma^{\rm exp}_i + \sum_j \Gamma^{\rm exp}_{ij} \beta_{j,\rm exp} - \sigma^{\rm th}_i - \sum_k \Gamma^{\rm th}_{ik}\beta_{k,\rm th} \right)^2}{\Delta_i^2} \\
&&   + \sum_j \beta_{j,\rm exp}^2 + \sum_k \beta_{k,\rm th}^2\,  .   \label{eq:chi2prof}
\end{eqnarray}
The correlated experimental and theoretical uncertainties are included
using the nuisance parameter vectors $\boldsymbol{\beta_{\rm exp}}$
and $\boldsymbol{\beta_{\rm th}}$, respectively. Their influence on
the data and theory predictions is described by the $\Gamma^{\rm
exp}_{ij}$ and $\Gamma^{\rm th}_{ik}$ matrices. The index $i$ runs
over all $N_{\rm data}$ data points, whereas the index $j$ ($k$)
corresponds to the experimental (theoretical) uncertainty nuisance
parameters. The measurements and the uncorrelated experimental
uncertainties are given by $\sigma^{\rm exp}_i$ and $\Delta_i$\,, respectively, and
the theory predictions are $\sigma_i^{\rm th}$. Following Ref.~\cite{Camarda:2015zba}, the profiling procedure is generalized
to account for asymmetric uncertainties:
\begin{eqnarray}
   \Gamma^{\rm th}_{ik} \to \Gamma^{\rm th}_{ik} +  \Omega^{\rm th}_{ik}\beta_{k, \rm th}\,, \label{eq:iter}
\end{eqnarray}
where $\Gamma^{\rm th}_{ik} = 0.5(\Gamma^{\rm th+}_{ik} - \Gamma^{\rm
th-}_{ik})$ and $\Omega^{\rm th}_{ik} = 0.5(\Gamma^{\rm th+}_{ik}
+ \Gamma^{\rm th-}_{ik})$ are determined from the shifts of
predictions corresponding to up ($\Gamma^{\rm th+}_{ik}$) and down
($ \Gamma^{\rm th-}_{ik}$) PDF uncertainty eigenvectors.

The values at the minimum of the nuisance parameters
$\beta^{\rm min}_{k,\rm th}$ can be interpreted as optimization
(``profiling'') of PDFs to describe the data. When profiling is
performed using pseudo-data, 
for which the data central values coincide with the 
prediction, the shifts of the PDF nuisance parameters vanish. 
However after the  profiling the  nuisance parameters have
reduced uncertainties which directly affects the uncertainty bands of the PDFs.

The predictions for  Drell-Yan production are obtained using the
{\small\sc FEWZ} program~\cite{Anastasiou:2003ds}. The predictions of $t\bar{t}$ production
are calculated using the {\sc\small top++} program~\cite{Czakon:2011xx}. All the calculations are performed at NNLO accuracy. The  PDF sets used for 
the profiling are CT10nnlo~\cite{Gao:2013xoa}, MMHT14~\cite{Harland-Lang:2014zoa} and a Hessian version of  
NNPDF3.0~\cite{Ball:2014uwa,Carrazza:2015aoa}.\footnote{
  Using the {\tt mc2hessian} algorithm developed in Ref.~\cite{Carrazza:2015aoa},
any Monte Carlo PDF set can be converted into a Hessian representation and thus the profiling method can be applied.
Usage of the profiling method on a hessian version of a Monte Carlo PDF set was checked on a MMHT2014\_hessian set, 
that was extracted from the hessian $\to$ MC $\to$ hessian transformation.
The size of the observed constraints was found to be similar to those on the original MMHT2014 PDF set.
} 
When needed, the PDF uncertainties are re-scaled to the $68\%$ confidence level.

The central values of the pseudo-data are 
taken to be equal to the central values of the predictions. The profiling uses $\Delta \chi^2=1$ criterion for the uncertainty
estimate, thus the impact of the data on the PDF uncertainties may differ compared to inclusion  in a full PDF fit, especially 
if there is a tension among different data sets.
Recall that Hessian global PDF fits
use alternative methods that produce PDF uncertainties that are larger than those for the $\Delta \chi^2 =1$ condition.\footnote{
  In particular, CT10 uses a two-tier method for the computation of the PDF uncertainty that is not equivalent to the $\Delta \chi^2 =100$ tolerance~\cite{Lai:2010vv}.} 

\subsubsection{Generation of pseudo-data}

The pseudo-measurements selected for the study satisfy the following criteria:
\begin{enumerate}
\item There are  NNLO predictions available. This 
requirement ensures that the  theoretical uncertainties  are smaller than 
the PDF uncertainties and comparable to the ultimate data uncertainties. 
In the following, other theoretical uncertainties such
as scale variations are neglected.
\item The data have $\sim 1\%$
 accuracy and can be described by a simple correlation model. This criterion excludes final states with jets, 
such as inclusive jet and vector boson plus jet production. 
With  recent developments of  NNLO calculations, these data may have the power to place strong constraints on the PDFs. However the impact of the data  depends strongly on measurement-specific correlation model, investigation of which
is beyond this study. 
\item The measurement can be expressed in a simplified phase-space region with well-defined particle to parton-level corrections. 
  This excludes observables such as $W$$+$ charm production.
\item Only data from the central detectors ATLAS and CMS are considered.
\end{enumerate}

\begin{table}
\begin{center}
\begin{tabular}{|l|cccc|}
\hline
                    &  $R_{W/Z}$ & $R_{\rm t\bar{t}/Z}$ & $A_{\ell}$ & $y_Z$ \\
\hline 
Kinematic range    &            &                &   $p_{t,\ell}>25\gev$, $|\eta_{\ell}|<2.5$ &       \\
Number of bins     &      1     &       1        &    10      &  12   \\
Baseline accuracy per bin   &   $1\%$         &      $2\%$          &       $\approx 1.5\%$     &  $\approx 1.5\%$ \\
\hline
\end{tabular}
\end{center}
\caption{Features of the pseudo-measurements considered for the $\sqrt{s}=13\tev$  profiling studies  \label{tab:psdata}}
\end{table}

The observables are also selected such that the correlations among them are reduced. This leads to a preference for ratio
measurements rather than absolute cross-section determinations. Measurements of absolute cross sections with full correlation
information may lead to better PDF constraints, however they depend on detector-specific correlation model, which is difficult to follow in this simplified investigation. 

Taking into account these requirements, the four pseudo-measurements used in
the present study of the PDF sensitivity of the LHC Run II at
$\sqrt{s}=13\tev$ data are the following:
\begin{itemize}
\item  Ratio of inclusive cross sections of $W$-boson to $Z$-boson production, $R_{W/Z}$. The reference measurements for this
observable are the ATLAS  measurement performed at $\sqrt{s}=7\tev$~\cite{Aad:2011dm} and the CMS measurement at $\sqrt{s}=8\tev$~\cite{Chatrchyan:2014mua}.
The ratio is  considered for the fiducial region defined by the lepton transverse momentum and pseudorapidity cuts,
$p_t>25\gev$ and $|\eta|<2.5$. The baseline uncertainty is taken to be $1\%$.
\item  Ratio of inclusive cross sections of $t\bar{t}$ to $Z$-boson production, $R_{\rm t\bar{t}/Z}$. The $t\bar{t}$ pseudo-data
are based on the ATLAS $7$ and $8\tev$ total cross-section measurement in $e\mu$ channel with $b$-tagged jets~\cite{Aad:2014kva}. This measurement reached $2\%$ 
accuracy, excluding the luminosity uncertainty. The luminosity uncertainty cancels for 
the $t\bar{t}$ to $Z$ cross-section ratio. If the $Z$ cross-section measurement is obtained using both $Z\to e^+e^-$ and $Z\to \mu^+\mu^-$ channels, a significant additional cancellation of uncertainties may be also  achieved for the  reconstruction of leptons. Thus $2\%$ uncertainty on $R_{\rm t\bar{t}/Z}$ is considered as a  baseline. The fiducial
definition for the $Z\to \ell\ell$ cross-section measurement is taken to be 
the same as for $R_{W/Z}$.
\item  Lepton charge asymmetry for $W$ decays, $A_{\ell}$. The pseudo-data are based on the CMS measurement of the muon charge asymmetry~\cite{Chatrchyan:2013mza}. The data are considered in fiducial region $p_T>25\gev$ and $|\eta_{\ell}|<2.5$. The data are binned in $10$ bins with bin width $\Delta|\eta_{\ell}|=0.25$. The baseline statistical
uncertainty  is taken to be $0.0005$ per bin, which roughly corresponds to integrated luminosity of $10$~fb$^{-1}$ of $\sqrt{s}=13\tev$ data. 
The baseline systematic uncertainty varies from 
$0.0020$ to $0.0036$ for the data from the most  central to the most forward bin. The bin-to-bin correlation model for the systematic uncertainties is taken similar to the CMS analysis, as implemented in the {\tt HERAfitter} package, 
with the correlation coefficient between $0.2$ and $0.3$.
\item Normalized inclusive $Z$-boson rapidity, $y_Z$. The pseudo-data are based on the CMS measurement of the Neutral-Current Drell-Yan production at $7\tev$~\cite{Chatrchyan:2013tia}. 
The data are considered in fiducial region $p_T>25\gev$ and $|\eta_{\ell}|<2.5$. 
The pseudo-data are binned in $12$ bins with bin width $\Delta|y_{Z}|=0.2$. 
The statistical uncertainty is expected to be negligible compared to the systematics for the Run II dataset.
The baseline total uncertainty varies between $0.00155$ and $0.00050$ for the central 
and the most forward regions, respectively.
The bin-to-bin correlation model for the systematic uncertainties is taken similar to the CMS analysis
with strong correlation for the neighboring bins, $\approx 0.7$, 
and some anti-correlation between far-apart bins, up to $-0.5$.
\end{itemize}
Basic properties of the pseudo-data samples are listed in Table~\ref{tab:psdata}.
The correlation model was kept unchanged between the baseline, aggressive and conservative 
scenarios for data uncertainties.

\subsubsection{Results}

\begin{figure}[t]
  \begin{center}
    {\includegraphics[width=0.32\textwidth]{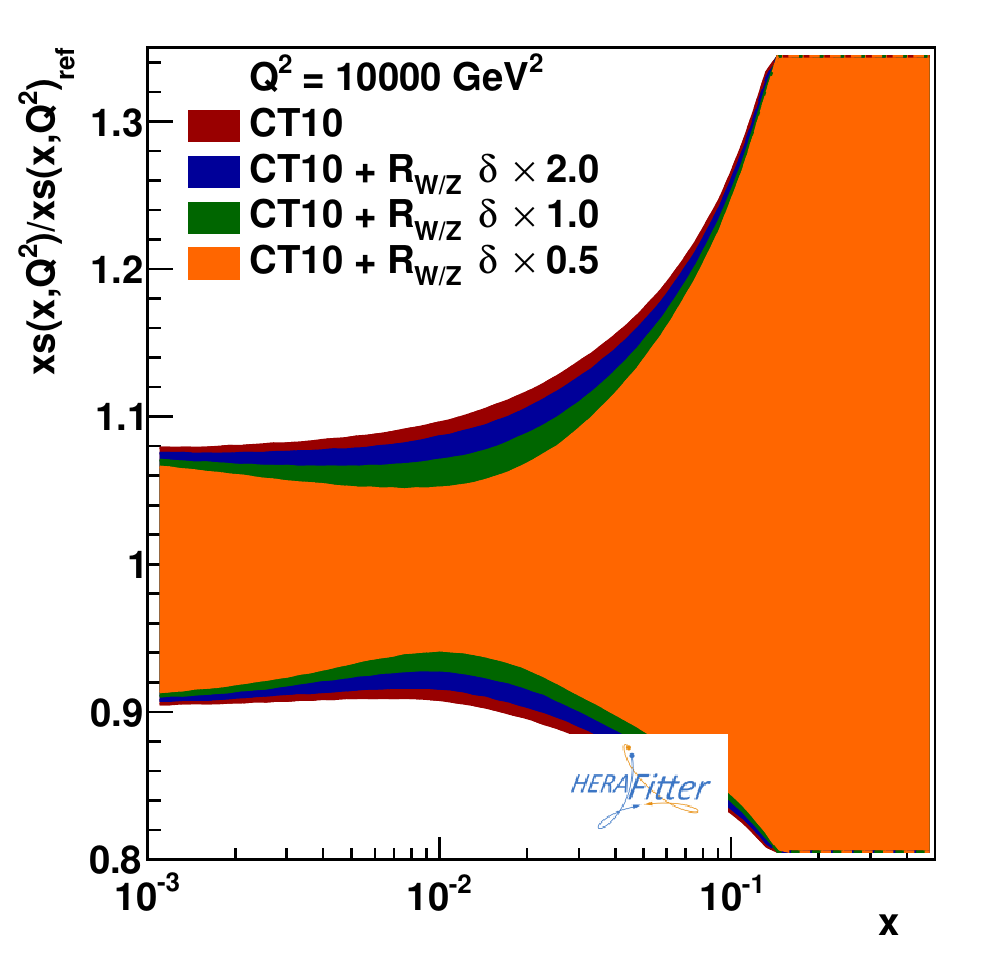}}
    {\includegraphics[width=0.32\textwidth]{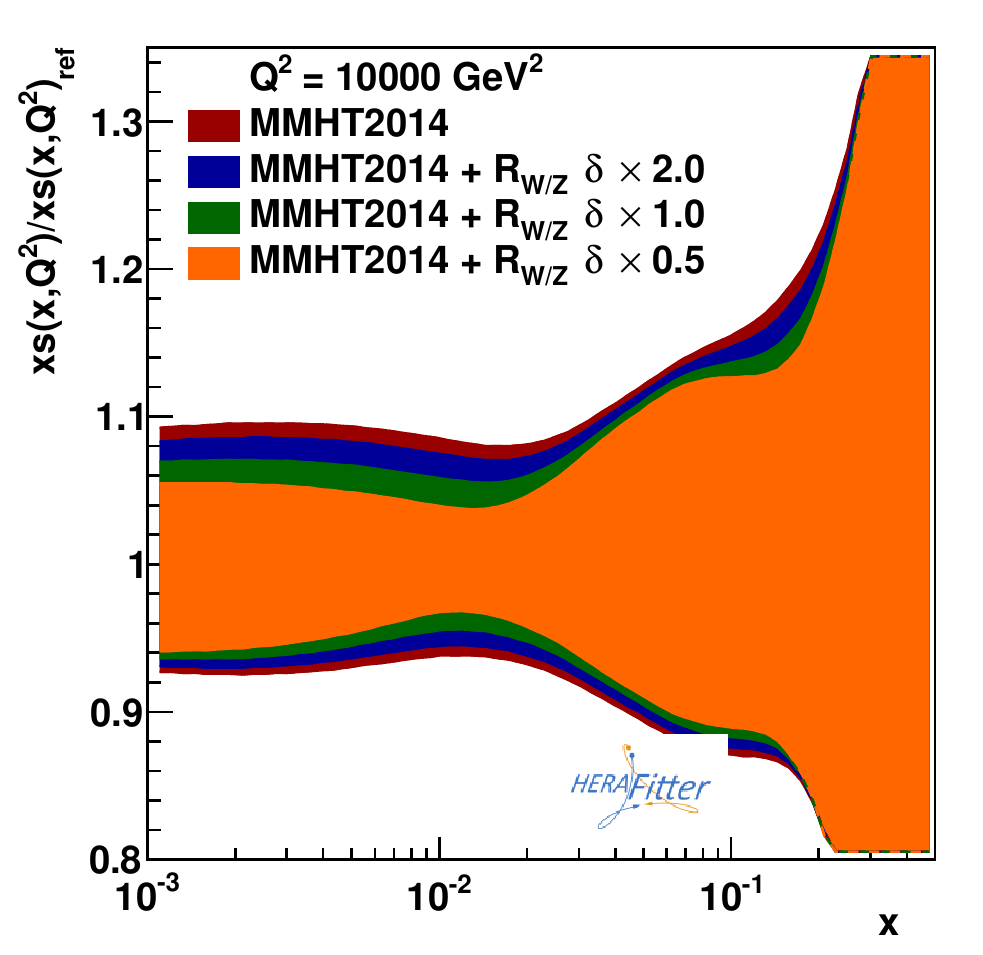}}
    {\includegraphics[width=0.32\textwidth]{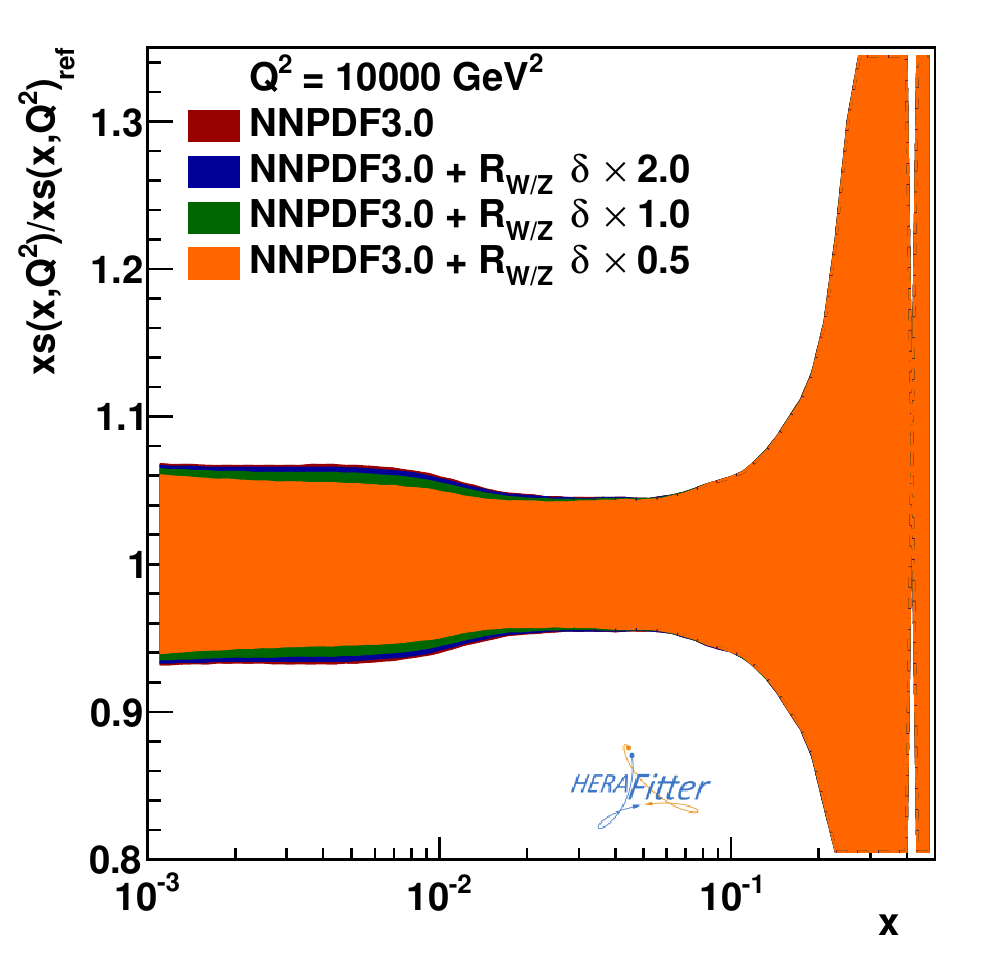}}
  \end{center}
\caption{\label{fig:rwz}Relative uncertainty of the strange-quark distribution as a function of $x$ for $Q^2=10^4\gev^2$ estimated
based on CT10nnlo (left), MMHT14 (middle) and NNPDF3.0 (right) PDF sets, respectively. The outer uncertainty band corresponds to the original
PDF uncertainty. The embedded bands represent results of the PDF profiling using $R_{W/Z}$ pseudo-data at $13$~TeV
corresponding to (from outermost to innermost band) conservative, baseline, aggressive model of the data uncertainties. }
\end{figure}

\begin{figure}[t]
  \begin{center}
    {\includegraphics[width=0.32\textwidth]{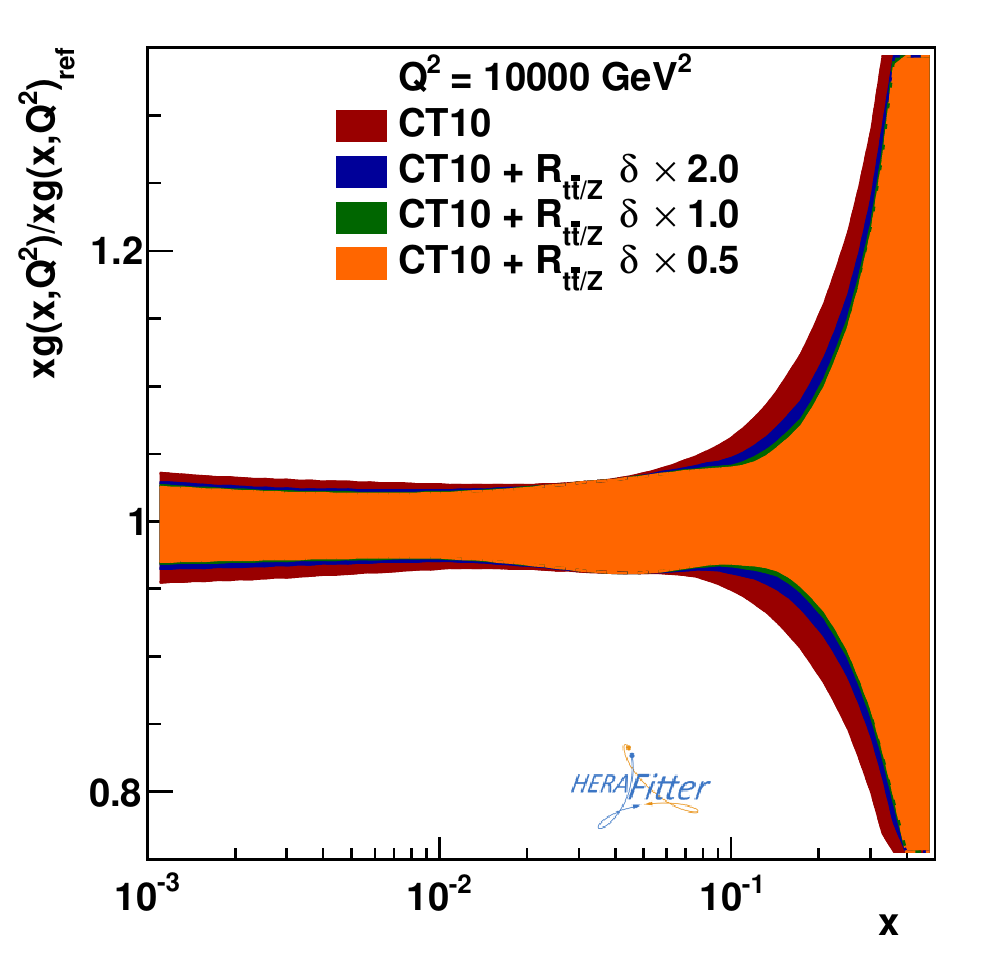}}
    {\includegraphics[width=0.32\textwidth]{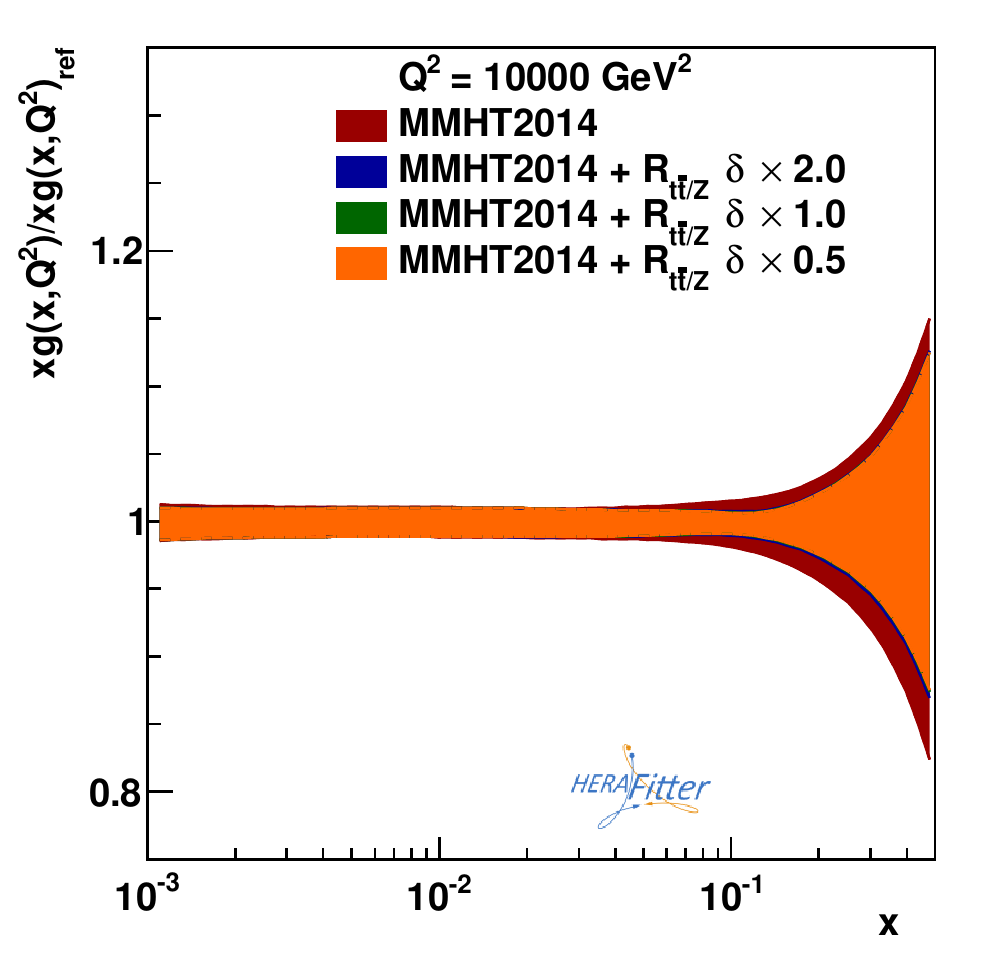}}
    {\includegraphics[width=0.32\textwidth]{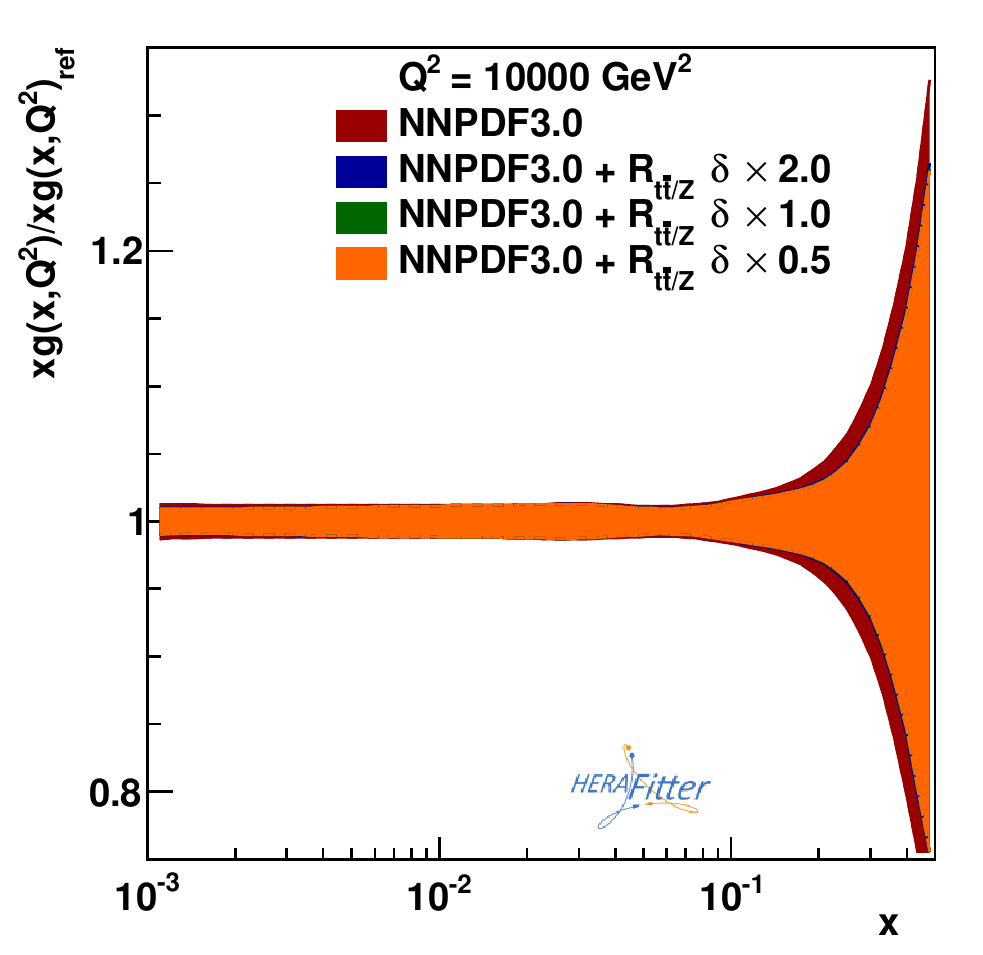}}
  \end{center}
  \caption{\label{fig:rttz} Same as Fig.~\ref{fig:rwz} this time for the gluon
    PDF, using the measurement of the $t\bar{t}/Z$ ratio as input to the
  profiling.}
\end{figure}

\begin{figure}[t]
  \begin{center}
    {\includegraphics[width=0.32\textwidth]{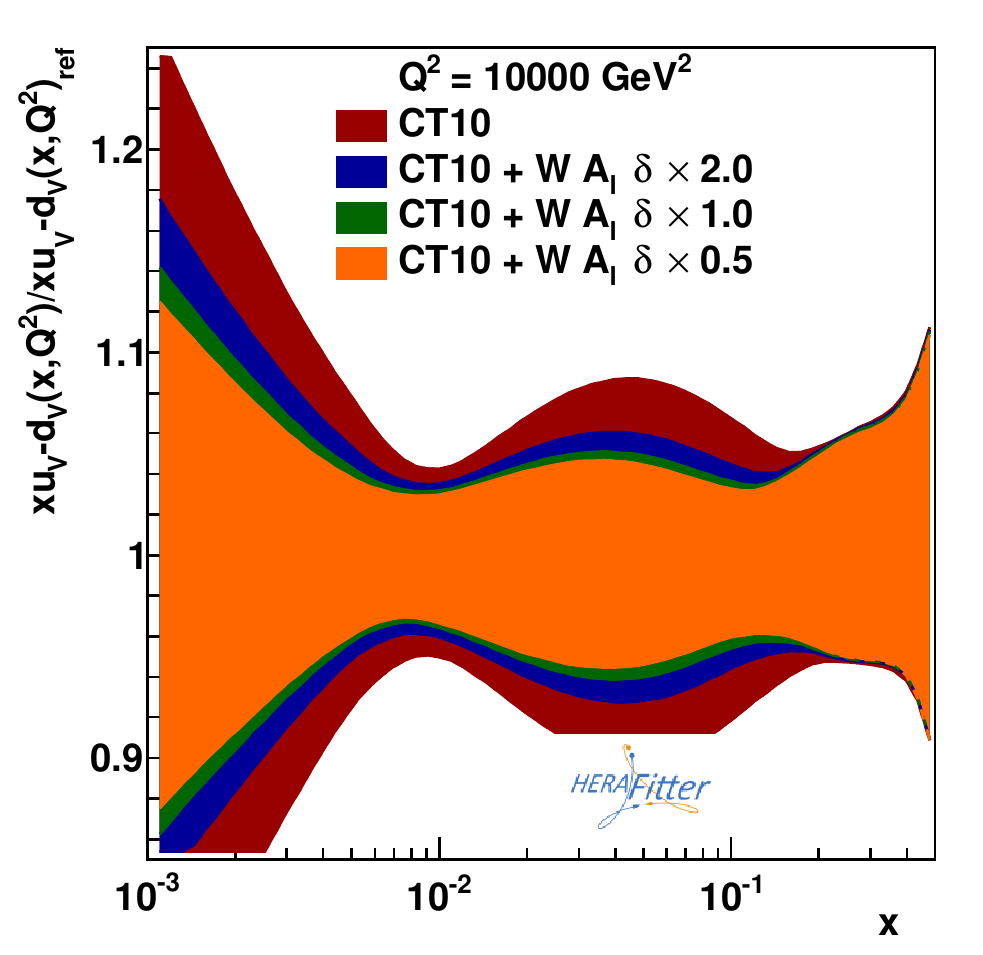}}
    {\includegraphics[width=0.32\textwidth]{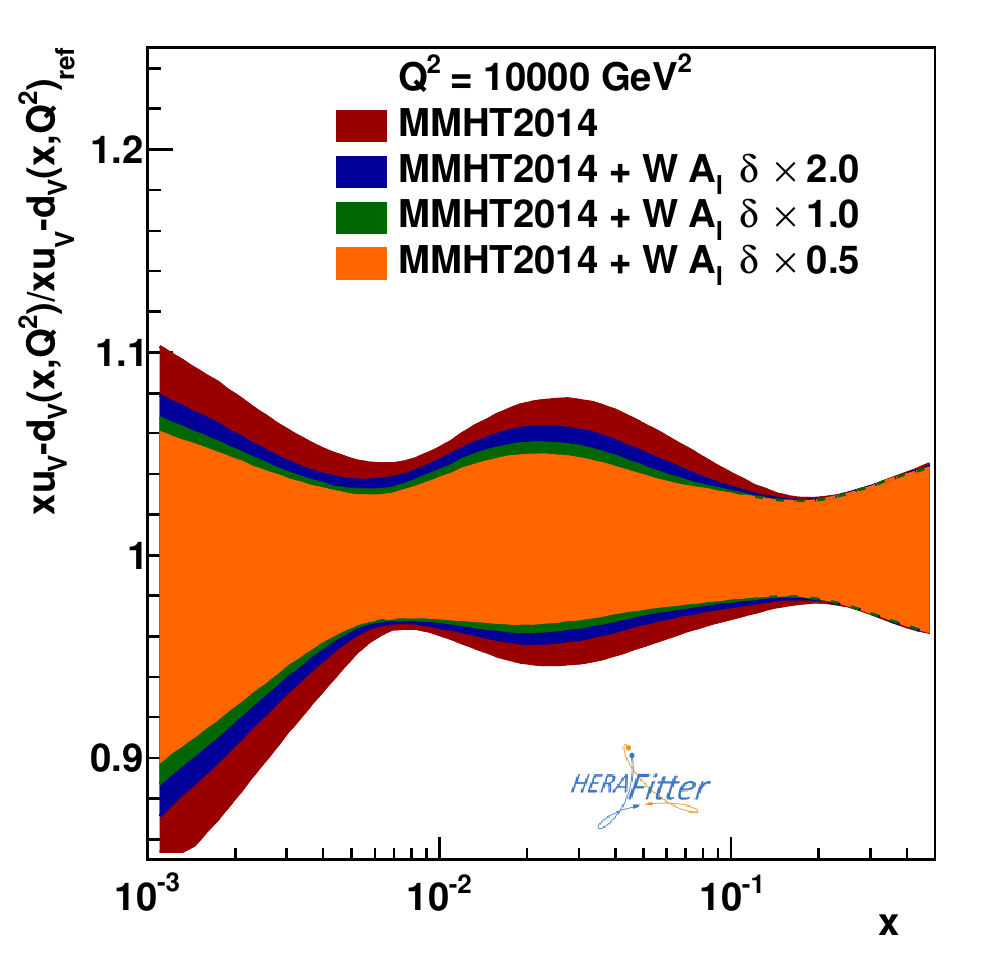}}
    {\includegraphics[width=0.32\textwidth]{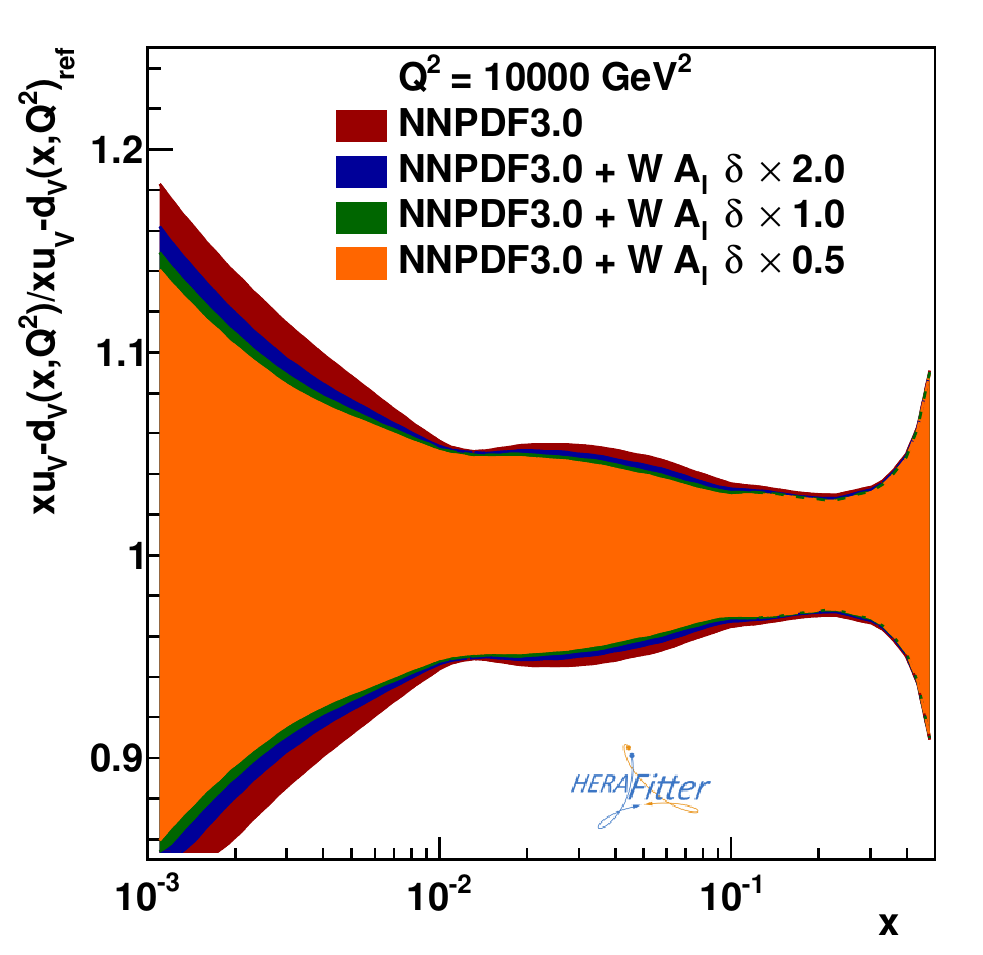}}
  \end{center}
  \caption{\label{fig:wasy} Same as Fig.~\ref{fig:rwz} this time for the
difference between $u_V$ and $d_V$ PDFs,
using the measurement of the $W$ lepton
asymmetry as input to the
  profiling. }
\end{figure}

\begin{figure}[t]
  \begin{center}
    {\includegraphics[width=0.32\textwidth]{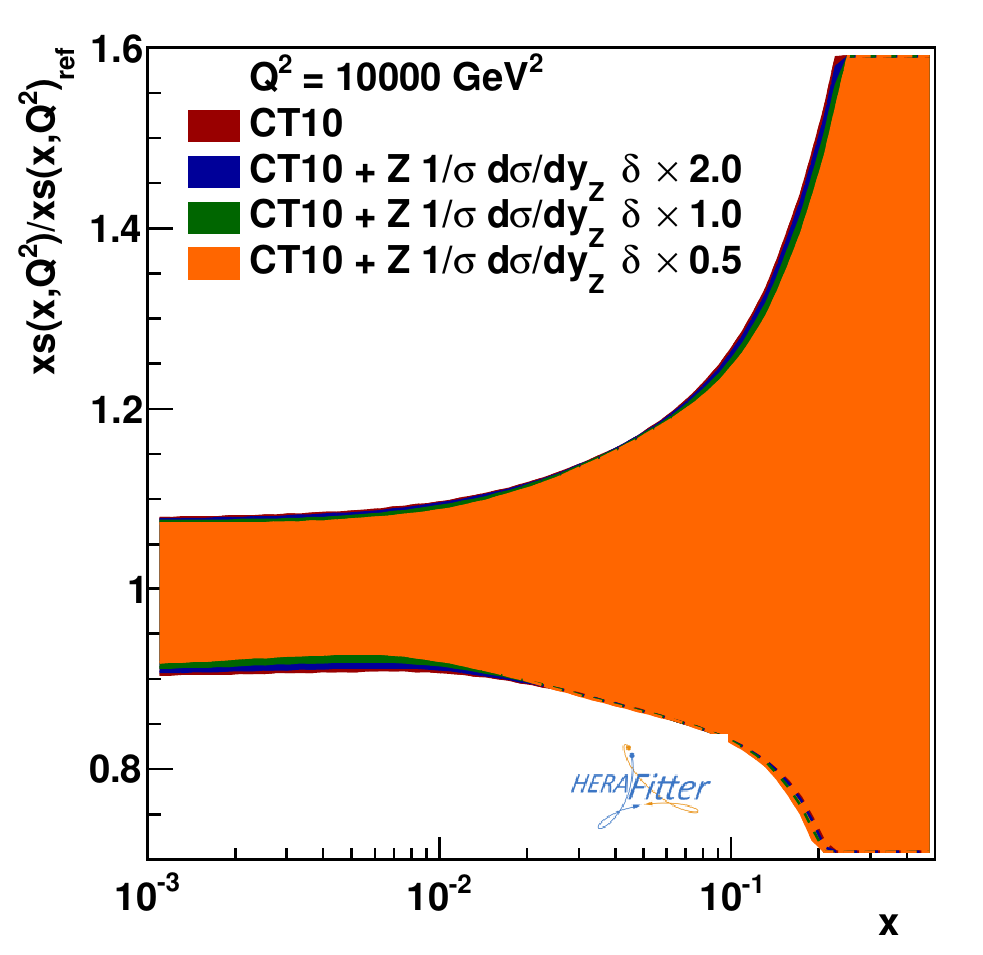}}
    {\includegraphics[width=0.32\textwidth]{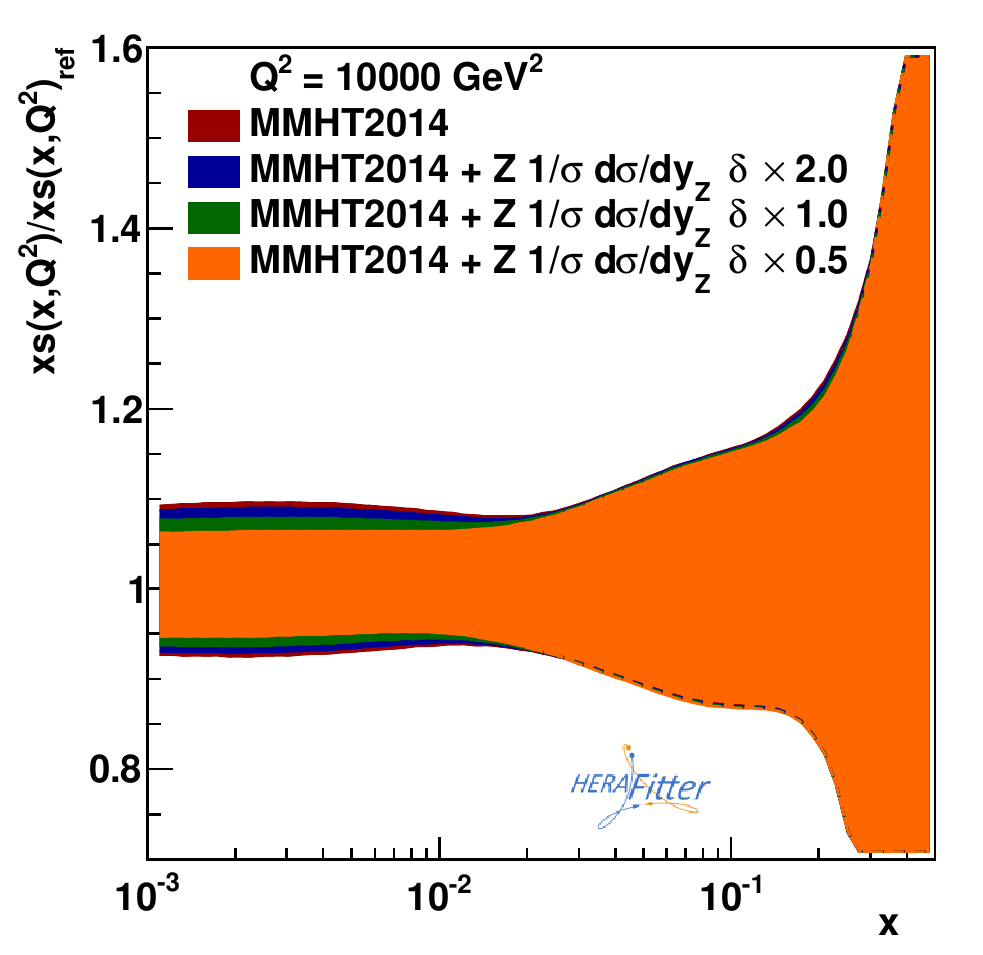}}
    {\includegraphics[width=0.32\textwidth]{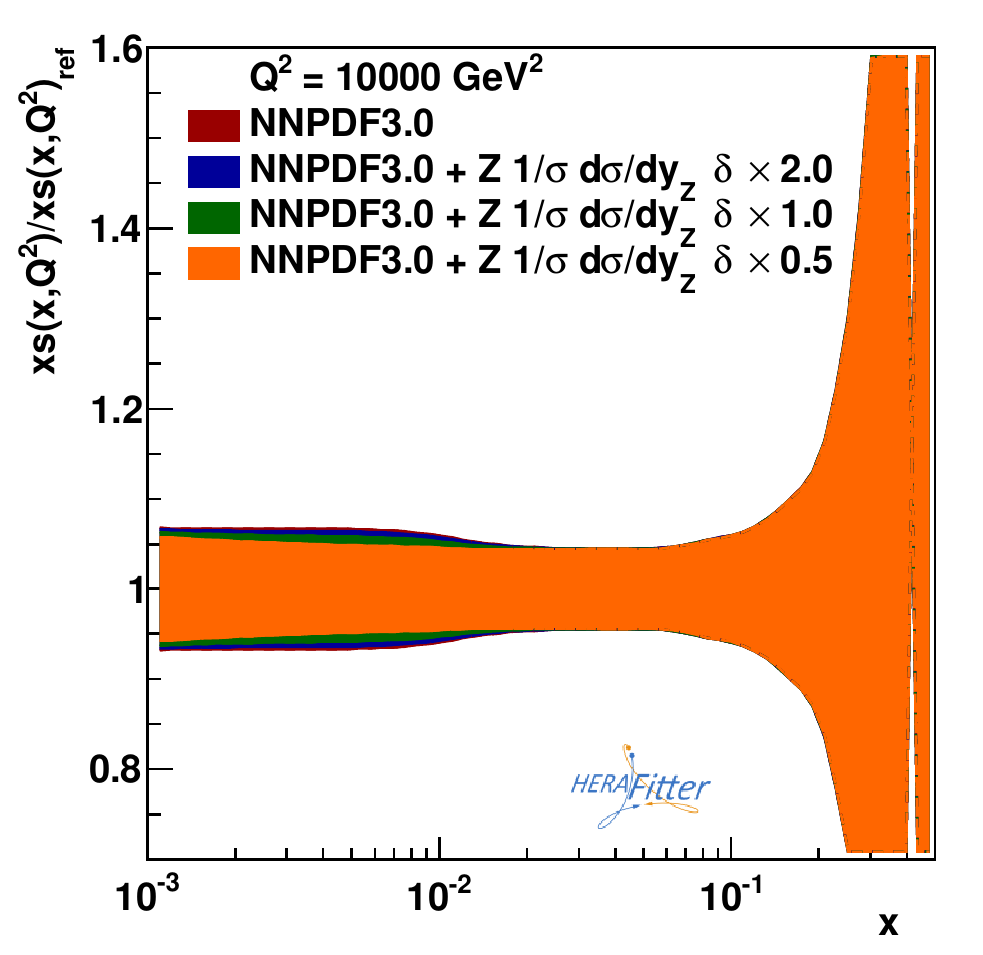}}
  \end{center}
  \caption{\label{fig:zrap}
 Same as Fig.~\ref{fig:rwz} this time for the strange PDF,
 using the measurement of the normalized rapidity
 distributions of $Z$ bosons as input for Run II.
  }
\end{figure}

Firstly, the effect of PDF profiling is studied separately
for each individual pseudo-data set. 
Profiling of different PDF sets show qualitatively similar behavior,
however the size of the constraints differs depending on 
how strongly the published PDF set was constrained by the input data used in the original fit.
For the main comparisons described below, the CT10nnlo set is used.  
It is however important to compare results for several sets. 
The MMHT14 and NNPDF3.0 sets are of particular interest since these two sets include the published Run I data from the LHC experiments.

The  PDF uncertainties are reported at momentum transfers squared $Q^2$ roughly corresponding to the data scale, $Q^2=10^4$~GeV$^2$.
For profiling of each pseudo-data sample the PDF or a combination of PDFs which are most affected by the measurement are reported.
The results are given for the baseline, conservative and aggressive scenarios of the data uncertainties.  

The profiling of the $R_{W/Z}$ pseudo-data has the largest impact on the strange-quark distribution which is shown in Fig.~\ref{fig:rwz}. 
For pseudo-data have maximal impact at $x\sim 0.01$ which agrees with the ATLAS observation~\cite{Aad:2012sb}. 
The uncertainty reduction improves significantly as the data accuracy improves;
with respect to the CT10 set the aggressive scenario leads to close to factor of $2$ reduction. The other PDFs are not affected significantly
by the inclusion of $R_{W/Z}$ data apart from moderate reduction in uncertainty for $\bar{d}$ and $\bar{u}$ distributions.

The profiling of the $R_{\rm t\bar{t}/Z}$ pseudo-data affects the gluon distribution the most, see Fig.~\ref{fig:rttz}.
The uncertainties
are reduced for $x\ge 0.1$ and $x<0.01$ regions. Contrary to the other observables, the difference in pseudo-data accuracy
does not affect the gluon density uncertainty significantly. The other PDF which has notable reduction
in uncertainty at low $x$ is the total light sea, $x\Sigma = x\bar{u} + x\bar{d} + x\bar{s}$. This reduction can be explained
by high degree of correlation between the gluon and sea distributions at low $x$.
Note also that constraints from measurements of
the $t\bar{t}$ cross sections depend strongly on the values
of the top mass and $\alpha_s(m_Z^2)$.

A study was performed to clarify the dependence of PDF uncertainty reduction as a function of the $R_{\rm t\bar{t}/Z}$ pseudo-data uncertainty.
Using the procedure described in Ref.~\cite{Pumplin:2009nk}  the PDFs eigenvectors were re-diagonalised to isolate a linear combination of them which affects the  $R_{\rm t\bar{t}/Z}$ 
observable the most.
For a single measurement such as  $R_{\rm t\bar{t}/Z}$ this procedure returns a single re-diagonalised eigenvector which affects the measurement while others have no impact.
This eigenvector has a significant contribution to the gluon density uncertainty at  $x=0.1$, however it does not saturate the uncertainty band. 
As a consequence, while the eigenvector is constrained progressively as the pseudo-data accuracy increases, the other irreducible uncertainty component prevents from further
improvement in the total gluon density uncertainty. 

The lepton-asymmetry measurement has the largest impact on the difference of the $u$- and $d$-valence distributions, $u_v - d_v$,
which is shown in Fig.~\ref{fig:wasy}. There is a sizable reduction in the uncertainty for $x\sim 0.03$ and
$x<0.003$ kinematic regions which becomes more significant as the pseudo-data accuracy increases. 

The data on $y_Z$ also has largest impact on the strange-quark distribution which is shown in Fig.~\ref{fig:zrap}. The effect is complementary to the
impact of the $W/Z$ cross-section ratio pseudo-data, compared which the reduction of the uncertainty is more concentrated in the small $x<0.01$ region. 
Similarly to $R_{W/Z}$, the data also constrain the $\bar{u}$ and $\bar{d}$ light sea-quark distributions.

It is interesting to notice that the level of uncertainty reduction due to inclusion of the pseudo-data is rather similar for the CT10nnlo and MMHT14 sets 
while it is significantly smaller for the NNPDF3.0 set. 
This behavior can be most likely explained by the difference of input data used in the sets and different level of parameterisation flexibility.

\begin{figure}
  \begin{center}
    {\includegraphics[width=0.32\textwidth]{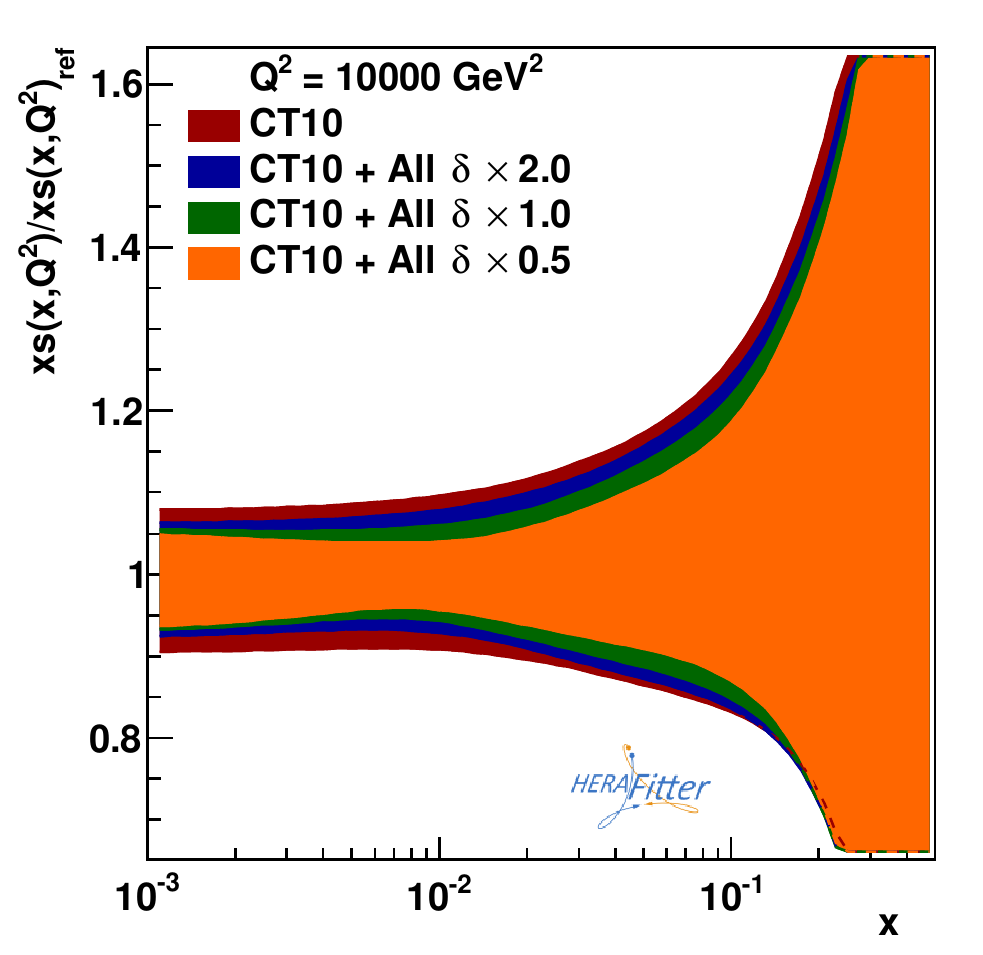}}
    {\includegraphics[width=0.32\textwidth]{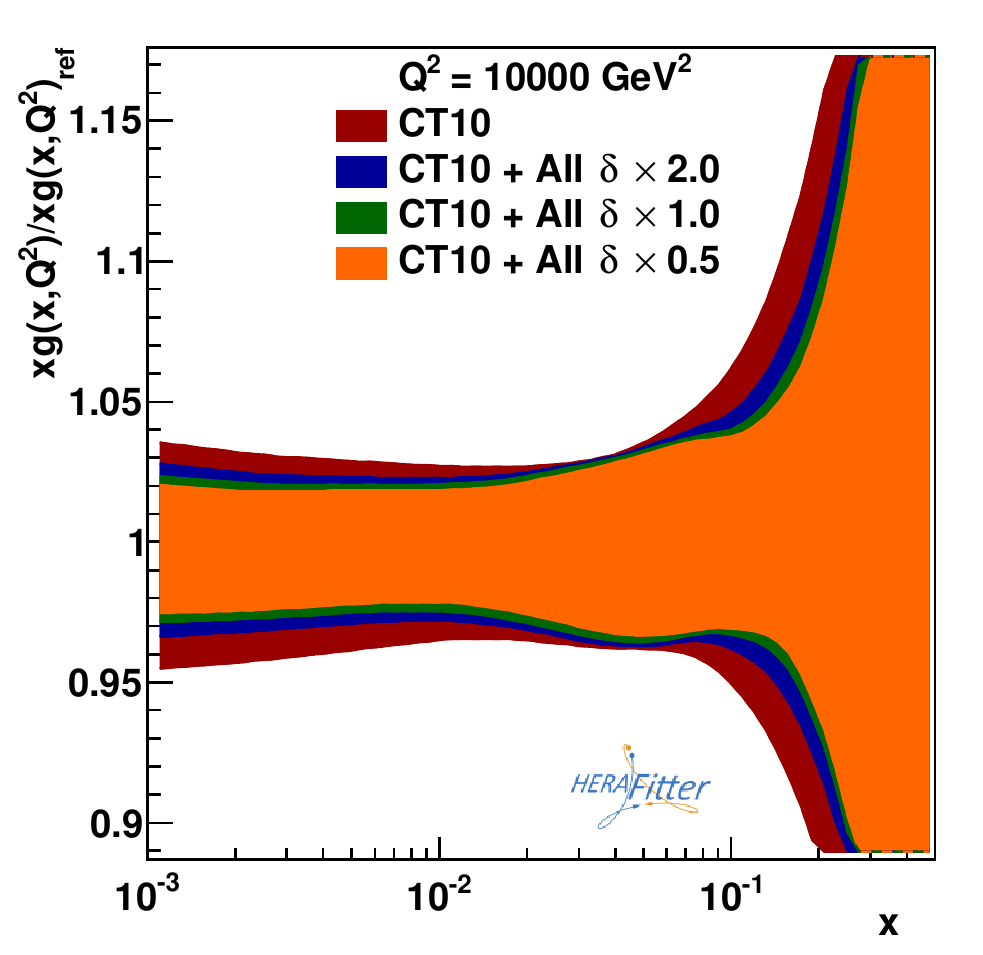}}
    {\includegraphics[width=0.32\textwidth]{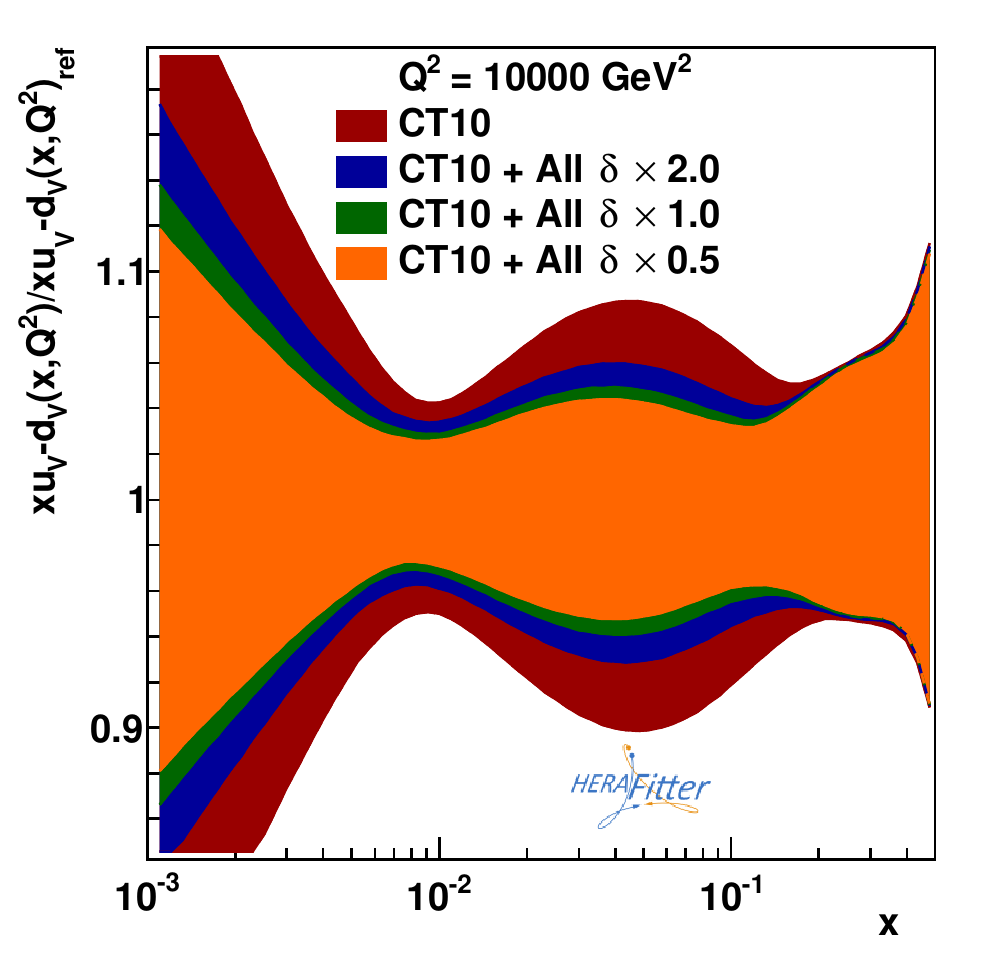}}
  \end{center}
\caption{\label{fig:profall}
Relative uncertainty of the strange-quark (left), gluon (center) 
and $u_v-d_v$ (right) distributions as a function of 
$x$ for $Q^2=10^4$~GeV$^2$ estimated
based on CT10nnlo PDF set. The outer uncertainty band corresponds to the original
PDF uncertainty. The embedded bands represent results of the PDF profiling using
the complete set of observables considered in this exercise:
$R_{W/Z}, R_{\rm t\bar{t}/Z}, A_{\ell}$ and $y_Z$ pseudo-data at $13$~TeV.
The various bands
correspond to (from outermost to innermost band) conservative, baseline, aggressive model of the data uncertainties. 
}
\end{figure}

Finally, all the pseudo-data samples are profiled together in a
simultaneous
 fit.
Fig.~\ref{fig:profall} shows result of this profiling for the CT10nnlo sample and for the most affected PDF distributions. 
The simultaneous fit yields to quantitatively similar reduction of PDF uncertainties compared to the fits to the individual observables.
This is not unexpected since with exception of $R_{W/Z}$ and $y_Z$, the observables are sensitive to different PDF combinations and they
are not correlated experimentally.   

To summarize, the $\sqrt{s}=13\tev$ LHC data will make a contribution for reduction of PDF uncertainties. 
Measurements of the cross-section ratios of the $W$- to $Z$-boson and $t\bar{t}$ to $Z$-boson production, 
$W$-boson lepton asymmetry and $Z$-boson rapidity distribution can be used to constrain strange-quark, gluon and valence-quark
distributions. 
Additional constraints from $13\tev$ LHC data will be provided from more differential distributions,
provided the statistical and systematic experimental uncertainties can
be kept under control.
The results of Fig.~\ref{fig:profall} also nicely illustrate the advantages
of achieving a reduction of experimental uncertainties in terms
of improved PDF constraints.

\subsection{Projected impact of the LHC inclusive jet data}

Single-inclusive jet production,
a key benchmark process at hadron-hadron colliders,
proceeds through multiple parton scattering channels.
Under LHC conditions, much of the PDF uncertainty of inclusive jet cross sections arises from the gluon PDF; hence they can constrain $g(x,Q)$
in a wide range of $x$. 
Potential impact of future LHC jet cross sections has been recently examined in the context of the CTEQ-TEA global analysis.
The CTEQ series of PDFs include single-inclusive jet cross sections from Tevatron D0
and CDF collaborations~\cite{Abazov:2010fr,Aaltonen:2008eq}, and,
starting with CT14, from ATLAS~\cite{Aad:2011fc}
and CMS~\cite{Chatrchyan:2012bja}.
Sect. 2 of ~\cite{Dulat:2015mca}
shows that the PDF uncertainty of inclusive jet data is correlated
with $g(x,Q)$ at $x\gsim 0.07$ at CDF and at
$x\gsim 0.005$ at ATLAS; i.e. the reach in $x$ of 
jet production is extended at least by an order of magnitude at the LHC.

\begin{figure}[t]
  \begin{center}
    \includegraphics[width=0.49\textwidth]{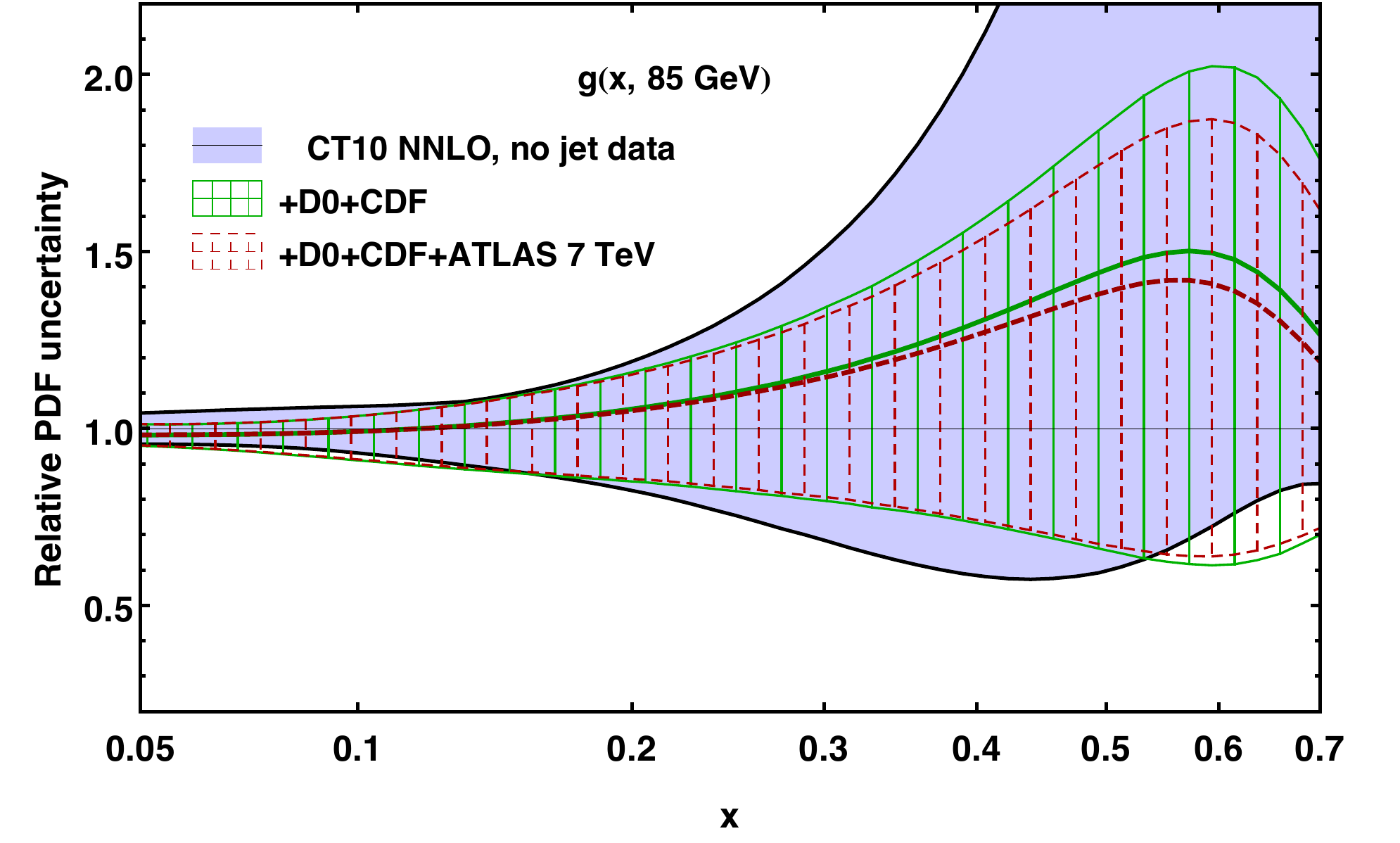}
    \includegraphics[width=0.49\textwidth]{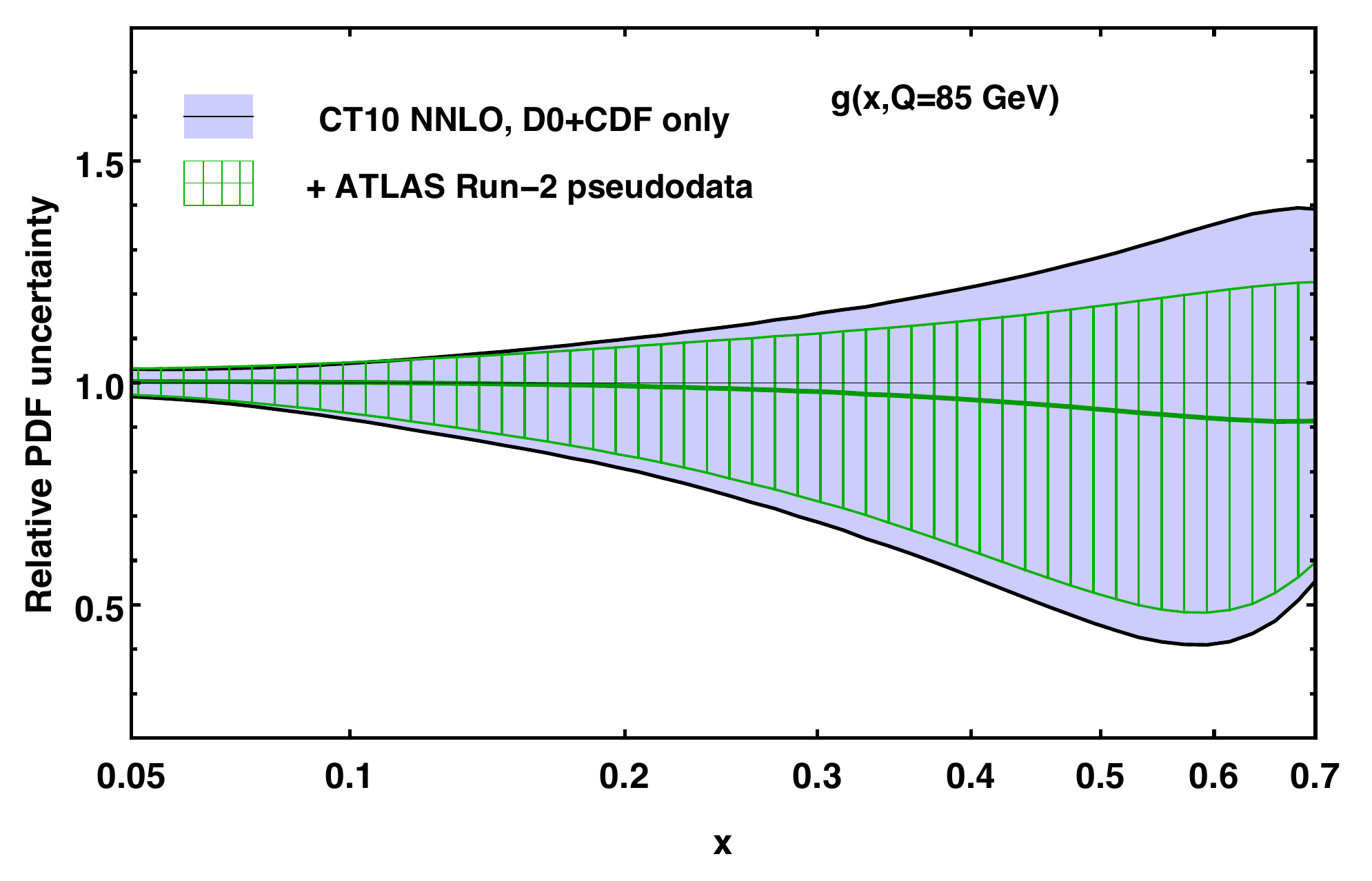}
  \end{center}
  \vspace{-0.5cm}
\caption{\label{fig:cteq-jet1} \small
  Left plot: Gluon PDF uncertainties at 90\% C.L. for the CT10-like
  fits without any jet data included, compared with the fits
  with the Run II D0 and CDF jet data, and with the Run I ATLAS jet data included.
  Right plot: same comparison, now for the fits including the Run 2 D0
  and CDF jet data, and in addition  with the ATLAS Run II simulated pseudo-data.
}
\end{figure}

Let us illustrate how the gluon PDF changes upon including various data sets on inclusive jet production, using the framework of the
CT10 NNLO QCD analysis~\cite{Gao:2013xoa} as an example.
We start by including all experiments used in CT10 NNLO, except for jet experiments, and assuming the world-average value of the QCD coupling constant,
$\alpha_s(m_Z)=0.118$.
The 90\% confidence level error PDFs are found by following the
Hessian approach, as summarized in~\cite{Lai:2010vv}.
Single-inclusive jet cross sections are evaluated 
using fast interpolation interfaces~\cite{Wobisch:2011ij,Carli:2010rw}
to the theoretical calculation at NLO in QCD~\cite{Ellis:1992en}.
We set the factorization and renormalization scales equal to $p_T$
of the jet in each experimental bin, which minimizes both the residual scale
dependence at NLO~\cite{Gao:2012he,Ball:2012wy} and the
NNLO/NLO correction~\cite{Carrazza:2014hra}
in the partial NNLO calculation in the $gg$ sub-channel~\cite{Currie:2013dwa,Ridder:2013mf}. Thus, the unknown NNLO corrections are believed to be inconsequential for the present study. 

We include the full ATLAS data sample (7 TeV, 37 pb$^{-1}$, cone size $R$ = 0.6). Similar outcomes are obtained with the ATLAS data set for $R$ = 0.4. As an option, we also estimate the possible impact of the NLO scale dependence and missing NNLO contributions on $g(x,Q)$ using a phenomenological approach that is similar to the ones proposed
in~\cite{Cacciari:2011ze,Olness:2009qd}. 
This is done by treating additional theoretical uncertainties as
correlated systematic errors and including corresponding columns in
the correlation matrix of each experiment, in addition to the usual
experimental systematic errors. We notice, in particular, that the estimated theoretical uncertainties do not exceed the experimental uncertainties in all bins on the ATLAS data set. 

If we compare the uncertainties on the gluon PDF in CT10-like fits without and with the Tevatron Run II and ATLAS  jet data, and not accounting for theoretical uncertainties, the resulting 90\% C.L. error bands on $g(x, Q = 85~{\rm GeV})$ in the range 0.05-0.7 are shown in the left sub-panel of Fig.~\ref{fig:cteq-jet1}.
The jet data modify the central gluon PDF and, as seen in the figure, greatly reduces the PDF uncertainty.
The early ATLAS jet data do not improve the constraints on the gluon PDF much, since they have large experimental errors.

To estimate effects of future LHC jet measurements,
we introduce two new pseudo-data sets
with the same kinematics as the ATLAS measurement~\cite{Aad:2011fc}, and $R=0.6$ and $0.4$.
We assume the statistical errors in the measurements to be reduced by a factor of 20, and the jet energy scale errors, which dominate the experimental systematic errors, to be reduced by a factor of 3.
The central values of the pseudo-data sets are generated randomly based on the theoretical predictions from one PDF set.
The right sub-panel of Fig.~\ref{fig:cteq-jet1} shows the effects that the pseudo-data sets have on the gluon PDF. The PDF uncertainty in this case is reduced by about 20 percentage points in the large-$x$ region.
At moderate $x \sim 0.05$, relevant to central Higgs boson production, the reduction in the uncertainty is less pronounced.

\begin{figure}[t]
  \begin{center}
    {\includegraphics[width=0.49\textwidth]{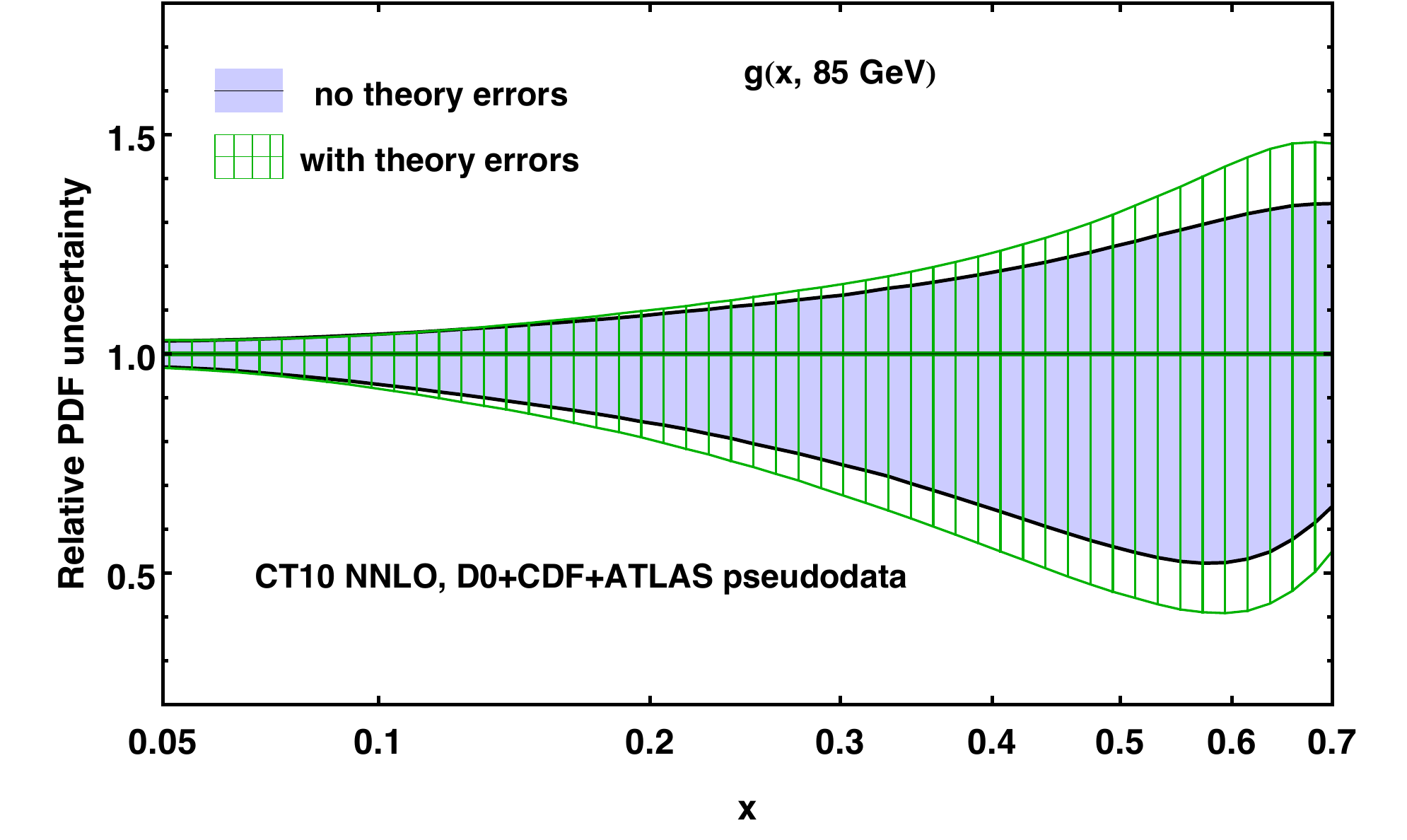}}
    {\includegraphics[width=0.49\textwidth]{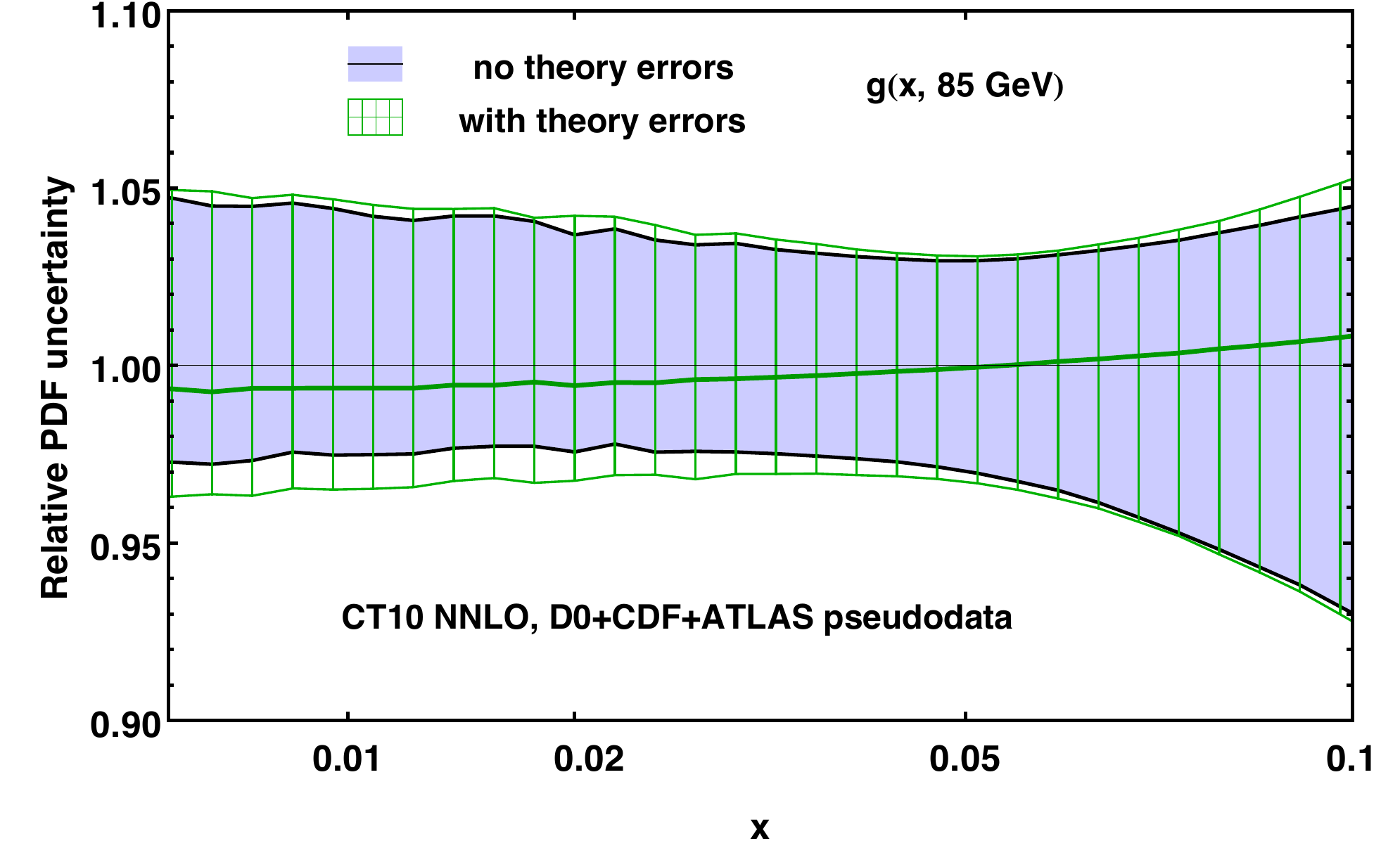}}
  \end{center}
   \vspace{-0.5cm}
\caption{\label{fig:cteq-jet2}
  \small Gluon PDF uncertainties at 90\% C.L. for the fits with and without theoretical errors.
  Left plot: large-$x$ region. Right plot: intermediate-$x$ region.
}
\end{figure}

Next, we add four correlated shifts in theoretical predictions that are allowed by NLO theoretical uncertainties according to our method.
The impact is illustrated in Fig.~\ref{fig:cteq-jet2},
comparing $g(x,Q)$ in the fits with and without the theoretical errors included.
With the additional correlated shifts due to theoretical errors, we obtain a slightly harder best-fit gluon in the large-$x$ region.
Most importantly, the PDF uncertainty increases by up to 15 percentage points at large $x$, and
by up to 1 percentage point at moderate $x$ (in the Higgs production region).
Needless to say, these preliminary estimates of the theoretical uncertainty at NLO (dominated by QCD scale dependence) will likely be reduced once the
full NNLO computation is completed.

To summarize, using pseudo-data sets of inclusive
jet measurements at the LHC, we have estimated the potential for reduction of the uncertainty in the gluon PDF upon inclusion in the future. Although the ATLAS pseudo-data sets in this exercise correspond to $\sqrt{s} = 7\mbox{ TeV}$, similar constraints are expected from the measurements at 8 TeV. Measurements at 13 TeV will probe the gluon PDF at even smaller $x$; their prospects will still be dependent on improvements in experimental systematic errors, as at the other two energies.

\section{On the presentation of LHC data for PDF fits}
\label{sec:recommendations}

We conclude this report with a number of practical
suggestions concerning the presentation and delivery
of LHC measurements to be used in global PDF analysis:

\begin{itemize}

\item It is important that measurements are provided with the
  complete information on correlated systematic uncertainties.
  While this has not always been the case in pre-LHC
  experiments, it is now common practice by ATLAS, CMS and LHCb,
  and thus it should be encouraged to continue.

  The preference is to make available the full breakdown of individual
  sources of experimental systematic uncertainties, in terms
  of nuisance parameters.
  If this is not possible, the full correlation matrix
  of the experimental data needs to be be provided.

\item It is common practice that experimental data is made
  publicly available through {\tt HepForge}, and it would be beneficial
  for all parties if this practice is continued.
  When available, also the corresponding {\tt Rivet} analysis
 could be made public there.

 \item Whenever possible, the LHC experiments should try to 
  provide information on cross-correlations between individual
  datasets, as well as between different data-taking years when
  available.
  These cross-correlations can be both of statistical and
  of systematic origin.
  This is necessary to consistently include at the same time in a common
  global analysis measurements like inclusive jets and dijets from
  the same dataset,
  which have both statistical and systematic correlations.

 \item In addition, it would be beneficial for PDF analysis
  if the LHC experiments could agree on which systematic
  uncertainties are correlated between ATLAS, CMS and LHCb,
  even partially, since this maximises the constraints
  that can be extracted from the LHC measurements.

\item The LHC experiments should give clear indications
  of which measurements are the most updated
  ones and should then be used on PDF
  analysis, and which ones are superseded and should thus not
  be used.
  This information can be make public for example
  in the analysis group webpages for each experiment.

  \item It would be advantageous if the LHC collaborations could
  agree on common settings for their PDF-sensitive measurements,
  for instance the jet radius $R$ for jet measurements, since
  this streamlines the comparisons between different measurements
  and their impact on PDFs.

\item With a similar motivation, it would be beneficial that the
  LHC experiments agree on a common set of observables and
  distributions to deliver in PDF-sensitive measurements.
  For instance, for immediate use in PDF fits,
  parton-level corrected data is needed, or, alternatively,
 experiments could supplement hadron-level data with the corresponding
  correction factors.

  This said, future fits should move towards using hadron level data,
 which is closer to the actual measurements and that cancels
 the hadron-to-parton modelling ambiguities.
 This is especially important for complicated processes like
 $W$+charm, where assessing the
 theoretical uncertainties in the hadron-to-parton correction factors
 is challenging.
 Note that the technology to include hadron-level theory
 calculations in PDF fits is already available~\cite{amcfast}.

\item Fully differential measurements in the fiducial
  regions are typically preferred over integrated cross-sections, due to the
  theoretical uncertainty induced by the extrapolation from the fiducial
  to the full phase space.
  This is now a common practice for both experiments, and should be
  encouraged to continue in future measurements.

  If presenting fiducial
  measurements is not possible, it would be important to provide the conversion factors used to extrapolate to the
  full phase space.
  In this case, as well as the conversion factors,
  additional useful information would be provided by the
   systematic uncertainties associated with the PDF dependence in this conversion,
   if it they are  at all significant.

\item When theory calculations are generated explicitly for
  PDF comparisons in a LHC analysis, using codes
  such as {\tt APPLgrid} or {\tt FastNLO}, it would
ease their inclusion in global PDF fits if the 
  corresponding
  fast grids could also be released together with the experimental
  data.
  This would also apply to other ingredients of the theoretical
  calculations, such as hadron-to-parton corrections or
  NNLO $K$-factors.
  This has already been done for a number of important analysis,
  and it will be important to ensure that releasing theory grids
  becomes a standard practice in the future.

\item  Ideally, it would be useful if experiments could  agree on the
  usage of common theory tools for constructing fast grids, such that
  theory calculations corresponding to
  measurements of one specific type
  are delivered as fast grids with a common format.

In the cases where different tools are used, for example for
jet production measurements, it would be important to ensure
that all theory settings, such as factorization and renormalization
scales, or heavy-quark schemes, are spelled out in detail
in the corresponding publications, allowing for
a posteriori comparison between the various theory codes.

\item Whenever possible, it would
  be useful to agree on a common treatment of theory corrections to be applied to the
data: for example, all gauge-boson
production data being presented with the same treatment of
final-state QED radiation and electroweak corrections.

  \end{itemize}

These suggestions should be helpful to streamline the process of adding new
LHC measurements into PDF fits, maximizing the constraining power of the data
and minimizing the theoretical and methodological uncertainties associated
to PDF-sensitive measurements.

\section{Summary and outlook}
\label{sec:outlook}

In this report we have summarized the constraints on PDFs that have been obtained from LHC measurements during Run I.
The impressive wealth and quality of PDF-sensitive measurements at the LHC Run II have been summarized in Tables~\ref{table:runIoverviewATLAS} 
for the ATLAS, Table~\ref{table:runIoverviewCMS} for the CMS and Table~\ref{table:runIoverviewLHCb} for the LHCb
collaborations.
Many of these measurements are already included in recent global PDF fits to date, where they provide
constraints in a wide range of $x$ and PDF flavours.
It is especially remarkable that the
high energy of the LHC also allows us to introduce new types of the processes in the PDF analyses for the first time,
from $t\bar{t}$
production and $W$+$c$ production to high-mass Drell-Yan data.

We have also reviewed the prospects for PDF studies with the recently started 13 TeV Run II data-taking.
A number of improvements are foreseen, thanks to the increase
on center-of-mass energy and in the integrated luminosity,
especially for those measurements that at Run I were limited
by statistical or statistics-related systematic uncertainties.
We have also performed a quantitative estimate of the impact on
PDFs based on Run II pseudo-data for $W$, $Z$ and $t\bar{t}$ production
using the profiling method.
Our results show that PDF uncertainties in a number of PDF flavors can be reduced with 13 TeV measurements
of these processes, and emphasize the importance of reducing  the systematic uncertainties in PDF-sensitive measurements.

This report addresses both the experimental collaborations,
in order to identify their priorities for PDF-sensitive measurements
at Run II, as well as the PDF fitting groups to have a clear
perspective of which measurements are already available and which ones
will also become available in the near future.
Exploiting the full potential of Run II data for PDF constrains
is essential for the LHC physics program for the next years,
and in turn feeds into many other analysis like improved determination
of the Higgs couplings or of precision SM measurements like the
$W$ mass.

In this report we have concentrated only on the experimental
constraints provided by the LHC data.
Equally important for PDF determinations is to use
state-of-the-art theoretical calculations, especially
exploiting the recent developments in NNLO calculations
for LHC processes.
The use of higher-order perturbative calculations is essential
for reducing sources of theoretical uncertainties in PDF fits,
which are presently not even estimated.
In this respect, a careful benchmarking of NLO and NNLO codes
for processes relevant for PDF studies would be certainly
interesting.
A collaborative effort is of particular importance for the progress of the benchmarking exercise, which was performed in the past~\cite{Butterworth:2014efa,Ball:2012wy,LHhq,Gao:2012he}
and is necessary to understand and reduce the differences between the results of different PDF groups. 

\section*{Acknowledgments}
We are grateful to all the participants of the various PDF4LHC meetings that
stimulated many enthusiastic discussions on the
 measurements and methodological developments reviewed in this report.
We are grateful to Jos\'e Ignacio Latorre and
all the {\it Pedro Pascual Benasque Center for Science} team
for their invaluable support during the workshop {\it Parton Distributions
  for the LHC} in February 2015, where many of the authors of
this paper were present.

The work of A.~A. was supported by the DOE contract No.~DE-AC05-06OR23177,
under which Jefferson Science Associates, LLC operates Jefferson Lab,
and by the DOE contract DE-SC008791.
The work of S.~Farry is supported by a research fellowship from the Royal Commission for the Exhibition of 1851.
The work of S.~Forte is supported in part by an Italian PRIN2010 grant
and by a European Investment Bank EIBURS grant.
The work of L.~H.-L. is supported by the
Science and Technology Facilities Council (STFC) for
via the grant award ST/L000377/1.
The research of J.~G. in the High Energy Physics Division at Argonne 
is supported by the U. S. Department of Energy, High Energy Physics,
Office of Science, under Contract No. DE-AC02-06CH11357.
The work of P.~N. is supported by the U.S. Department of Energy under grants DE-SC0003870 and DE-SC0013681.
The work of J.~R. is supported by an STFC Rutherford Fellowship ST/K005227/1
and by an European Research Council Starting Grant "PDF4BSM".
The work of R.~T.
is supported partly by the London Centre for Terauniverse Studies (LCTS),
using funding from the European Research Council via the Advanced
Investigator Grant 267352.
R.~T. would also like to thank the
Science and Technology Facilities Council (STFC) for support via grant
awards ST/J000515/1 and ST/L000377/1, and the IPPP, Durham, for
the award of a Research Associateship.
%

\providecommand{\href}[2]{#2}\begingroup\raggedright\endgroup

\end{document}